\documentclass[12pt]{article}
\usepackage{graphicx}
\usepackage[utf8]{inputenc}
\usepackage{mathtools}
\usepackage{graphicx,psfrag,epsf,color}
\usepackage{amsmath,amssymb,amsfonts}
\usepackage{array}
\usepackage{cite}
\usepackage{float}
\usepackage{enumitem}
\usepackage{appendix}
\usepackage{bm}
\usepackage{ascii}
\usepackage{dsfont}
\usepackage{caption}
\usepackage{subcaption}
\usepackage[colorlinks]{hyperref}
\usepackage[dvipsnames]{xcolor}
\hypersetup{pageanchor=false,linkcolor=NavyBlue,citecolor=NavyBlue,urlcolor=RoyalPurple}

\renewcommand{\Im}{\textrm{Im}}
\renewcommand{\vec}[1]{\bm{#1}}
\newcommand{\cor}[2]{#1\langle #2 #1\rangle}
\newcommand{\p}{\partial}
\newcommand{\vp}{\varphi}
\newcommand{\ve}{\varepsilon}
\newcommand{\der}{\textrm{d}}
\newcommand{\Lo}{\mathcal{O}}

\newcommand{\Z}{\mathbb{Z}}

\newcommand{\tmu}{\tilde\mu}

\newcommand{\un}{\mathds{1}}
\newcommand{\secref}[1]{Sec.~\ref{#1}}
\newcommand{\figref}[1]{Fig.~\ref{#1}}
\newcommand{\appref}[1]{App.~\ref{#1}}
\newcommand{\nn}{\nonumber}
\renewcommand{\log}{\ln}

\numberwithin{equation}{section}

\begin{document}
\allowdisplaybreaks

\begin{titlepage}

\begin{flushright}
{\small
TUM-HEP-1599/26\\
CERN-TH-2026-084\\
15 April 2026\\
}
\end{flushright}

\vskip1cm
\begin{center}
{\Large \bf Quantum correction to the diffusion term in\\[0.01cm]
stochastic inflation from composite-operator matching\\[0.2cm] 
in Soft de Sitter Effective Theory}
\end{center}
  \vspace{0.5cm}
\begin{center}
{\sc Martin~Beneke,$^{a}$ \sc Patrick~Hager,$^{b}$ and Andrea~F.~Sanfilippo$^{c}$} 
\\[6mm]
{\it ${}^a$Physik Department T31,\\
James-Franck-Stra\ss e~1,
Technische Universit\"at M\"unchen,\\
D--85748 Garching, Germany}\\[0.2cm]
{\it ${}^b$CERN, Theoretical Physics Department, CH-1211 Geneva 23, Switzerland}\\[0.2cm]
{\it ${}^c$Departamento de F\'isica Te\'orica y del Cosmos, Universidad de Granada,}\\
{\it Campus de Fuentenueva, E–18071 Granada, Spain}
\end{center}
\vskip1cm

\begin{abstract}
\noindent 
In the framework of Soft de Sitter Effective Theory (SdSET), 
the Fokker-Planck equation for the late-time dynamics of the massless minimally coupled scalar field and its extension to the Kramers-Moyal equation are obtained from operator mixing of composite operators of the effective superhorizon field. We construct the formalism for composite-operator renormalisation, mixing and matching in dimensional regularisation, allowing for computations beyond the leading order. The general formalism is illustrated in free SdSET, which already features non-trivial structures including the well-known diffusion coefficient for stochastic inflation. As explicit examples in the interacting theory, we renormalise the one-loop bispectrum and the two-loop one-point function of the composite operator $\vp_+^2$, and match them onto their full-theory counterparts. These results allow us to determine the next-to-leading order (two-loop) correction to the diffusion term of the Fokker-Planck equation of stochastic inflation for the first time.
\end{abstract}

\end{titlepage}

\pagenumbering{roman}
{\hypersetup{hidelinks}
\pdfbookmark[1]{Contents}{ToC}
\setcounter{tocdepth}{2}
\small  \tableofcontents}
\vspace{6mm}
\newpage
\pagenumbering{arabic}


\section{Introduction}

Perturbatively computed correlation functions of massless, minimally coupled scalar fields in de Sitter (dS) space suffer from infrared (IR) divergences due to the accumulation of superhorizon modes.
At late times, these effects manifest themselves through secularly growing terms containing the scale factor, indicating a breakdown of fixed-order perturbation theory, which can be traced to the non-existence of a dS-invariant vacuum state in the {\em non-interacting} theory \cite{Allen:1985ux}. As the resolution of these difficulties is tied to interactions and non-perturbative effects, such as ``dynamical mass generation'' observed in Euclidean de Sitter space~\cite{Rajaraman:2010xd,Beneke:2012kn}, a systematic treatment of late-time correlation functions requires a framework that describes the dynamics of long-wavelength modes.

A long-standing proposal to capture these infrared effects is the stochastic formalism~\cite{Starobinsky:1982ee,Starobinsky:1986fx,Starobinsky:1994bd}.
In this approach, the long-wavelength physics 
of one-point expectation values $\langle \phi^n(t,\vec{x})\rangle$ 
is generated by a probability distribution $P(t,\varphi)$ obeying the Fokker-Planck equation 
\begin{equation}
\frac{\p}{\p t}P(t,\vp)=\frac{1}{3H}\frac{\p}{\p\vp}\Big[V'(\vp)P(t,\vp)\Big]+\frac{H^3}{8\pi^2}\frac{\p^2}{\p\vp^2}P(t,\vp)\,.
\label{eq::FP}
\end{equation}
In the present work, we focus on the massless scalar field with $\frac{\kappa}{4!}\phi^4$ interaction.
In perturbation theory in $\kappa$, the IR-divergent expectation values receive corrections of order $(\kappa t^2)^n$ when evaluated with a comoving IR regulator, which exhibit the secular logarithms as $\eta\to 0$, if $t=-1/H\times \ln(-H\eta)$ is expressed in terms of conformal time $\eta$.
The Fokker-Planck equation of the stochastic formalism has been derived in various other ways \cite{Prokopec:2007ak,Garbrecht:2014dca,Moss:2016uix,Collins:2017haz,Mirbabayi:2019qtx,Baumgart:2019clc,Andersen:2021lii,Cespedes:2023aal} and resums the leading logarithms in conformal time to a stationary, IR finite value as $t\to \infty$, successfully describing the late-time behaviour of correlation functions in a variety of cosmological settings. 
The Fokker-Planck equation is non-perturbative in nature and can be regarded as the leading term in an expansion in $\sqrt{\kappa}$ rather than the original $\phi^4$ coupling $\kappa$. Its structure remains valid at next-to-leading order  \cite{Gorbenko:2019rza,Mirbabayi:2020vyt,Cohen:2021fzf} subject only to the replacement of the classical potential $V(\varphi)$ by an effective one. Nevertheless, its systematic extension beyond leading late-time logarithms requires clarification. 

As shown in~\cite{Cohen:2020php,Cohen:2021fzf,Cohen:2021jbo} the stochastic description admits a natural interpretation within the framework of Soft de Sitter Effective Theory (SdSET) proposed in these articles, an effective description that isolates and controls the dynamics of superhorizon modes. 
In this formulation, the long-wavelength physics is described by effective fields whose composite operators encode the late-time dynamics.
A key observation of~\cite{Cohen:2020php} is that the Fokker-Planck equation can be understood as the renormalisation-group equation (RGE) governing the mixing of composite operators of the superhorizon field $\varphi_+$ under renormalisation. 
In particular, the time-dependence associated with large logarithms of the scale factor $a(t)$ is reinterpreted as RG flow in the effective theory.
Consequently, in the subsequent work \cite{Cohen:2021fzf} it was argued that the Fokker-Planck equation is only the next-to-leading-order truncation of the more general Kramers-Moyal equation in a perturbative expansion in $\sqrt{\kappa}$. 
In \cite{Cohen:2021fzf}, the structurally new effects at next-to-next-to-leading order 
in powers of $\sqrt{\kappa}$ were then computed by determining the pole terms of correlation functions involving insertions of~$\vp_+^n$ composite operators. 

Given these important conceptual insights, a systematic framework for the renormalisation of composite operators in SdSET and for determining their anomalous dimensions beyond their leading non-vanishing terms is highly desirable. Beyond leading order in $\kappa$, the full-theory operators $\phi^n$ and their SdSET counterparts $\vp^n_+$ are expected to differ due to short-distance effects which enter in the form of operator-matching coefficients that relate them. 
Since physical observables are defined in terms of renormalised full-theory correlation functions, a consistent extension of the stochastic description beyond the leading order requires a precise specification of the matching procedure between full-theory and effective composite operators and the resolution of renormalisation-scheme dependencies.

The purpose of this paper is to address these points. 
In \secref{sec:prerequisites} we collect required expressions for the SdSET action, initial-condition functional, and  RGEs. Then, building on the formalism set up in~\cite{Beneke:2026rtf}, we construct  in \secref{sec:EFTopmatch} the general framework for the renormalisation of composite effective operators of the form $\vp^n_+$ in dimensional regularisation, as well as for the matching onto their renormalised full-theory counterparts. We illustrate the consequences of the dimensionlessness of the field $\vp_+$, which leads to non-trivial operator renormalisation and mixing already in the free theory, in \secref{sec:freemix}. In this language, the diffusion term with coefficient $D=H^3/(8 \pi^2)$ in \eqref{eq::FP} arises from one-loop operator mixing in the free theory, which explains the absence of a coupling factor. 
Focusing on the effective operator $\vp^2_+$, we then show how to renormalise and match it onto the corresponding full-theory operator $\phi^2$ at the one-loop order in \secref{sec::22q0EFT} and at two loops in \secref{sec::vp20}. 
These computations allow us to determine three of the required operator-matching coefficients relating the renormalised full-theory and SdSET operators. These coefficients are expected to contribute to resummed late-time correlation functions involving $\phi^2$ from next-to-leading order (NLO) in the $\sqrt{\kappa}$ expansion. 
Finally, we use the found results to compute for the first time the two-loop $\mathcal{O}(\kappa)$ correction to the diffusion coefficient of the Fokker-Planck equation, which is a next-to-next-to-leading order  (NNLO) effect that was not considered in \cite{Cohen:2021fzf}. 

\section{SdSET prerequisites}
\label{sec:prerequisites}

In this article, we employ the formalism, regularisation schemes, notation, and conventions set up in~\cite{Beneke:2026rtf}. 
We use dimensional regularisation for the UV divergences, while IR divergences are regulated by means of a comoving regulator $\Lambda$. In the full theory, this regulator is implemented at the level of the action, while in SdSET it enters via the Gaussian and non-Gaussian initial conditions. For practical computations, its presence amounts to shifting the absolute values of all three-momenta in both theories as
\begin{equation}
k\rightarrow k_{\Lambda}\equiv\sqrt{\vec k^2+\Lambda^2}\,.
\end{equation}
For completeness, we recall (from \cite{Beneke:2026rtf}) that in $d$-dimensional de~Sitter spacetime, SdSET for the minimally coupled, massless scalar field\footnote{We add an evanescent mass term \cite{Melville:2021lst}, which keeps the index of the Hankel mode functions at value $\frac{3}{2}$ independent of $d$.} 
is defined at leading order in the power-counting parameter $\lambda\sim k_{\textrm{phys}}/H$ by an action of the form
\begin{align}
    S &= \int\der^{d-1}x\der t\;\bigg[-3\vp_{-,0}\dot\vp_{+,0} + c^0_{1,1}\bigg(\frac{\nu}{a(t)H}\bigg)^{\!-2\delta}\vp_{+,0}(t,\vec x)\vp_{-,0}(t,\vec x)\nonumber\\
&\quad+\sum_{n=1}^{\infty}a(t)^{2n\ve}\frac{c^0_{2n+1,1}}{(2n+1)!}\,\vp^{2n+1}_{+,0}(t,\vec x)\vp_{-,0}(t,\vec x)\bigg]\,,
\label{eq:SdSETLagrangian}
\end{align}
supplemented by the non-Gaussian initial conditions via
\begin{flalign}
F[\vp_{\pm}]=&\int\frac{\der^{d-1}k}{(2\pi)^{d-1}}\;\bigg(\frac{\nu}{a_*H}\bigg)^{\!-2\delta}\Xi^0_{1,1}(\vec k,-\vec k)\,\vp_{+,0}(t_*,\vec k)\vp_{-,0}(t_*,-\vec k)\nonumber\\
&+\,\sum_{n=1}^{\infty}\int\bigg[\prod_{j=1}^{2n+2}\frac{\der^{d-1}k_j}{(2\pi)^{d-1}}\bigg]\;(2\pi)^{d-1}\delta^{(d-1)}\bigg(\sum_{l=1}^{2n+2}\vec k_l\bigg)
\frac{a^{2n\ve}_*}{(2n+1)!}\,\Xi^0_{2n+1,1}(\vec k_1,...,\vec k_{2n+2})
\nonumber\\
&\quad\times\,\bigg[\prod_{l=1}^{2n+1}\vp_{+,0}(t_*,\vec k_l)\bigg]\,\vp_{-,0}(t_*,\vec k_{2n+2})\,,
\label{eq::ICFmom}
\end{flalign}
which encodes the effects of early-time subhorizon evolution. 
The couplings and initial conditions (ICs) must be renormalised, and we restate here for convenience the relations between bare and renormalised quantities~\cite{Beneke:2026rtf}
\begin{align}
c^0_{2n+1,1}
&=\tmu^{2n\ve}\,\Big[c_{2n+1,1}+
\delta c_{2n+1,1}(\{c_{2 l+1,1}\},\varepsilon)\Big]\,,
\label{eq::renCoup}
\\
\Xi^0_{2n+1,1}(\vec k_1,...,\vec k_{2n+2})&=\tmu^{2n\ve}\Big[\xi_{2n+1,1}(\vec k_1,...,\vec k_{2n+2})+\Xi_{2n+1,1}(\vec k_1,...,\vec k_{2n+2})\Big]\,,
\label{eq::renICs}
\end{align}
where $c_{2n+1,1}$ ($\Xi_{2n+1,1}$) are the renormalised couplings (IC functions) and
$\delta c_{2n+1,1}$ ($\xi_{2n+1,1}$) denotes the coupling (IC) counterterms.

The renormalisation of the effective couplings introduces a running of $c_{2n+1,1}$ with $\mu$, as is standard, which is controlled by a set of renormalisation-group equations (RGEs). Since the bare quantities $c^0_{2n+1,1}$ are independent of $\mu$, one finds
\begin{equation}
\frac{\der c_{2n+1,1}}{\der\log(\mu)}=-2n\ve\Big[c_{2n+1,1}+
\delta c_{2n+1,1}\Big]-\frac{\der \delta c_{2n+1,1}}{\der\log(\mu)}\,.
\label{eq::cRGE}
\end{equation}
The renormalised IC functions satisfy an analogous equation
\begin{equation}
\frac{\der\Xi_{2n+1,1}}{\der\log(\mu)}=-2n\ve\Big[\xi_{2n+1,1}+\Xi_{2n+1,1}\Big]-\frac{\der\xi_{2n+1,1}}{\der\log(\mu)}\,,
\label{eq::XimuRGE}
\end{equation}
where we suppress the momentum arguments for conciseness.
Additionally, the renormalisation of time UV-divergences introduces a running of the $\Xi_{2n+1,1}$ with the reference scale factor $a_*$.
At leading order in $\lambda$ and the full-theory coupling $\kappa$, the combination
\begin{equation}
a^{2n\ve}_*\Xi^0_{2n+1,1}(\vec k_1,...,\vec k_{2n+1})
\end{equation}
is independent of $a_*$, leading to the RGE
\begin{equation}
\frac{\der\Xi_{2n+1,1}}{\der\log(a_*)}=-2n\ve\Big[\xi_{2n+1,1}+\Xi_{2n+1,1}\Big]-\frac{\der\xi_{2n+1,1}}{\der\log(a_*)}\,.
\label{eq::XiasRGE}
\end{equation}
Since $a_*$ only appears in the IC functional, the renormalised effective couplings do not depend on it.

The expressions above are sufficient to compute the correlation functions and operator mixings that account for the late-time asymptotic behaviour of correlation functions up to power corrections in conformal time that vanish as $\eta\to 0$. This corresponds to the leading power of the EFT expansion in $\lambda$, while including all orders in the coupling $\sqrt{\kappa}$ and all the secular logarithms at late times.

In the Schwinger-Keldysh formalism~\cite{Schwinger:1960qe,Keldysh:1964ud}, the action and the IC functional are doubled and appear in the exponential of the path integral as
\begin{equation}
S_{\textrm{int}}[\vp^+_{\pm}]-S_{\textrm{int}}[\vp^-_{\pm}] + F[\vp^+_{\pm}]-F[\vp^-_{\pm}]\,,
\end{equation}
where $+$ and $-$ denote the closed-time-path (CTP) indices. All details pertaining to the detailed construction, renormalisation, and example matching calculations can be found in~\cite{Beneke:2026rtf}.
For the composite-operator matching performed in this paper, the necessary full-theory computations, as well as further technical details on the SdSET computations, are collected in Appendices \ref{app::Zinv}-\ref{app::Phi20loopsEFT}.

\section{Composite operator renormalisation in SdSET}
\label{sec:EFTopmatch}

To set the stage, we present the framework for the renormalisation of  composite SdSET operators and discuss generic properties of the renormalisation factors.
We proceed by defining the matching equations onto the renormalised full-theory composite operators, introducing the matching coefficients and their RGEs in SdSET.

\subsection{Renormalisation of \texorpdfstring{$\vp^n_+$}{φ₊ⁿ}}

We consider composite operators built from {\em renormalised} SdSET fields $\vp_{+}$ of the form
\begin{equation}
\vp^n_{+}(t,\vec x)\,.
\end{equation}
When such operators are inserted into correlation functions, they generate, in general, new UV divergences which are not accounted for by the renormalisation of $\vp_{+}$.\footnote{Field renormalisation in SdSET is in fact absent at leading power in the power-counting parameter~$\lambda$~\cite{Beneke:2026rtf}.} 
Therefore, these composite operators require additional renormalisation to define their renormalised counterparts 
$[\vp^n_+]$, which yield  UV-finite correlation functions. Furthermore, since in $d=4$ dimensions the effective field $\vp_+$ has scaling and mass dimension zero, there is an infinite number of relevant operators $\vp^n_+$ which may mix under renormalisation, analogously to the  renormalisation of the effective couplings $c_{2n+1,1}$~\cite{Beneke:2026rtf}. 

The relation between the bare operators $\vp^n_+$ and their renormalised counterparts reads 
\begin{equation}
(a(t)\tmu)^{n\ve}\vp^n_+(t,\vec x)=\sum_{m=0}^{\infty}\,(a(t)\tmu)^{m\ve}\,Z_{nm}\,[\vp^m_+](t,\vec x)\,,
\label{eq::vpnren}
\end{equation}
where $\ve=(4-d)/2$, $\tmu = \mu \sqrt{e^{\gamma_E}/(4\pi)}$ is  the $\overline{\textrm{MS}}$ scale, and $Z_{nm}$ denotes the matrix of operator renormalisation factors. Since the left- and right-hand sides of this equation both generate UV-divergences when inserted into correlation functions, it is only well-defined for $\ve\neq0$, and all quantities appearing should be understood as $d$-dimensional. The various elements of \eqref{eq::vpnren} arise as follows:\ both the $d$-dimensional bare operators $\vp^n_+$ and renormalised operators $[\vp^n_+]$ have mass dimension $-n\ve$, so to relate operators with different exponents to each other, the dimensional mismatch must be compensated by appropriate powers of the renormalisation scale $\mu$.
Furthermore, the effective fields transform non-trivially under infinitesimal dilatations and special conformal transformations, so there is also a mismatch between the transformation behaviour of operators with different exponents.
This can be compensated by including appropriate factors of $a(t)$, since the combination $a(t)^{\ve}\vp_+=a(t)^{-\alpha}\vp_+$ is invariant under both transformations~\cite{Beneke:2026rtf}, which leads to the factors of $a(t)\tmu$ in \eqref{eq::vpnren}. Lastly, the vanishing scaling- and mass dimensions of the renormalised operators in $d=4$ imply that one must allow for the mixing of all $\vp^n_+$ with each other under renormalisation. This results in the infinite sum on the right-hand side of \eqref{eq::vpnren}. 

Renormalisation induces a dependence of the renormalised quantities on the scale $\mu$. Since the bare operator $\vp^n_+$ is $\mu$-independent,
\begin{equation}
\frac{\der}{\der\log(\mu)}\,\vp^n_+(t,\vec x)=0\,.
\label{eq::vpnnorun}
\end{equation}
Substituting \eqref{eq::vpnren} into \eqref{eq::vpnnorun} results in the following RGE for the composite operator $[\vp^n_+]$ in $d$ dimensions
\begin{equation}
\frac{\der[\vp^n_+]}{\der\log(\mu)}=(a(t)\tmu)^{-n\ve}\sum_{m=0}^{\infty}\bigg[-n\ve\delta_{nm}+\sum_{l=0}^{\infty}Z^{-1}_{nl}\bigg(l\ve Z_{lm}-\frac{\der Z_{lm}}{\der\log(\mu)}\bigg)\bigg](a(t)\tmu)^{m\ve}\,[\vp^m_+]\,.\label{eq::ddimRGE}
\end{equation}
This equation suggests defining the anomalous-dimension matrix (ADM) $\gamma^{\mu}$ as
\begin{equation}
\gamma^{\mu}_{nm}\equiv -n\ve\delta_{nm}+\sum_{l=0}^{\infty}Z^{-1}_{nl}\bigg[l\ve Z_{lm}-\frac{\der Z_{lm}}{\der\log(\mu)}\bigg]\,.
\label{eq::muADM}
\end{equation}
Even though divergent $Z$-factors enter the definition of $\gamma^{\mu}$, it must be finite for \eqref{eq::ddimRGE} to be valid.
This is not obvious from its definition and implies non-trivial relations between the $Z$-factors. Therefore, one can safely let $\ve\rightarrow0$ to find the four-dimensional RGE
\begin{equation}
\frac{\der[\vp^n_+]}{\der\log(\mu)}=\sum_{m=0}^{\infty}\gamma^{\mu}_{nm}\,[\vp^m_+]\,.
\label{eq::vpnRGE1}
\end{equation}

As a novel feature of SdSET, the renormalisation of UV-divergences originating from time integrals introduces a reference scale factor $a_*$~\cite{Beneke:2026rtf} and a residual logarithmic $a_*$-dependence of the SdSET correlation functions.
This dependence is cancelled by the renormalised initial-condition functions $\Xi_{2n+1,1}$.
In the following, we demonstrate that the same occurs for composite-operator correlation functions, where now the $a_*$-dependence is cancelled by the operator matching coefficients $C_{nm}$, which we discuss below. 
Since the bare operators are $a_*$-independent, \eqref{eq::vpnren} then implies the four-dimensional RGE
\begin{equation}
\frac{\der[\vp^n_+]}{\der\log(a_*)}=\sum_{m=0}^{\infty}\gamma^{a_*}_{nm}[\vp^m_+]\,,
\label{eq::vpnRGE2}
\end{equation}
where the elements of the ADM $\gamma^{a_*}$ are defined as
\begin{equation}
\gamma^{a_*}_{nm}\equiv-\sum_{l=0}^{\infty}Z^{-1}_{nl}\frac{\der Z_{lm}}{\der\log(a_*)}\,.
\label{eq::asADM}
\end{equation}

We use the $\overline{\textrm{MS}}$ scheme to renormalise the $\vp^n_+$, where the $Z_{nm}$ only subtract the $\ve$-poles of composite-operator correlation functions. The choice of $\tmu$ automatically subtracts the associated factors of $\gamma_E-\log(4\pi)$.

\subsection{Properties of the renormalisation factors}

Before turning to concrete examples of composite operator renormalisation in SdSET, it is instructive to reflect on the general properties of the $Z_{nm}$.

Since $\vp_+^n$ can mix with any $\vp_+^m$, the $Z$-factor matrix is infinite-dimensional. It contains two trivial rows with only one non-zero entry, to all orders in the effective couplings, namely
\begin{equation}
Z_{0m}=\delta_{0m}\,,\qquad Z_{1m}=\delta_{1m}\,,
\end{equation}
since the unit operator $\vp^0_+=\un$ requires no renormalisation, and we assume that the elementary field $\vp^1_+=\vp_+$ has already been fully renormalised. This remark holds for the SdSET matched to any UV theory. Specialising now to the UV completion of interest, the massless $\kappa\phi^4$-theory, the remaining matrix elements for $n\geq 2$ are given in terms of infinite series in powers of $\ve$, as
\begin{equation}
    Z_{nm} = \delta_{nm} + \sum_{k=0}^\infty \kappa^{k} \sum_{l=1}^{\infty}\frac{z_{nm}^{(k,l)}}{\ve^l}\,.
\end{equation}
For given $m,n$ and order $k$ in the coupling expansion, the sum over $l$ always terminates.
The two infinite sums can be further constrained depending on $n$, $m$ as follows. Due to the underlying $\Z_2$ symmetry of the EFT under $\vp_{\pm}\rightarrow-\vp_{\pm}$ inherited from the $\phi^4$ theory, the difference between the powers of $\vp_+$ that mix among each other must be an integer multiple of 2, so $n-m$ is even. If $n-m$ is odd, the corresponding renormalisation-factor matrix element $Z_{nm}$ vanishes. 
Thus $(n-m)/2$ is always integer.

An operator $\vp^n_+$ can only mix with a higher power of $\vp_+$ via an interaction term of some kind, and the first type of diagram which can generate a divergence is a one-loop diagram with insertions of the quartic coupling $c_{3,1}\sim\kappa$ and quartic initial condition $\Xi_{3,1}\sim\kappa$. This constrains the lowest possible $\kappa$-order, and consequently one has for $n\leq m$ 
\begin{equation}\label{eq::zass1}
    \quad Z_{nm}=\delta_{nm}+\sum_{k=\frac{m-n}{2}+1}^{\infty}\hspace*{-0.35cm}\kappa^k \sum_{l=1}^{\infty}\frac{z^{(k,l)}_{nm}}{\ve^l}\,,
\end{equation}
where the $\delta_{nm}$ arises since at tree level bare and renormalised operators are equivalent. 

For $n>m$, the composite operator $\vp^n_+$ can mix with an operator of a lower exponent by tying pairs of fields together in ``elementary" loops, such as
\begin{align}
\hspace{1cm}(a(t)&\tmu)^{2\ve}\cor{}{\vp^2_+(t,\vec x)}=\frac{(a(t)\tmu)^{2\ve}}{2}\int\frac{\der^{d-1}l}{(2\pi)^{d-1}}\frac{1}{l_{\Lambda}^3}
\nonumber\\
&=\frac{1}{8\pi^2}\bigg(\frac{\Lambda}{a(t)\mu}\bigg)^{-2\ve}e^{\ve\gamma_E}\Gamma(\ve)
=\frac{1}{8\pi^2}\bigg[\frac{1}{\ve}-2\log\bigg(\frac{\Lambda}{a(t)\mu}\bigg)+\mathcal{O}(\ve)\bigg]\,,
\label{eq::vp2free}
\end{align}
without having to pay a price in powers of $\kappa$. 
Therefore, since each such elementary loop contributes a factor $\frac{1}{\ve}$, the $Z_{nm}$ with $n>m$ already start at $\Lo(\kappa^0)$ with pole terms of order $\frac{n-m}{2}$. Consequently, for $n>m$
\begin{equation}\label{eq::zass2}
    Z_{nm}=
    \sum_{k=0}^{\infty}\kappa^k\hspace*{-0.2cm}
    \sum_{l=\frac{n-m}{2}}^{\infty} \frac{z^{(k,l)}_{nm}}{\ve^l}\,.
\end{equation}
This is a consequence of the vanishing mass- and scaling dimension of the $\vp_+$ in the massless SdSET, and it has profound consequences for the effective dynamics of the theory. 
In particular, this implies that, unlike for coupling renormalisation, already the free theory features a matrix $Z$ containing infinitely many non-vanishing entries. 
Indeed, from \eqref{eq::zass1}, \eqref{eq::zass2}, one observes that when turning off the effective couplings, the infinite-dimensional matrix $Z$ reduces to a lower-triangular form with 1 on the diagonal entries
\begin{equation}
Z_{nm}=
\begin{dcases}
\;\delta_{nm} & \text{for } n\leq m\,,\\
\;\frac{z^{(0,\frac{n-m}{2})}_{nm}}{\ve^{\frac{n-m}{2}}} & \text{for } n>m\,.
\end{dcases}
\label{eq::zfree}
\end{equation}
This lower-triangular form of $Z$ in the free theory is the crucial property that makes the problem tractable, even in the absence of a perturbative parameter. Once  the small effective couplings are switched on, the above form persists at leading order in the couplings and only receives perturbative corrections.

\subsection{Matching of \texorpdfstring{$\vp^n_+$}{φ₊ⁿ}}
\label{sec::compmatch}

We now turn to the matching of renormalised SdSET operators $[\vp^n_+]$ to renormalised operators $[\phi^n]$ in the full theory. We make the following ansatz for the four-dimensional matching equation at leading power in $\lambda$
\begin{equation}
[\phi^n](t,\vec x)=H^n\sum_{m=0}^{\infty}C_{nm}(t)[\vp^m_+](t,\vec x)\,.
\label{eq::opmatch}
\end{equation}
The overall factor $H^n$ appears due to the relation between $\phi$ and $\vp_+$  \cite{Cohen:2020php}. 
We anticipate that operator-matching coefficients $C_{nm}$ will be required to correctly reproduce correlators involving insertions of composite operators in the full theory. Physically, the $C_{nm}$ originate from dynamics outside of the regime of validity of SdSET. From the region structure of full-theory correlation functions \cite{Beneke:2023wmt}, one expects that these matching coefficients reproduce the physics of the late-time, hard-momentum regions. The $C_{nm}$ must therefore be spatially local, and thus independent of position arguments, but they may be time-dependent. 

Even though the matching equation \eqref{eq::opmatch} is supposed to hold to all orders in perturbation theory, in practice we determine the operator matching coefficients perturbatively by matching UV-renormalised but IR-divergent correlation functions involving full-theory operators $[\phi^n]$ onto analogous SdSET correlation functions featuring $[\vp^n_+]$. Since SdSET by construction reproduces the full theory in the regions where the IR divergences arise, they cancel in the matching coefficients. 
Therefore, the $C_{nm}$ are computed as power series in the full-theory coupling $\kappa$. The analogous considerations made above to infer the general structure of the $Z_{nm}$ apply here as well, and the $C_{nm}$ have the following form
\begin{equation}
    C_{nm}= \begin{dcases}
       \; \delta_{nm}+\hspace*{-0.2cm}\sum_{k=\frac{m-n}{2}+1}^{\infty}\hspace*{-0.2cm}\kappa ^k C^{(k)}_{nm} & \text{for } n\leq m\,,\\
       \: \sum_{k=0}^{\infty}\kappa^kC^{(k)}_{nm} & \text{for } n>m\,.
    \end{dcases}
\end{equation}
This mirrors the structure of the $Z_{nm}$, see \eqref{eq::zass1}, \eqref{eq::zass2}.

Renormalisation introduces a logarithmic dependence on the quantities $\mu$ and $a_*$ into the SdSET correlation functions~\cite{Beneke:2026rtf}, which play the roles of both renormalisation and factorisation scales. 
Since the full-theory correlation functions do not depend on these scales, this dependence must be compensated by the $C_{nm}$
\begin{equation}
[\phi^n](\mu_f;t,\vec x)=H^n\sum_{m=0}^{\infty}C_{nm}\bigg(\mu_f,\mu,\frac{a_*}{a(t)}\bigg)\,[\vp^m_+](\mu,a_*;t,\vec x)\,.
\label{matchdiffmu}
\end{equation}
Here, we explicitly distinguished the full-theory renormalisation scale $\mu_f$ from the SdSET one $\mu$.
The requirement that the dependence on $\mu$ and $a_*$ cancels between the $C_{nm}$ and the $[\vp^n_+]$ implies the following two RGEs for the renormalised EFT-operators
\begin{equation}
\frac{\der[\vp^n_+]}{\der\log(\mu,a_*)}=-\sum_{m=0}^{\infty}\bigg[\sum_{l=0}^{\infty}C^{-1}_{nl}\frac{\der C_{lm}}{\der\log(\mu,a_*)}\bigg][\vp^m_+]\,.
\label{eq::CRGE}
\end{equation}
Consistency of SdSET then imposes that the coefficients of the $[\vp^n_+]$ on the right-hand side of \eqref{eq::CRGE} can be expressed in terms of the same ADMs $\gamma^{\mu,a_*}$ computed from the operator-renormalisation factors $Z_{nm}$ \eqref{eq::muADM} and \eqref{eq::asADM} as
\begin{equation}
\gamma^{\mu,a_*}_{nm}=-\sum_{l=0}^{\infty}C^{-1}_{nl}\frac{\der C_{lm}}{\der\log(\mu,a_*)}\,.
\label{eq::CADM}
\end{equation}
This is the standard situation.
The formula \eqref{eq::CADM} can be used to cross-check the dependence of the $C_{nm}$ on $\mu$ and $a_*$ against the results for the ADMs obtained from the operator-renormalisation factors.


\section{Operator mixing in the free SdSET}
\label{sec:freemix}

We first address the renormalisation of composite operators in the free theory. We begin by analysing four concrete cases covering the renormalised free-theory one-point functions that will be needed in the following sections. We then generalise to the mixing of any $\vp^n_+$ with any other operator $\vp^m_+$, and derive closed expressions for the $Z$-factor and anomalous dimension matrices in the free theory.
When drawing diagrams for the correlation functions containing composite operators, we depict the insertion of $\vp^n_+(t,\vec x)$ by the symbol $\otimes$. We omit any labeling of this symbol in this section, since the inserted operator will always be at the point $(t,\vec x)$. The exponent $n$ of the inserted operator can be inferred from the number of legs attached to this symbol.

\subsection{Specific cases}

\subsubsection{Mixing of \texorpdfstring{$\vp^2_+\rightarrow\un$}{φ₊² → 1}}

\begin{figure}[t]
\centering
\begin{subfigure}{0.33\textwidth}
\centering
\includegraphics[width=0.45\textwidth]{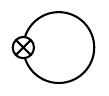}
\caption{}
\end{subfigure}%
\begin{subfigure}{0.33\textwidth}
\centering
\raisebox{0.13cm}{\includegraphics[width=0.95\textwidth]{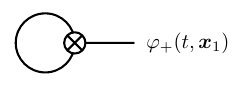}}
\caption{}
\end{subfigure}%
\begin{subfigure}{0.33\textwidth}
\centering
\includegraphics[width=0.65\textwidth]{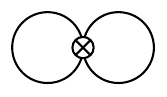}
\caption{}
\end{subfigure}
\caption{The relevant diagrams for the mixing of $\vp^2_+\rightarrow\un$ (left), $\vp^3_+\rightarrow\vp_+$ (middle), and $\vp^4_+\rightarrow\un$ (right panel).}
\label{fig:freemix}
\end{figure}

The mixing of $\vp^2_+$ with the unit operator is the simplest case. From \eqref{eq::vpnren} and the properties of the free-theory matrix $Z$ collected in~\eqref{eq::zfree}, we find the operator-renormalisation equation
\begin{equation}
(a(t)\tmu)^{2\ve}\vp^2_+(t,\vec x)=(a(t)\tmu)^{2\ve}Z_{22}[\vp^2_+](t,\vec x)+Z_{20}\un\,.
\label{eq::vp2freeren}
\end{equation}
The $Z$-factor is extracted from the diagram shown in the left panel of \figref{fig:freemix}. From~\eqref{eq::vp2free} we find
\begin{equation}
(a(t)\tmu)^{2\ve}\cor{}{\vp^2_+(t,\vec x)}
=\frac{1}{8\pi^2}\bigg[\frac{1}{\ve}-2\log\bigg(\frac{\Lambda}{a(t)\mu}\bigg)+\Lo(\ve)\bigg]
\label{g20result}
\end{equation}
and, since in the free theory $Z_{22}=1$, the divergent piece of \eqref{g20result} is subtracted by choosing
\begin{equation}
Z_{20}=\frac{1}{8\pi^2\ve}\,.
\end{equation}
Eq.~\eqref{eq::vp2freeren} then defines the four-dimensional renormalised one-point function
\begin{equation}
\cor{}{[\vp^2_+](t,\vec x)}=-\frac{1}{4\pi^2}\log\bigg(\frac{\Lambda}{a(t)\mu}\bigg)\,.
\label{eq::ren20}
\end{equation}

\subsubsection{Mixing of \texorpdfstring{$\vp^3_+\rightarrow\vp_+$}{φ₊³ → φ₊}}

This case is almost identical to the previous one, since
\begin{equation}
\cor{}{\vp^3_+(t,\vec x)\vp_+(t,\vec x_1)}=3\cor{}{\vp^2_+(t,\vec x)}\cor{}{\vp_+(t,\vec x)\vp_+(t,\vec x_1)}\,,
\end{equation}
which corresponds to the diagram in the middle panel of \figref{fig:freemix}. Using the results from the previous subsection, we find
\begin{equation}
(a(t)\tmu)^{2\ve}\cor{}{\vp^3_+(t,\vec x)\vp_+(t,\vec x_1)}=\frac{3}{8\pi^2}\bigg[\frac{1}{\ve}-2\log\bigg(\frac{\Lambda}{a(t)\mu}\bigg)+\Lo(\ve)\bigg]\cor{}{\vp_+(t,\vec x)\vp_+(t,\vec x_1)}\,.
\label{eq::vp3vpfree}
\end{equation}
The renormalised operator $\vp^3_+$ is defined by
\begin{equation}
(a(t)\tmu)^{3\ve}\vp^3_+(t,\vec x)=(a(t)\tmu)^{3\ve}Z_{33}[\vp^3_+](t,\vec x)+(a(t)\tmu)^{\ve}Z_{31}\vp_+(t,\vec x)\,.
\end{equation}
Since $Z_{33}=1$ in the free theory, the pole in \eqref{eq::vp3vpfree} is subtracted by choosing
\begin{equation}
Z_{31}=\frac{3}{8\pi^2\ve}=3Z_{20}\,,
\end{equation}
which defines the renormalised four-dimensional two-point function
\begin{equation}
\cor{}{[\vp^3_+](t,\vec x)\vp_+(t,\vec x_1)}=-\frac{3}{4\pi^2}\log\bigg(\frac{\Lambda}{a(t)\mu}\bigg)\cor{}{\vp_+(t,\vec x)\vp_+(t,\vec x_1)}\,.
\label{eq:phi3tophi1}
\end{equation}

\subsubsection{Mixing of \texorpdfstring{$\vp^4_+\rightarrow\un$}{φ₊⁴ → 1}}

Next, we compute the mixing of $\vp^4_+$ with the unit operator, which is determined from 
\begin{align}
(a(t)\tmu)^{4\ve}\cor{}{\vp^4_+(t,\vec x)}&=3(a(t)\tmu)^{4\ve}\cor{}{\vp^2_+(t,\vec x)}^2\nonumber\\[-.17cm]
&\hspace*{-2.5cm} =\frac{3}{64\pi^4}\bigg(\frac{\Lambda}{a(t)\mu}\bigg)^{-4\ve}e^{2\ve\gamma_E}\Gamma(\ve)^2\nonumber\\[-.05cm]
&\hspace*{-2.5cm} =\frac{3}{64\pi^4}\bigg\{\frac{1}{\ve^2}-\frac{4}{\ve}\log\bigg(\frac{\Lambda}{a(t)\mu}\bigg)+\frac{\pi^2}{6}+8\log^2\bigg(\frac{\Lambda}{a(t)\mu}\bigg)+\Lo(\ve)\bigg\}\,,
\label{eq::vp4free}
\end{align}
corresponding to the diagram shown in the right panel of \figref{fig:freemix}. 

This result involves a pole term with a $\Lambda$-dependent coefficient, which seems problematic, as the renormalisation factors are UV quantities that must not depend on the IR regularisation. We now demonstrate that this pole is a consequence of operator mixing and not part of a $Z$-factor.
The renormalised operator $\vp^4_+$ is defined by
\begin{equation}
(a(t)\tmu)^{4\ve}\vp^4_+(t,\vec x)=(a(t)\tmu)^{4\ve}Z_{44}[\vp^4_+](t,\vec x)+(a(t)\tmu)^{2\ve}Z_{42}[\vp^2_+](t,\vec x)+Z_{40}\un\,.
\label{eq::vp4ren}
\end{equation}
Using that $Z_{44}=1$ in the free theory, as well as the renormalised result \eqref{eq::ren20}, which contains an explicit factor of $\log(\Lambda/(a(t)\mu)),$ we see that the poles are subtracted by
\begin{flalign}
Z_{40}&=\frac{3}{64\pi^4\ve^2}\,,\\
Z_{42}&=\frac{3}{4\pi^2\ve}\,,
\end{flalign}
independent of $\Lambda$. Then, from
\begin{flalign}
Z_{42}(a(t)\tmu)^{2\ve}\cor{}{[\vp^2_+](t,\vec x)}&=\frac{3}{4\pi^2\ve}\bigg[\frac{1}{8\pi^2}\bigg(\frac{\Lambda}{a(t)\mu}\bigg)^{-2\ve}e^{\ve\gamma_E}\Gamma(\ve)-\frac{1}{8\pi^2\ve}\bigg]\nonumber\\
&=\frac{3}{16\pi^4}\bigg[-\frac{1}{\ve}\log\bigg(\frac{\Lambda}{a(t)\mu}\bigg)+\log^2\bigg(\frac{\Lambda}{a(t)\mu}\bigg)+\frac{\pi^2}{24}\bigg]\,,
\label{eq:Z42phisq}
\end{flalign}
where the $d$-dimensional expression for $\cor{}{[\vp^2_+](t,\vec x)}$ must be used, 
one finds that \eqref{eq::vp4ren} defines the renormalised one-point function
\begin{equation}
\cor{}{[\vp^4_+](t,\vec x)}=\frac{3}{16\pi^4}\log^2\bigg(\frac{\Lambda}{a(t)\mu}\bigg)\,.
\label{eq::ren40}
\end{equation}

This example emphasises that the operator-renormalisation equation \eqref{eq::vpnren} is only well-defined in $d$ dimensions. The $d$-dimensional correlation functions of renormalised operators still contain terms involving positive powers of $\ve$, which  generate non-vanishing terms in the limit $\ve\rightarrow0$ when multiplied with the divergent $Z_{nm}$.
This observation will be especially important below, when we compute composite-operator correlation functions in the interacting theory.

\subsubsection{Mixing of \texorpdfstring{$\vp^4_+\rightarrow\vp^2_+$}{φ₊⁴ → φ₊²}}

\begin{figure}[t]
\centering
\includegraphics[width=0.3\textwidth]{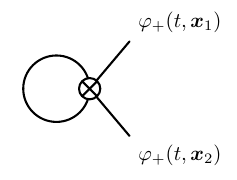}
\caption{The relevant diagram for the mixing of the operator $\vp^4_+$ with $\vp^2_+$.}
\label{fig:42freemix}
\end{figure}

As a consistency check of the above result, we compute
\begin{flalign}
&(a(t)\tmu)^{2\ve}\cor{}{\vp^4_+(t,\vec x)\vp_+(t,\vec x_1)\vp_+(t,\vec x_2)}_{|\,\textrm{conn.}}\nonumber\\[0.2cm]
&=12(a(t)\tmu)^{2\ve}\cor{}{\vp^2_+(t,\vec x)}\cor{}{\vp_+(t,\vec x)\vp_+(t,\vec x_1)}\cor{}{\vp_+(t,\vec x)\vp_+(t,\vec x_2)}\nonumber\\
&=\frac{3}{2\pi^2}\bigg[\frac{1}{\ve}-2\log\bigg(\frac{\Lambda}{a(t)\mu}\bigg)\bigg]\,\cor{}{\vp_+(t,\vec x)\vp_+(t,\vec x_1)}\cor{}{\vp_+(t,\vec x)\vp_+(t,\vec x_2)}\,,
\end{flalign}
where the subscript ``conn." denotes the connected component of the correlation function, which is depicted in~\figref{fig:42freemix}.
Employing \eqref{eq::vp4ren}, we see that the counterterm
\begin{equation}
Z_{42}=\frac{3}{4\pi^2\ve}
\end{equation}
indeed absorbs the divergence, since at tree level one has
\begin{equation}
\cor{}{[\vp^2_+](t,\vec x)\vp_+(t,\vec x_1)\vp_+(t,\vec x_2)}=2\cor{}{\vp_+(t,\vec x)\vp_+(t,\vec x_1)}\cor{}{\vp_+(t,\vec x)\vp_+(t,\vec x_2)}\,.
\end{equation}
The renormalised correlator then reads
\begin{align}
\cor{}{[\vp^4_+](t,\vec x)\vp_+(t,\vec x_1)\vp_+(t,\vec x_2)}&=-\frac{3}{\pi^2}\log\bigg(\frac{\Lambda}{a(t)\mu}\bigg)\,\cor{}{\vp_+(t,\vec x)\vp_+(t,\vec x_1)}\cor{}{\vp_+(t,\vec x)\vp_+(t,\vec x_2)}\nn\\
&=-\frac{3}{2\pi^2}\log\bigg(\frac{\Lambda}{a(t)\mu}\bigg)\,\cor{}{[\vp^2_+](t,\vec x)\vp_+(t,\vec x_1)\vp_+(t,\vec x_2)}\,.
\end{align}

\subsection{General case: mixing of \texorpdfstring{$\vp^n_+\rightarrow\vp^m_+$}{φ₊ⁿ → φ₊ᵐ} and anomalous dimensions}
\label{sec::freemix}

In all of the above computations, one observes that the pole terms are generated by powers of the elementary loop $\cor{}{\vp^2_+(t,\vec x)}$, and this feature would persist if one were to compute more such free-theory correlators.
This allows us to generalise the above considerations to determine all $Z_{nm}$ in the free theory in closed form. 
From this result, one obtains the corresponding full ADM $\gamma^{\mu}$ in closed form, which already possesses non-trivial entries. On the other hand, since in the free theory there is no $a_*$-dependence, the ADM $\gamma^{a_*}$ vanishes.

Consider the mixing of $\vp^n_+$ with $\vp^m_+$, where $n-m$ must be even to obtain a non-trivial result, by analysing the correlation function 
\begin{flalign}
&\cor{}{\vp^n_+(t,\vec x)\vp_+(t,\vec x_1)...\vp_+(t,\vec x_m)}\nonumber\\
&=\frac{n!}{(n-m)!}\frac{(n-m)!}{2^{\frac{n-m}{2}}(\frac{n-m}{2})!}\,\cor{}{\vp^2_+(t,\vec x)}^{\frac{n-m}{2}}\prod_{j=1}^m\cor{}{\vp_+(t,\vec x)\vp_+(t,\vec x_j)}\nn\\
&=\frac{n!}{2^{\frac{n-m}{2}}(\frac{n-m}{2})!}\bigg[\frac{1}{8\pi^2}\bigg(\frac{\Lambda}{a(t)\mu}\bigg)^{-2\ve}e^{\ve\gamma_E}\Gamma(\ve)\bigg]^{\frac{n-m}{2}}\prod_{j=1}^m\,\cor{}{\vp_+(t,\vec x)\vp_+(t,\vec x_j)}\,.
\end{flalign}
The combinatorial factor arises as follows: there are $n!/(n-m)!$ ways of contracting $m$ external (distinguishable) fields with $n$ indistinguishable fields at $\vec x$, and there are
$(2l)!/(2^l l)$
different ways of tying the remaining $2l=n-m$ indistinguishable fields into $l$ loops.

The pole part of the above correlator is
\begin{equation}
\cor{}{\vp^n_+(t,\vec x)\vp_+(t,\vec x_1)...\vp_+(t,\vec x_m)}_{|\,\textrm{pole}}=\frac{n!}{2^{\frac{n-m}{2}}(\frac{n-m}{2})!}\bigg(\frac{1}{8\pi^2\ve}\bigg)^{\frac{n-m}{2}}\prod_{j=1}^m\cor{}{\vp_+(t,\vec x)\vp_+(t,\vec x_j)}
\end{equation}
and it needs to be absorbed by the renormalisation factor $Z_{nm}$ via
\begin{equation}
\cor{}{\vp^n_+(t,\vec x)\vp_+(t,\vec x_1)...\vp_+(t,\vec x_m)}_{|\,\textrm{pole}}\equiv Z_{nm}\cor{}{[\vp^m_+](t,\vec x)\vp_+(t,\vec x_1)...\vp_+(t,\vec x_m)}\,.
\end{equation}
In the free theory, 
\begin{equation}
\cor{}{[\vp^m_+](t,\vec x)\vp_+(t,\vec x_1)...\vp_+(t,\vec x_m)}=m!\,\prod_{j=1}^m\,\cor{}{\vp_+(t,\vec x)\vp_+(t,\vec x_j)}\,.
\end{equation}
Comparing the coefficients of the product of the $m$ two-point functions, we find
\begin{equation}
Z_{nm}=\frac{n!}{2^{\frac{n-m}{2}}m!(\frac{n-m}{2})!}\bigg(\frac{1}{8\pi^2\ve}\bigg)^{\frac{n-m}{2}}
=\frac{n!}{2^{\frac{n-m}{2}}m!(\frac{n-m}{2})!}Z_{20}^{\frac{n-m}{2}}\,.
\label{eq::Znmfree}
\end{equation}
As expected, all $Z_{nm}$ are related to $Z_{20}$. This formula reproduces the specific examples considered above, as well as the general case where $n=m$ and $Z_{nn}=1$.
In the free theory, we thus have for $n-m$ even
\begin{equation}
Z_{nm}=\begin{cases}
\;0&\quad n<m\,,\\
\;\displaystyle{\frac{n!}{2^{\frac{n-m}{2}}m!(\frac{n-m}{2})!}\bigg(\frac{1}{8\pi^2\ve}\bigg)^{\frac{n-m}{2}}}&\quad n\geq m\,,
\label{eq::Z0}
\end{cases}
\end{equation}
and for $n-m$ odd
\begin{equation}
Z_{nm}=0\,.
\end{equation}
These results match our expectation from the previous general considerations about the $Z_{nm}$, see \eqref{eq::zfree}.

From the renormalisation factors, one can now determine the anomalous dimensions $\gamma^{\mu}_{nm}$. 
Since in the free theory both $Z_{nm}$ and $Z^{-1}_{nm}$ are lower-triangular matrices, the only non-vanishing summands in \eqref{eq::muADM} are those for which the summation index $l$ satisfies
\begin{equation}
m\leq l\leq n\,.
\end{equation}
Therefore, one only needs to evaluate the finite sum
\begin{equation}
\gamma^{\mu}_{nm}=-n\ve\delta_{nm}+\ve\sum_{l=m}^{n}Z^{-1}_{nl}lZ_{lm}\,.
\label{eq::ADM0}
\end{equation}
This immediately implies that
\begin{equation}
\gamma^{\mu}_{nm}=0\quad\textrm{ for }n\leq m\,,
\end{equation}
where the result for $n=m$ holds since in the free theory, $Z_{nn}=Z^{-1}_{nn}=1$. 
The computation of the remaining ADM elements requires the computation of the entries of the inverse matrix $Z^{-1}$. Following \eqref{eq::Z0}, we decompose $Z$ as
\begin{equation}
Z=\un+N\,,
\label{eq::Zfreedecomp}
\end{equation}
where $N$ has zeroes on its main diagonal, and the lower triangle under the main diagonal contains all the non-vanishing entries of $Z$. We can write the formal inverse of $Z$ as
\begin{equation}
Z^{-1}=[\un+N]^{-1}=\un+\sum_{l=1}^{\infty}(-1)^lN^l\,.
\label{eq::Zfreeinv}
\end{equation}
The matrix $N^l$ has zeroes on its main diagonal and on its first $2l-1$ lower off-diagonals. This means that to determine the entry $Z^{-1}_{n,n-k}$ of $Z^{-1}$, we can truncate the sum after $l=k/2$, since all the subsequent summands will only contribute zeroes, hence
\begin{equation}
Z^{-1}_{n,n-k}=\delta_{n,n-k}+\sum_{l=1}^{k/2}(-1)^l(N^l)_{n,n-k}\,.
\label{eq::Z0inv}
\end{equation}
 Notice that $k/2$ is always an integer since $k$ must be a multiple of $2$ to get a non-vanishing entry of $Z^{-1}$. The finite sum in \eqref{eq::Z0inv} is the essential feature that makes the present problem tractable. Even though the matrix $Z$ has infinitely many entries, the entries above any $n$-th row, which are always finitely many, can be treated independently of the infinitely many entries below this row, if one is interested in computing the corresponding entries of the inverse matrix. Remarkably, this structure makes it possible to compute the matrix $Z^{-1}$ in closed form, and the result reads
\begin{equation}
Z^{-1}_{nm}=(-1)^{\frac{n-m}{2}}Z_{nm}\,.
\label{eq::Zinvfree}
\end{equation}
The derivation of this result is presented in \appref{app::Zinv}. Due to the non-trivial structure of $Z$ in the free theory, the inversion of the operator-renormalisation matrix in the interacting SdSET, which is needed in order to compute anomalous dimensions in the interacting theory, is more involved as well. The procedure is described in \appref{app:intZinv}. 
Eq.~\eqref{eq::Zinvfree} now allows us to compute the ADM in closed form using \eqref{eq::ADM0} and the identity 
\begin{equation}
\sum_{l=0}^{\frac{n-m}{2}}\frac{(-1)^l(2l+m)}{(\frac{n-m}{2}-l)!\, l!}=n\delta_{nm}-2\delta_{n-2,m}\,,
\end{equation}
and we find
\begin{equation}
\gamma^{\mu}_{nm}=\frac{n(n-1)}{8\pi^2}\delta_{n-2,m}
\label{eq::freegammamu}
\end{equation}
for any $n$ and $m$. 

\subsection{Matching coefficients in the free theory}

Next, we consider the matching of full-theory composite operators $\phi^n$ to the SdSET operators $\vp_+^m$ in the free theory, which can be non-trivial given that there is operator mixing.  
In the free theory, the reference scale factor $a_*$ is absent, so the matching coefficients cannot depend on it. 
Moreover, if the matching is performed at a common scale $\mu_f=\mu$, then the correlation functions of $[\phi^n]$ and $[\vp^n_+]$ will agree by construction. In this case, 
\begin{equation}
C_{nm}=\delta_{nm}\,,
\end{equation}
resulting in the direct operator correspondence
\begin{equation}
[\phi^n](t,\vec x)=H^n[\vp^n_+](t,\vec x)\,.
\end{equation}
However, if $\mu_f$ and $\mu$ are not set equal, the $C_{nm}$ differ from the above result by logarithms of $\mu_f/\mu$, while the matching still works consistently. The anomalous dimensions \eqref{eq::freegammamu} can then be alternatively computed from $C_{nm}$ by making use of \eqref{eq::CADM}. To illustrate this, a few free-theory matching examples are presented in \appref{app:freematch}.


\section{Matching the one-loop bispectrum of the composite operator \texorpdfstring{$\phi^2$}{φ²}}
\label{sec::22q0EFT}

After setting up the general formalism, we now start to match interacting SdSET correlation functions involving composite operators to their full-theory counterparts. As a first example, we consider the late-time limit of the one-loop bispectrum of the operator $\phi^2$,\footnote{
This correlation function was already determined in~\cite{Beneke:2023wmt}, where the IR was regulated in dimensional regularisation.
Here, we instead use the cutoff $\Lambda$ to directly disentangle the UV and IR divergences.}
\begin{equation}
\cor{}{[\phi^2](t,\vec q)\phi(t,\vec k_1)\phi(t,\vec k_2)}'_{|\,\Lo(\kappa)}
\label{eq:fullthphi2bispectrumcorrelator}
\end{equation}
in the special kinematic configuration $\vec q=0$ (which implies $\vec{k}_1=-\vec{k}_2\equiv \vec k$).  This choice simplifies the following computations while still allowing for the extraction of the subhorizon physics encoded in the operator matching coefficients, which link full-theory and SdSET composite operators. 

\begin{figure}[t]
    \centering
    \begin{subfigure}{\textwidth}
    \centering
    \includegraphics[width=0.43\textwidth]{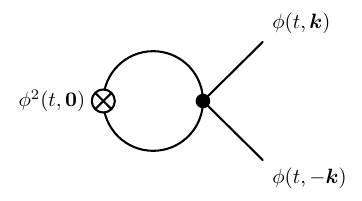}
\caption{}
\end{subfigure}\\
\begin{subfigure}{0.47\textwidth}
\centering
\includegraphics[width=\textwidth]{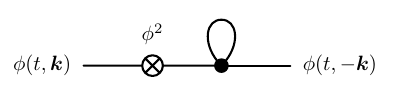}
\caption{}
\end{subfigure}%
\hspace{0.5cm}\begin{subfigure}{0.47\textwidth}
\centering
\includegraphics[width=\textwidth]{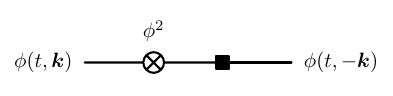}
\caption{}
\end{subfigure}
\caption{1PI (a) and the two 1PR (b), (c) diagram topologies contributing to the bispectrum of $\phi^2$ at one-loop order. The circled cross denotes the operator insertion, the black square refers to the mass counterterm in the full theory. The symmetric 1PR diagrams are not shown explicitly.}
\label{fig:Phi2PhiPhiall}
\end{figure}

The tree-level expression for the full-theory correlator \eqref{eq:fullthphi2bispectrumcorrelator} in the late-time limit reads
\begin{equation}
\cor{}{[\phi^2](t,\vec q)\phi(t,\vec k_1)\phi(t,\vec k_2)}'|_{\Lo(\kappa^0)}=\frac{H^4}{2k^3_1k^3_2}\,.
\label{eq::fulltreephi22}
\end{equation}
The one-loop correction to this expression 
can be split into a one-particle irreducible (1PI) and a one-particle reducible (1PR) part as displayed in Fig.~\ref{fig:Phi2PhiPhiall}.  The renormalised 1PI contribution with the comoving IR regulator $\Lambda$ 
is computed in \appref{app:22q0} with the cosmological method-of-regions \cite{Beneke:2023wmt} and reads
\begin{flalign}
&\cor{}{[\phi^2](t,\vec 0)\phi(t,\vec k)\phi(t,-\vec k)}'_{|\,\Lo(\kappa)\textrm{, 1PI}}\nonumber\\
&=\frac{\kappa H^4}{96\pi^2k^6}\,\Bigg\{-3\log\bigg(\frac{e^{\gamma_E}\mu_f}{H}\bigg)+2\log\bigg(\frac{2e^{\gamma_E}k}{a(t)H}\bigg)\bigg[-\log\bigg(\frac{2e^{\gamma_E}k}{a(t)H}\bigg)+2\log\bigg(\frac{2k}{\Lambda}\bigg)+2\,\bigg]\nonumber\\
&\quad-\frac{11}{3}\log\bigg(\frac{2k}{\Lambda}\bigg)-\frac{\pi k^3}{4\Lambda^3}\bigg[2-\log\bigg(\frac{2e^{\gamma_E}k}{a(t)H}\bigg)\bigg]-\frac{9\pi k}{4\Lambda}+\frac{1}{9}+\frac{\pi^2}{6}\Bigg\}\,,
\label{eq:full2221PI}
\end{flalign}
keeping only the non-vanishing terms in the double limit $\ve\rightarrow0$, $\Lambda\rightarrow0$, where the expansion in $\ve$ is performed first and positive powers of $\ve$ and $\Lambda$ are dropped.\footnote{Since the dependence on the external momentum $\vec k$ of the full-theory result is reproduced exactly by the EFT, it is not necessary to expand the regularised external momentum $k_\Lambda$ in $\Lambda$, and we write $k$ instead of $k_{\Lambda}$ for brevity in the final results.
} 
The 1PR-part of the correlator factorises. Making use of the result for the one-loop power spectrum computed in the same regularisation scheme in \cite{Beneke:2026rtf}, it is given by
\begin{flalign}
&\cor{}{\phi^2(t,\vec 0)\phi(t,\vec k)\phi(t,-\vec k)}'_{|\,\Lo(\kappa)\textrm{, 1PR}}\nonumber\\[0.2cm]
&=2\,\cor{}{\phi(t,\vec k)\phi(t,-\vec k)}'_{|\,\Lo(\kappa^0)}\times\cor{}{\phi(t,\vec k)\phi(t,-\vec k)}'_{|\,\Lo(\kappa)}\nonumber\\
&=\frac{\kappa H^4}{24\pi^2k^6}\,\Bigg\{\bigg[2-\log\bigg(\frac{2e^{\gamma_E}k}{a(t)H}\bigg)\bigg]\bigg[\log\bigg(\frac{\Lambda}{a(t)\mu_f}\bigg)-\delta\hat m^2_{\textrm{fin}}\bigg]+\frac{1}{2}\log^2\bigg(\frac{2e^{\gamma_E}k}{a(t)H}\bigg)\nonumber\\
&\quad-\frac{7}{3}\log\bigg(\frac{2e^{\gamma_E}k}{a(t)H}\bigg)+\frac{8}{3}-\frac{\pi^2}{24}\Bigg\}\,.
\label{eq:full2221PR}
\end{flalign}

In the remainder of this section, we match the sum of these two terms to SdSET and extract renormalisation factors, anomalous dimensions, and matching coefficients. Both the 1PI and 1PR terms must be considered, since we match correlation functions and not scattering amplitudes.

\subsection{Set-up of the matching computation}
\label{sec::Phi2match}

The matching equation for the renormalised composite operator $\phi^2$ up to $\Lo(\kappa)$ reads
\begin{equation}
[\phi^2](t,\vec x)=H^2\Big[C_{20}\un+C_{22}[\vp^2_+](t,\vec x)+C_{24}[\vp^4_+](t,\vec x)+\Lo(\kappa^2)\Big]\,,
\label{eq::Phi2opmatch}
\end{equation}
where 
\begin{equation}
C_{20}=\Lo(\kappa^0)\,,\quad C_{22}=1+\Lo(\kappa)\,,\quad C_{24}=\Lo(\kappa)\,.
\end{equation}
We refer to \eqref{eq:C22C20free} for the $\mathcal{O}(\kappa^0)$ results. Here, we are interested in the next order in $\kappa$.
The coefficient $C_{24}$ at $\mathcal{O}(\kappa)$ can be found 
without any explicit matching computation by the following reasoning: 
the relation between the bare full-theory and EFT-fields up to $\Lo(\kappa)$ and at leading power in $\lambda$ reads~\cite{Beneke:2026rtf}
\begin{equation}
\phi(t,\vec x)=H(a(t)H)^{\ve}\bigg[\bigg(1+\frac{c^0_{1,1}}{9}\bigg)\vp_+(t,\vec x)+\frac{c^0_{3,1}a(t)^{2\ve}}{18(3+2\ve)}\,\vp^3_+(t,\vec x)+\Lo(\kappa^2)\bigg]\,.
\label{eq::phivprelation}
\end{equation}
This equation implies that at $\Lo(\kappa)$ the operator $[\phi^2]$ is not only related to $[\vp^2_+]$, but also to $[\vp^4_+]$, due to the $\vp^3_+$ term  in \eqref{eq::phivprelation}. By squaring both sides of \eqref{eq::phivprelation} and inspecting the resulting coefficient of $\vp^4_+$ in four dimensions, we infer
\begin{equation}
C_{24}=\frac{\kappa}{27}+\Lo(\kappa^2)\,,
\end{equation}
where $c^0_{3,1}=\kappa+\Lo(\kappa^2)$ has been used.
Below, we confirm this result by explicit matching.

From the operator-matching equation \eqref{eq::Phi2opmatch} and the relation \eqref{eq::phivprelation}, we derive the matching equation for the connected momentum-space correlation functions, up to $\Lo(\kappa)$
\begin{flalign}
&\cor{}{[\phi^2](t,\vec 0)\phi(t,\vec k)\phi(t,-\vec k)}'_{|\,\Lo(\kappa)}\nonumber\\
&=H^4\,\Bigg\{\Big[C_{22}\,\cor{}{[\vp^2_+](t,\vec 0)\vp_+(t,\vec k)\vp_+(t,-\vec k)}'+\,C_{24}\cor{}{[\vp^4_+](t,\vec 0)\vp_+(t,\vec k)\vp_+(t,-\vec k)}'\Big]\Big|_{\,\Lo(\kappa)}\nonumber\\
&\quad+\frac{2c^0_{1,1}}{9}\,\cor{}{[\vp^2_+](t,\vec 0)\vp_+(t,\vec k)\vp_+(t,-\vec k)}'_{|\,\Lo(\kappa^0)}+\frac{c^0_{3,1}a(t)^{2\ve}}{18(3+2\ve)}\,\Big[\cor{}{[\vp^2_+](t,\vec 0)\vp^3_+(t,\vec k)\vp_+(t,-\vec k)}'\nonumber\\
&\quad+\cor{}{[\vp^2_+](t,\vec 0)\vp_+(t,\vec k)\vp^3_+(t,-\vec k)}'\Big]\Big|_{\,\Lo(\kappa^0)}\Bigg\}\,.
\label{eq::phi2cormatch2}
\end{flalign}
 In the above equation, we use the same definition as in~\cite{Beneke:2026rtf} for the momentum-space composite operators, 
\begin{equation}
\vp^n_+(t,\vec k)\equiv\int\der^{d-1}x\;e^{-i\vec k\cdot\vec x}\vp^n_+(t,\vec x)\,.
\end{equation}

The terms appearing in the last two lines of the matching equation \eqref{eq::phi2cormatch2} account for the terms generated by the relation \eqref{eq::phivprelation} between the full theory and SdSET field 
associated to the external fields, but not for the ones associated to the renormalised operator $[\phi^2]$. These must be reproduced by the matching coefficient $C_{22}$, which therefore receives contributions from both the 1PI- and 1PR-components of the correlation function. 
To this end, we also need to compute the additional composite-operator correlation functions appearing on the right-hand side of \eqref{eq::phi2cormatch2}. Note that while the operators $\vp^2_+$ and $\vp^4_+$ have been renormalised, the correlation functions in the matching equation \eqref{eq::phi2cormatch2} also feature the bare operator $\vp^3_+$, which generates new UV divergences, and bare effective couplings containing explicit UV poles, which arise from \eqref{eq::phivprelation}.
Since the left-hand side of \eqref{eq::phi2cormatch2} is UV-finite, all remaining UV divergences on the right-hand side must cancel once all the pieces are combined. Indeed, analogously to the matching of the one-loop power spectrum in \cite{Beneke:2026rtf}, the pole contained in $c^0_{1,1}$ cancels the one generated by the insertion of the bare composite operators $\vp^3_+$ into the correlation functions in the last two lines of \eqref{eq::phi2cormatch2}. We explicitly verify this below. 

We finally note that no term proportional to $C_{20}$ appears in \eqref{eq::phi2cormatch2}, since such a term only contributes to the disconnected component of the correlation function, 
which is not relevant for the matching. However, the coefficient $C_{20}$ will play a central role in~\secref{sec::vp20}.

\subsection{Bispectrum of \texorpdfstring{$\vp^2_+$}{φ₊²}}

We start by computing the SdSET counterpart of the correlation function, defined as
\begin{equation}
\cor{}{\vp_+^2(t,\vec 0)\vp_+(t,\vec k)\vp_+(t,-\vec k)}=\int\frac{\der^{d-1}l}{(2\pi)^{d-1}}\cor{}{\vp_+(t,\vec l)\vp_+(t,-\vec l)\vp_+(t,\vec k)\vp_+(t,-\vec k)}'\,,
\label{eq::SdSET22}
\end{equation}
which appears in the first line of the matching equation \eqref{eq::phi2cormatch2}.  

This correlator was already considered in \cite{Cohen:2021fzf}, where the authors focused on the extraction of the pole terms from its initial-condition contribution as required to determine the Kramers-Moyal coefficients. In the following, we present the full computation including the finite parts required for extracting matching coefficients. 

As was the case in the full theory, the correlator \eqref{eq::SdSET22} can be split into a 1PI and a 1PR part. To renormalise the operator, only the former is relevant, while for the matching computation also the latter is needed. We begin by computing the 1PI topologies. They feature UV-divergent terms, which must be removed by renormalising the operator $\vp^2_+$. 

\subsubsection{EFT-vertex and initial-condition counterterm insertion contributions}

\begin{figure}[t]
\centering
\begin{subfigure}{0.44\textwidth}
\centering
\includegraphics[width=\textwidth]{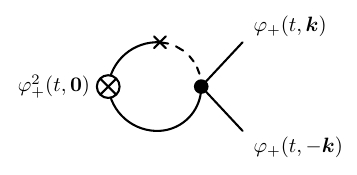}
\caption{}
\end{subfigure}%
\hspace{1cm}\begin{subfigure}{0.44\textwidth}
\centering
\includegraphics[width=\textwidth]{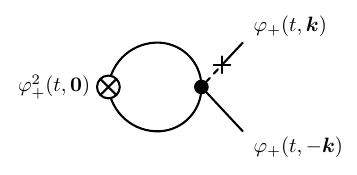}
\caption{}
\end{subfigure}
\caption{The two 1PI diagrams topologies resulting from the insertion of the quartic EFT vertex into the correlation function $\cor{}{\vp^2_+(t,\vec 0)\vp_+(t,\vec k)\vp_+(t,-\vec k)}$. Symmetric diagrams with crossed propagators on the respective lower lines are not shown explicitly.}
\label{fig:Phi2PhiPhiEFT1}
\end{figure}

It is convenient to combine the quartic EFT vertex insertion  proportional to $c_{3,1}$ with the $\xi_{3,1}$ initial-condition counterterm contribution, since in the sum their poles partially cancel.\footnote{Recall that the initial-condition counterterm is constructed to remove the pole from the UV-divergent time integral~\cite{Beneke:2026rtf}.} 
The topologies corresponding to the insertion of the quartic EFT vertex are shown in~\figref{fig:Phi2PhiPhiEFT1}. The same topologies arise when inserting the initial-condition counterterm, with the vertex symbol replaced by a small full square.
We find
\begin{flalign}
&\cor{}{\vp_+^2(t,\vec 0)\vp_+(t,\vec k)\vp_+(t,-\vec k)}'_{|\,c_{3,1}+\xi_{3,1}}\nonumber\\
&=-\frac{\tmu^{2\ve}\kappa}{24k_{\Lambda}^6}\frac{a(t)^{2\ve}-a^{2\ve}_*}{2\ve}\,\Bigg[2\int\frac{\der^{d-1}l}{(2\pi)^{d-1}}\frac{1}{l^3_{\Lambda}}+2k^3_{\Lambda}\int\frac{\der^{d-1}l}{(2\pi)^{d-1}}\frac{1}{l^6_{\Lambda}}\,\Bigg]\nonumber\\
&=\frac{\kappa}{96\pi^2k^6}\bigg(\frac{\Lambda}{a(t)\mu}\bigg)^{-2\ve}\log\bigg(\frac{a_*}{a(t)}\bigg)\,\Bigg[\frac{2}{\ve}+2\log\bigg(\frac{a_*}{a(t)}\bigg)+\frac{\pi k^3}{4\Lambda^3}\,\Bigg]\,,
\label{eq::22vertexins}
\end{flalign}
where we kept the full $\ve$-dependence only in the prefactor for compactness. 
This result is both UV- and IR-divergent.
As expected, the $\ve$-pole originating from the time integral in the vertex-insertion diagrams is cancelled by the initial-condition counterterm $\xi_{3,1}$.
Therefore, only the UV divergence generated by the momentum integral remains. The IR divergences are logarithmic, manifested by the $\log \Lambda$ term generated by the $\ve$-expansion, and  power-like, as evidenced by the  $1/\Lambda^3$ term.

\subsubsection{Initial-condition insertion}

\begin{figure}[t]
\centering
\begin{subfigure}{0.44\textwidth}
\centering
\includegraphics[width=\textwidth]{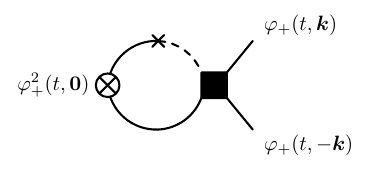}
\caption{}
\end{subfigure}%
\hspace{1cm}\begin{subfigure}{0.44\textwidth}
\centering
\includegraphics[width=\textwidth]{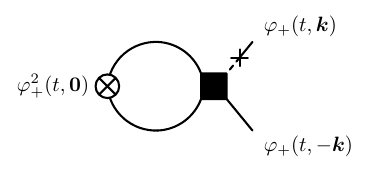}
\caption{}
\end{subfigure}
\caption{The two 1PI diagrams topologies resulting from the insertion of the quartic initial condition into the correlation function $\cor{}{\vp^2_+(t,\vec 0)\vp_+(t,\vec k)\vp_+(t,-\vec k)}$.}
\label{fig:Phi2PhiPhiEFT2}
\end{figure}

The second contribution to the 1PI part of the correlation function is the insertion of the renormalised initial-condition function $\Xi_{3,1}$.
This function is determined in~\cite{Beneke:2026rtf} and reads
\begin{equation}
\Xi_{3,1}(\vec k_1,\vec k_2,\vec k_3,\vec k_4)=3\kappa\,\Bigg\{\frac{1}{3}\log\bigg(\frac{e^{\gamma_E}k_t}{a_*H}\bigg)+\bigg(\sum_{j=1}^4k^3_j\bigg)^{-1}\bigg[-\frac{k_1k_2k_3k_4}{k_t}+k_t\bigg(\sum_{j<l}k_jk_l-\frac{4}{9}k_t^2\bigg)\bigg]\Bigg\}\,.
\end{equation}
The respective topologies for its insertion are shown in~\figref{fig:Phi2PhiPhiEFT2}. Replacing $k_i$ by $k_{i\Lambda}$, leads to the expression
\begin{flalign}
&\cor{}{\vp^2_+(t,\vec 0)\vp_+(t,\vec k)\vp_+(t,-\vec k)}'_{|\,\Xi_{3,1}}\nonumber\\
&=-\frac{i(a_*\tmu)^{2\ve}}{8}\int\frac{\der^{d-1}l}{(2\pi)^{d-1}}\frac{2(k_{\Lambda}^3+l_{\Lambda}^3)}{k_{\Lambda}^6l_{\Lambda}^6}\,\frac{i}{3}\Xi_{3,1}(\vec k,-\vec k,\vec l,-\vec l)\nonumber\\
&=\frac{\kappa(a_*\tmu)^{2\ve}}{8k_{\Lambda}^6}\int\frac{\der^{d-1}l}{(2\pi)^{d-1}}\frac{1}{l_{\Lambda}^6}\,\Bigg[\frac{2(l_{\Lambda}^3+k_{\Lambda}^3)}{3}\log\bigg(\frac{2e^{\gamma_E}(l_{\Lambda}+k_{\Lambda})}{a_*H}\bigg)-\frac{l_{\Lambda}^2k_{\Lambda}^2}{2(l_{\Lambda}+k_{\Lambda})}\nonumber\\
&\hspace{5cm}-\frac{2}{9}\Big(7k_{\Lambda}^3+3k_{\Lambda}^2l_{\Lambda}+3k_{\Lambda}l_{\Lambda}^2+7l_{\Lambda}^3\Big)\Bigg]\,.
\end{flalign}
This expression features simple integrals involving only powers of $l_\Lambda$ and non-trivial integrals containing the combination $(l_{\Lambda}+k_{\Lambda})$.
The details of their evaluation are allocated to~\appref{app::22q0loops}. We obtain
\begin{flalign}
&\cor{}{\vp^2_+(t,\vec 0)\vp_+(t,\vec k)\vp_+(t,-\vec k)}'_{|\,\Xi_{3,1}}\nonumber\\
&=\frac{\kappa}{96\pi^2k^6}\,\Bigg\{\frac{1}{\ve^2}+\frac{2}{\ve}\bigg[\log\bigg(\frac{e^{\gamma_E}\mu}{H}\bigg)-\frac{4}{3}\bigg]+\log\bigg(\frac{e^{\gamma_E}\Lambda}{a_*H}\bigg)\bigg[5-4\log\bigg(\frac{2e^{\gamma_E}k}{a_*H}\bigg)\bigg]\nonumber\\
&\phantom{=}+\log\bigg(\frac{2e^{\gamma_E}k}{a_*H}\bigg)\bigg[\frac{1}{3}+2\log\bigg(\frac{2e^{\gamma_E}k}{a_*H}\bigg)\bigg]-2\log\bigg(\frac{e^{\gamma_E}\mu}{H}\bigg)\bigg[\frac{8}{3}-\log\bigg(\frac{e^{\gamma_E}\mu}{H}\bigg)\bigg]\nonumber\\
&\phantom{=}+\frac{1}{9}+\frac{5\pi^2}{12}-\frac{9\pi k}{4\Lambda}-\frac{\pi k^3}{4\Lambda^3}\bigg[\frac{7}{3}-\log\bigg(\frac{2e^{\gamma_E}k}{a_*H}\bigg)\bigg]\Bigg\}\,.
\label{eq::22ICins}
\end{flalign}
This result features a double pole, despite formally being of one-loop order.
It originates from the UV-divergence of the integral containing the logarithm.
As a consequence, the single pole term contains a logarithm of the factorisation scale $\mu$ which vanishes when setting $\mu=e^{-\gamma_E}H$.
Crucially, however, this is still a local term which can be renormalised using a standard $Z$-factor.

\subsubsection{Renormalisation of \texorpdfstring{
$\vp^2_+$}{φ₊²}}

Collecting the pole terms of both contributions, we get
\begin{flalign}
&\Big[\cor{}{\vp^2_+(t,\vec 0)\vp_+(t,\vec k)\vp_+(t,-\vec k)}'_{|\,c_{3,1}+\xi_{3,1}+\Xi_{3,1}}\Big]\Big|_{\,\textrm{poles}}\nonumber\\
&=\frac{\kappa}{96\pi^2k^6}\,\Bigg\{\frac{1}{\ve^2}-\frac{2}{\ve}\bigg[\frac{4}{3}-\log\bigg(\frac{e^{\gamma_E}\mu}{H}\bigg)\bigg]\Bigg\}+\frac{c_{3,1}}{48\pi^2k^6}\frac{1}{\ve}\log\bigg(\frac{a_*}{a(t)}\bigg)\,.
\label{eq:phi2phiphi1PI}
\end{flalign}
To renormalise the operator $\vp^2_+$ we use \eqref{eq::vpnren} in the form
\begin{equation}
(a(t)\tmu)^{2\ve}\vp^2_+(t,\vec x)=
(a(t)\tmu)^{2\ve}\underbrace{Z_{22}\,[\vp^2_+](t,\vec x)}_{\rm 1PI}+
\underbrace{Z_{20}\,\un}_{\rm disconnected}+
(a(t)\tmu)^{4\ve}\underbrace{Z_{24}\,[\vp^4_+](t,\vec x)+
\ldots}_{\mathcal{O}(\kappa^2)}\,.
\label{eq::vp2reneq}
\end{equation}
The underbraces refer to properties of the respective summands when inserted into the correlator \eqref{eq::SdSET22}.
Eq.~\eqref{eq:phi2phiphi1PI}  therefore allows us to determine $Z_{22}$ in this equation. Indeed, the $k$ dependence of the pole terms matches the structure of the free-theory correlation function
\begin{equation}
\cor{}{[\vp^2_+](t,\vec 0)\vp_+(t,\vec k)\vp_+(t,-\vec k)}'_{|\,\Lo(\kappa^0)}=\frac{1}{2k^6}\,,
\label{eq:freevp2vpvp}
\end{equation}
cf. \eqref{eq::fulltreephi22}, from which we infer that the appropriate renormalisation factor to absorb the UV poles is
\begin{equation}
Z_{22}=1+\frac{\kappa}{48\pi^2}\,\Bigg\{\frac{1}{\ve^2}+\frac{2}{\ve}\bigg[\log\bigg(\frac{e^{\gamma_E}\mu}{H}\bigg)-\frac{4}{3}\bigg]\Bigg\}+\frac{c_{3,1}}{24\pi^2}\frac{1}{\ve}\log\bigg(\frac{a_*}{a(t)}\bigg)+\Lo(\kappa^2)\,.
\label{eq::z22}
\end{equation}

Similarly to the case of the initial-condition counterterm $\xi_{1,1}$ computed in \cite{Beneke:2026rtf}, we used the matched expression for $\Xi_{3,1}$ in the above computation, which leads to the full-theory coupling $\kappa$ appearing in \eqref{eq::z22}. 
This is not necessary, and $Z_{22}$ could be written more generally as containing the pole part of the integral over the generic IC function $\Xi_{3,1}$.
The effective couplings $c_{3,1}$ and $\Xi_{3,1}$ differ in their RG evolution, and when computing anomalous dimensions later, it is important to distinguish their contributions to the $Z$-factors. 
The above result~\eqref{eq::z22} has a double pole as well as a single pole multiplied by a logarithm due to the combined time- and momentum-integral divergence. However, it remains local as required for operator renormalisation.
In addition, it contains an interesting feature: the pole term from the vertex insertion~\eqref{eq::22vertexins} vanishes under the scale choice $a_* = a(t)$. For this choice, the renormalisation originates completely from the IC insertion~\eqref{eq::22ICins}.

For matching to the full-theory correlator, we may use the matched value $c_{3,1} = \kappa + \mathcal{O}(\kappa^2)$, to find the renormalised expression 
\begin{flalign}
&\cor{}{[\vp^2_+](t,\vec 0)\vp_+(t,\vec k)\vp_+(t,-\vec k)}'_{|\,c_{3,1}+\xi_{3,1}+\Xi_{3,1}}\nonumber\\
&=\frac{\kappa}{96\pi^2k^6}\,\Bigg\{2\log\bigg(\frac{e^{\gamma_E}\mu}{H}\bigg)\bigg[\log\bigg(\frac{e^{\gamma_E}\mu}{H}\bigg)+2\log\bigg(\frac{a_*}{a(t)}\bigg)-\frac{8}{3}\bigg]\nonumber\\
&\quad+2\log\bigg(\frac{2e^{\gamma_E}k}{a(t)H}\bigg)\bigg[\log\bigg(\frac{2e^{\gamma_E}k}{a(t)H}\bigg)-2\log\bigg(\frac{e^{\gamma_E}\Lambda}{a(t)H}\bigg)-2\log\bigg(\frac{a_*}{a(t)}\bigg)-\frac{4}{3}\bigg]\nonumber\\
&\quad+\frac{28}{9}+\frac{5\pi^2}{12}-\frac{3\pi k}{\Lambda}-\frac{\pi k^3}{4\Lambda^3}\bigg[\frac{7}{3}-\log\bigg(\frac{2e^{\gamma_E}k}{a(t)H}\bigg)\bigg]\Bigg\}\,.
\label{eq::22q0EFT}
\end{flalign}
This result constitutes the first consistency check in the interacting theory of the SdSET operator-renormalisation framework set up in \secref{sec:EFTopmatch}, and it serves as one of the building blocks for the matching computation below. 

\subsubsection{1PR contribution to the bispectrum of \texorpdfstring{$\vp^2_+$}{φ₊²}}
\label{sec::22q01PR}

The computation of the 1PR component necessitates no further calculations, since it factorises into the product of two power spectra
\begin{flalign}
&\cor{}{\vp^2_+(t,\vec 0)\vp_+(t,\vec k)\vp_+(t,-\vec k)}'_{|\,\Lo(\kappa)\textrm{, 1PR}}\nonumber\\
&=2\cor{}{\vp_+(t,\vec k)\vp_+(t,-\vec k)}'_{|\,\Lo(\kappa^0)}\times\cor{}{\vp_+(t,\vec k)\vp_+(t,-\vec k)}'_{|\,\Lo(\kappa)}\,.
\end{flalign}
We can therefore directly use the known results for the tree-level and renormalised one-loop SdSET power spectra~\cite{Beneke:2026rtf} to obtain
\begin{flalign}
&\cor{}{\vp^2_+(t,\vec 0)\vp_+(t,\vec k)\vp_+(t,-\vec k)}'_{|\,\Lo(\kappa)\textrm{, 1PR}}\nonumber\\
&=\frac{\kappa}{24\pi^2k^6}\,\Bigg\{-\log\bigg(\frac{2e^{\gamma_E}k}{a(t)H}\bigg)\bigg[\log\bigg(\frac{\Lambda}{a(t)\mu}\bigg)-\hat c_{1,1}\bigg]+4\log\bigg(\frac{\Lambda}{a(t)\mu}\bigg)\nonumber\\
&\quad+\frac{5\hat c_{1,1}}{3}+\frac{1}{2}\log^2\bigg(\frac{2e^{\gamma_E}k}{a(t)H}\bigg)-\frac{7}{3}\log\bigg(\frac{2e^{\gamma_E}k}{a(t)H}\bigg)+\frac{10}{3}-\frac{\pi^2}{24}\,\Bigg\}\,,
\label{eq:22q0EFT1PR}
\end{flalign}
where $\hat c_{1,1}$ denotes the finite part of the SdSET mass counterterm~\cite{Beneke:2026rtf}.

\subsection{Remaining SdSET correlators}

We now focus on the remaining terms in the matching equation \eqref{eq::phi2cormatch2}. 

\subsubsection{Contribution from \texorpdfstring{$[\vp^4_+](t,\vec x)$}{[φ₊⁴](t,x)}}

In \secref{sec:freemix} above, we already defined the renormalised position-space correlator
\begin{equation}
\cor{}{[\vp^4_+](t,\vec x)\vp_+(t,\vec x_1)\vp_+(t,\vec x_2)}=-\frac{3}{2\pi^2}\log\bigg(\frac{\Lambda}{a(t)\mu}\bigg)\cor{}{[\vp^2_+](t,\vec x)\vp_+(t,\vec x_1)\vp_+(t,\vec x_2)}\,,
\end{equation}
and it is depicted diagrammatically in \figref{fig:42freemix}. In momentum space, setting $\vec{q}=0$, $\vec k_1=-\vec k_2=\vec k$, it translates into
\begin{equation}
\cor{}{[\vp^4_+](t,\vec 0)\vp_+(t,\vec k)\vp_+(t,-\vec k)}'=-\frac{3}{4\pi^2k^6}\log\bigg(\frac{\Lambda}{a(t)\mu}\bigg)\,.
\label{eq::22q0vp4}
\end{equation}

\subsubsection{Contributions from full-to-SdSET field relation}

\begin{figure}[t]
\centering
\begin{subfigure}{0.46\textwidth}
\centering
\raisebox{0.2cm}{\includegraphics[width=\textwidth]{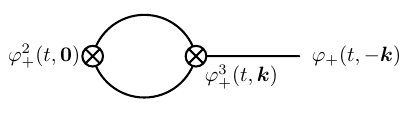}}
\caption{}
\end{subfigure}%
\hspace{1cm}\begin{subfigure}{0.46\textwidth}
\centering
\includegraphics[width=\textwidth]{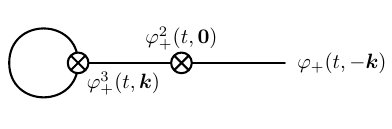}
\caption{}
\end{subfigure}
\caption{The 1PI (a) and 1PR (b) diagram topologies resulting from the free-theory correlation function $\cor{}{\vp^2_+(t,\vec 0)\vp^3_+(t,\vec k)\vp_+(t,-\vec k)}'$.}
\label{fig:vp221PR}
\end{figure}

Next, we consider 
\begin{flalign}
&\cor{}{\vp^2_+(t,\vec q)\vp^3_+(t,\vec k_1)\vp_+(t,\vec k_2)}_{|\,\Lo(\kappa^0)}\nonumber\\
&=6\,(2\pi)^{d-1}\delta^{(d-1)}(\vec k_1+\vec k_2+\vec q)\,\frac{1}{2k^3_{2\Lambda}}\int\frac{\der^{d-1}l}{(2\pi)^{d-1}}\bigg[\frac{1}{4l^3_{\Lambda}|\vec l-\vec q|^3_{\Lambda}}+\frac{1}{4k^3_{1\Lambda}l^3_{\Lambda}}\bigg]\,,
\end{flalign}
with the kinematic configuration $\vec{q}=0$, $\vec k_1=-\vec k_2=\vec k$, as well as the correlator with $\vp^3_+$ and $\vp_+$ exchanged. It is sufficient to compute the above correlation function, since the other simply adds a factor of two. 
 We also note that we are allowed to replace the renormalised operator $[\vp_+^2]$, which appears in this correlator in \eqref{eq::phi2cormatch2}, by the bare one, since the difference at $\mathcal{O}(\kappa^0)$ amounts to an irrelevant disconnected contribution.  It remains to compute the two diagrams shown in  \figref{fig:vp221PR}. We find 
\begin{align}
\cor{}{\vp^2_+(t,\vec 0)\vp^3_+(t,\vec k)\vp_+(t,-\vec k)}'_{|\,\Lo(\kappa^0),\,\textrm{(a)}}&=\frac{3}{4k_{\Lambda}^3}\int\frac{\der^{d-1}l}{(2\pi)^{d-1}}\frac{1}{l_{\Lambda}^6}
=\frac{3}{4k^3}\frac{1}{32\pi\Lambda^3}\,,
\label{eq:phi3diaga}\\
\cor{}{\vp^2_+(t,\vec 0)\vp^3_+(t,\vec k)\vp_+(t,-\vec k)}'_{|\,\Lo(\kappa^0),\,\textrm{(b)}}&=\frac{3}{4k_{\Lambda}^6}\int\frac{\der^{d-1}l}{(2\pi)^{d-1}}\frac{1}{l^3_{\Lambda}}
=\frac{3}{16\pi^2k^6}\bigg(\frac{e^{\gamma_E}\Lambda^2}{4\pi}\bigg)^{\!-\ve}\,\frac{1}{\ve}\,,
\label{eq:phi3diagb}
\end{align} 
for diagrams (a) and (b), respectively.

Finally, the tree-level correlator
\begin{equation}
\cor{}{[\vp^2_+](t,\vec 0)\vp_+(t,\vec k)\vp_+(t,-\vec k)}'_{|\,\Lo(\kappa^0)}=\frac{1}{2k^6}
\end{equation}
is already given in \eqref{eq:freevp2vpvp}.

Putting everything together in \eqref{eq::phi2cormatch2}, and using the known $d$-dimensional results for the bare effective couplings \cite{Beneke:2026rtf}
\begin{flalign}
c^0_{1,1}&=\;\frac{\kappa}{8\pi^2}\bigg[-\frac{1}{2\ve}+\underbrace{\delta\hat m^2_{\textrm{fin}}+\log\bigg(\frac{\mu_f}{\mu}\bigg)}_{=\,\hat c_{1,1}}\bigg]+\Lo(\kappa^2)\,,\label{eq::c11}\\
c^0_{3,1}&=\tmu^{2\ve}c_{3,1}+\Lo(\kappa^2)\,,
\end{flalign}
we obtain 
\begin{flalign}
&\frac{2c^0_{1,1}}{9}\cor{}{[\vp^2_+](t,\vec 0)\vp_+(t,\vec k)\vp_+(t,-\vec k)}'_{|\,\Lo(\kappa^0)}+\frac{c^0_{3,1}a(t)^{2\ve}}{18(3+2\ve)}\Big[\cor{}{\vp^2_+(t,\vec 0)\vp^3_+(t,\vec k)\vp_+(t,-\vec k)}'\nonumber\\
&\quad+\cor{}{\vp^2_+(t,\vec 0)\vp_+(t,\vec k)\vp^3_+(t,-\vec k)}'\Big]\Big|_{\,\Lo(\kappa^0)}\nonumber\\
&=\frac{c_{3,1}}{96\pi^2k^6}\frac{4}{3}\,\bigg[\hat c_{1,1}-\frac{1}{3}-\log\bigg(\frac{\Lambda}{a(t)\mu}\bigg)+\frac{\pi k^3}{16\Lambda^3}\bigg]\,,
\label{eq:fieldrelationcontributions}
\end{flalign}
As anticipated above, even though these combinations involve bare effective couplings, the specific combinations entering the matching equation \eqref{eq::phi2cormatch2} are UV finite.

\subsection{Determination of the operator matching coefficients}
\label{sec::C22det}

Up to $\Lo(\kappa)$, the relevant terms of the operator matching equation \eqref{eq::Phi2opmatch} are
\begin{equation}
[\phi^2](t,\vec x)=H^2\Big[\Big(1+C^{(\kappa)}_{22}\Big)[\vp^2_+](t,\vec x)+C^{(\kappa)}_{24}[\vp^4_+](t,\vec x)+\Lo(\kappa^2)\Big]\,.
\end{equation}
To determine $C^{(\kappa)}_{22}$, $C^{(\kappa)}_{24}$, we compare the left-hand side of \eqref{eq::phi2cormatch2}, i.e. 
\begin{equation}
    H^{-4}\cor{}{[\phi^2](t,\vec 0)\phi(t,\vec k)\phi(t,-\vec k)}'_{|\,\Lo(\kappa)}\,,
\end{equation}
given by the sum of \eqref{eq:full2221PI} and \eqref{eq:full2221PR}
to the right-hand side
\vskip-0.4cm
\begin{flalign}
&\cor{}{[\vp^2_+](t,\vec 0)\vp_+(t,\vec k)\vp_+(t,-\vec k)}'_{|\,\Lo(\kappa)}+C^{(\kappa)}_{22}\cor{}{[\vp^2_+](t,\vec 0)\vp_+(t,\vec k)\vp_+(t,-\vec k)}'_{|\,\Lo(\kappa^0)}\nonumber\\
&\quad
+C^{(\kappa)}_{24}\cor{}{[\vp^4_+](t,\vec 0)\vp_+(t,\vec k)\vp_+(t,-\vec k)}'_{|\,\Lo(\kappa^0)}\\
&\quad+\frac{c^0_{3,1}a(t)^{2\ve}}{18(3+2\ve)}\Big[\cor{}{[\vp^2_+](t,\vec 0)\vp^3_+(t,\vec k)\vp_+(t,-\vec k)}'+\cor{}{[\vp^2_+](t,\vec 0)\vp_+(t,\vec k)\vp^3_+(t,-\vec k)}'\Big]\Big|_{\,\Lo(\kappa^0)}\,.\nonumber
\label{eq::1PIphi2match}
\end{flalign}
Inserting the sum of \eqref{eq::22q0EFT} and \eqref{eq:22q0EFT1PR} for the first term and \eqref{eq:phi3diaga} plus \eqref{eq:phi3diagb} to evaluate the last line, we solve \eqref{eq::phi2cormatch2} for 
\vskip-0.4cm
\begin{flalign}
&C^{(\kappa)}_{22}\cor{}{[\vp^2_+](t,\vec 0)\vp_+(t,\vec k)\vp_+(t,-\vec k)}'_{|\,\Lo(\kappa^0)}+C^{(\kappa)}_{24}\cor{}{[\vp^4_+](t,\vec 0)\vp_+(t,\vec k)\vp_+(t,-\vec k)}'_{|\,\Lo(\kappa^0)\textrm{, 1PI}}\nonumber\\
&=\frac{\kappa}{96\pi^2k^6}\,\bigg\{\log\bigg(\frac{e^{\gamma_E}\mu}{H}\bigg)\bigg[\frac{7}{3}-4\log\bigg(\frac{a_*}{a(t)}\bigg)-2\log\bigg(\frac{e^{\gamma_E}\mu}{H}\bigg)\bigg]+4\log\bigg(\frac{a_*}{a(t)}\bigg)-3\log\bigg(\frac{\mu_f}{\mu}\bigg)\nonumber\\
&\quad-\frac{4}{3}\log\bigg(\frac{e^{\gamma_E}\Lambda}{a_*H}\bigg)-\frac{\pi^2}{4}\bigg\}\nonumber\\
&\quad+\frac{\kappa}{36\pi^2k^6}\bigg[\delta\hat m^2_{\textrm{fin}}-\log\bigg(\frac{\Lambda}{a(t)\mu_f}\bigg)-\frac{1}{3}\bigg]
-\frac{c_{3,1}}{72\pi^2k^6}\bigg[\hat c_{1,1}-\frac{1}{3}-\log\bigg(\frac{\Lambda}{a(t)\mu}\bigg)\bigg]\,.
\end{flalign}
Inserting $\hat{c}_{1,1}$ from \eqref{eq::c11} and $c_{3,1}=\kappa$, the right-hand side simplifies to
\begin{flalign}
&C^{(\kappa)}_{22}\cor{}{[\vp^2_+](t,\vec 0)\vp_+(t,\vec k)\vp_+(t,-\vec k)}'_{|\,\Lo(\kappa^0)}+C^{(\kappa)}_{24}\cor{}{[\vp^4_+](t,\vec 0)\vp_+(t,\vec k)\vp_+(t,-\vec k)}'_{|\,\Lo(\kappa^0)\textrm{, 1PI}}\nonumber\\
&=\frac{\kappa}{96\pi^2k^6}\bigg\{-4\log\bigg(\frac{a_*}{a(t)}\bigg)\bigg[\log\biggl(\frac{e^{\gamma_E}\mu}{H}\biggr)-\frac{4}{3}\bigg]-2\log^2\!\bigg(\frac{e^{\gamma_E}\mu}{H}\bigg)+\log\bigg(\frac{e^{\gamma_E}\mu}{H}\bigg)\nonumber\\
&\phantom{=}-\frac{5}{3}\log\bigg(\frac{\mu_f}{\mu}\bigg)+\frac{4}{3}\delta\hat m^2_{\textrm{fin}}-\frac{4}{9}-\frac{\pi^2}{4}\bigg\}-\frac{\kappa}{36\pi^2k^6}\log\bigg(\frac{\Lambda}{a(t)\mu}\bigg)\,.
\end{flalign}
Evaluating the free correlation functions in the first line with the help of \eqref{eq:freevp2vpvp} and 
\begin{flalign}
\cor{}{[\vp^4_+](t,\vec 0)\vp_+(t,\vec k)\vp_+(t,-\vec k)}'_{|\,\Lo(\kappa^0)}&=-\frac{3}{4\pi^2k^6}\log\bigg(\frac{\Lambda}{a(t)\mu}\bigg)\,,\label{eq::vp40tree}
\end{flalign}
we observe that the $\log\Lambda$-term is indeed matched by
\begin{equation}
C^{(\kappa)}_{24}=\frac{\kappa}{27}+\Lo(\kappa^2)\,,
\label{eq::C24}
\end{equation}
as anticipated. The remaining terms yield the desired result for the one-loop correction to the operator matching coefficient $C_{22}$. Including its tree-level value, we obtain 
\begin{flalign}
C_{22}&=1+\frac{\kappa}{48\pi^2}\bigg\{-4\log\bigg(\frac{a_*}{a(t)}\bigg)\bigg[\log\biggl(\frac{e^{\gamma_E}\mu}{H}\biggr)-\frac{4}{3}\bigg]-2\log^2\!\bigg(\frac{e^{\gamma_E}\mu}{H}\bigg)+\log\bigg(\frac{e^{\gamma_E}\mu}{H}\bigg)\nonumber\\
&\phantom{=}-\frac{5}{3}\log\bigg(\frac{\mu_f}{\mu}\bigg)+\frac{4}{3}\delta\hat m^2_{\textrm{fin}}-\frac{4}{9}-\frac{\pi^2}{4}\bigg\}+\Lo(\kappa^2)\,,
\label{eq::C22final}
\end{flalign}
which is the main result of this section.

The fact that the complete $\Lambda$ and $k$ dependence of the full-theory correlation function is reproduced by matching local coefficients $C_{22}$ and $C_{24}$ constitutes a highly non-trivial consistency check on the general framework of SdSET. The coefficient $C_{22}$ is local in position, but depends on the scale $\mu$, as well as on the scale factors $a_*$ and $a(t)$. Its logarithmic terms vanish when the matching is carried out at a common scale $\mu=\mu_f$, and, furthermore, when choosing $\mu=H e^{-\gamma_E}$ and $a_*=a(t)$, which defines the natural scales at which $C_{22}$ should be evaluated. This reflects the physics encoded in the operator-matching coefficients, since they contain the information about the late-time, hard-momentum contributions to the composite-operator correlation functions in the full theory \cite{Beneke:2023wmt}. Therefore, for $\mu$ chosen at the scale of hard momenta ($\mu\sim H$) and $a_*$ at late times ($a_*\sim a(t)$), the matching coefficients should be free of large logarithms. This expectation is confirmed by the above result for $C_{22}$. 

\subsection{Anomalous dimensions}
\label{sec::22ADMs}

With the result for $Z_{22}$ we can compute the operator anomalous dimensions $\gamma^{\mu}_{22}$ and $\gamma^{a_*}_{22}$ at $\Lo(\kappa)$ using \eqref{eq::muADM} and \eqref{eq::asADM}. For the case of interest, they read 
\begin{flalign}
\gamma^{\mu}_{22}&=-2\ve+\sum_{l=0}^{\infty}Z^{-1}_{2l}\bigg[l\ve Z_{l2}-\frac{\der Z_{l2}}{\der\log(\mu)}\bigg]\,,\label{eq::gamma22Zmu}\\
\gamma^{a_*}_{22}&=-\sum_{l=0}^{\infty}Z^{-1}_{2l}\frac{\der Z_{l2}}{\der\log(a_*)}\label{eq::gamma22Zas}\,.
\end{flalign}
From the general form of the $Z$ factors for 
$n\leq m$~\eqref{eq::zass1}, one finds that $Z^{-1}_{2l}$ is at least $\mathcal{O}(\kappa^2)$ for $l>2$,\footnote{See App.~\ref{app:intZinv} for details on the perturbative inversion.} so the infinite sums in both anomalous dimensions can be truncated at $l=2$ for the desired perturbative order, yielding for the $\mu$-anomalous dimension
\begin{equation}
    \gamma^{\mu}_{22}=-\frac{\der Z^{(\kappa)}_{22}}{\der\log(\mu)}+\Lo(\kappa^2)\,.
\end{equation}
To calculate the derivative, one must include the tree-level running
\begin{flalign}
\frac{\der c_{3,1}}{\der\log(\mu)}&=-2\ve c_{3,1}\,,
\label{eq::c31Running}
\\
\frac{\der}{\der\log(\mu)}\,\Xi_{3,1}(\vec k_1,...,\vec k_4)&=-2\ve \Xi_{3,1}(\vec k_1,...,\vec k_4)\label{eq::Xi31Running}
\end{flalign}
of the renormalised couplings and IC functions. The first equation follows from \eqref{eq::cRGE} dropping $\delta c_{3,1}=\Lo(\kappa^2)$, the second one from \eqref{eq::XimuRGE} and the tree-level relation \cite{Beneke:2026rtf}
\begin{equation}
\xi_{3,1}=\frac{c_{3,1}}{2\ve}\,.
\label{eq::xi31}
\end{equation}
Together with the expression of the $Z_{22}$~\eqref{eq::z22} and the matched values for $c_{3,1}$ and $\Xi_{3,1}$, this gives
\begin{equation}
\gamma^{\mu}_{22}=\frac{\kappa}{12\pi^2}\bigg[\log\bigg(\frac{a_*}{a(t)}\bigg)+\log\bigg(\frac{e^{\gamma_E}\mu}{H}\bigg)-\frac{4}{3}\bigg]+\Lo(\kappa^2)\,.
\label{eq::gamma22muZ}
\end{equation}
Similarly, \eqref{eq::gamma22Zas} yields
\begin{flalign}
\gamma^{a_*}_{22}&=-\frac{\der Z^{(\kappa)}_{22}}{\der\log(a_*)}+\Lo(\kappa^2)\nonumber\\
&=\frac{\kappa}{12\pi^2}\bigg[\log\bigg(\frac{e^{\gamma_E}\mu}{H}\bigg)-\frac{4}{3}\bigg]+\Lo(\kappa^2)\,,
\label{eq::gamma22asZ}
\end{flalign}
where we used the $a_*$-running of the $d$-dimensional $\Xi_{3,1}$
\begin{equation}
\frac{\der}{\der\log(a_*)}\,\Xi_{3,1}(\vec k_1,...,\vec k_4)=-c_{3,1}-2\ve \Xi_{3,1}(\vec k_1,...,\vec k_4)\,,
\label{eq::asRunningXi31}
\end{equation}
which follows from \eqref{eq::XiasRGE} and \eqref{eq::xi31}. 

The operator anomalous dimensions also govern the dependence of the operator matching coefficients on the factorisation scale $\mu$ and on the reference scale factor $a_*$, respectively
\begin{flalign}
\gamma^{\mu}_{22}&=-\sum_{l=0}^{\infty}C^{-1}_{2l}\frac{\der C_{l2}}{\der\log(\mu)}\label{eq::gamma22Cmu}\,,\\
\gamma^{a_*}_{22}&=-\sum_{l=0}^{\infty}C^{-1}_{2l}\frac{\der C_{l2}}{\der\log(a_*)}\label{eq::gamma22Cas}\,.
\end{flalign}
The requirement that the two sets of equations yield the same result for $\gamma^{\mu,a_*}_{22}$ provides a self-consistency check of the framework.
At $\Lo(\kappa)$ \eqref{eq::gamma22Cmu} reduces to
\begin{equation}
\gamma^{\mu}_{22}=-\bigg[\frac{\der C^{(\kappa)}_{22}}{\der\log(\mu)}-C^{(\kappa)}_{24}\frac{\der C^{(\kappa^0)}_{42}}{\der\log(\mu)}\bigg]+\Lo(\kappa^2)\,.
\end{equation}
From the free-theory result \eqref{eq::C42free} for $C^{(\kappa^0)}_{42}$, we obtain
\begin{equation}
\frac{\der C^{(\kappa^0)}_{42}}{\der\log(\mu)}=-\frac{3}{2\pi^2}\,,
\end{equation}
and then find the same result for $\gamma^{\mu}_{22}$ as given in \eqref{eq::gamma22muZ}. Similarly, \eqref{eq::gamma22Cas} at $\Lo(\kappa)$ reads
\begin{equation}
\gamma^{a_*}_{22}=-\frac{\der C^{(\kappa)}_{22}}{\der\log(a_*)}+\Lo(\kappa^2)\,.
\end{equation} 
Taking the derivative of \eqref{eq::C22final}, we find agreement with \eqref{eq::gamma22asZ}. 
We therefore explicitly verified that the $\mu$- and $a_*$-dependence of the SdSET correlation functions involved in the matching of their full-theory counterpart is correctly compensated by $C_{22}$. 

It is interesting to note that the anomalous dimensions feature logarithms of $\mu$ and $a_*$, and, as a consequence of the latter, are explicitly time dependent.
Logarithmic dependence on $\mu$ is not untypical for theories featuring double poles, and arises e.g.~in the case of the cusp anomalous dimension, relevant to the soft-collinear limit of scattering amplitudes \cite{Korchemskaya:1992je}.
The time-dependence, on the other hand, appears unconventional at first. 
However, recall that the correlation time $t$ enters in the EFT construction as an external ``kinematic'' parameter.
It appears in the anomalous dimensions, which are related to UV poles of the EFT, as the only possible quantity that can compensate the reference scale factor $a_*$.
The physical interpretation of the anomalous dimensions in the context of the Kramers-Moyal equation is discussed below in \secref{sec:KM}.


\section{Matching the two-loop one-point function of the operator \texorpdfstring{$\phi^2$}{φ₊²}}
\label{sec::vp20}

We now turn to  the matching of the renormalised full-theory one-point function $\cor{}{[\phi^2](t,\vec x)}$ at $\Lo(\kappa)$.
This is the simplest example of a two-loop calculation. The full-theory computation is presented in \appref{app:Phi20}. In this section, we focus on the SdSET side of the matching and begin by evaluating the EFT-counterpart of $\cor{}{[\phi^2](t,\vec x)}$, which is the renormalised one-point function $\cor{}{[\vp^2_+](t,\vec x)}$.
This also serves to further analyse the renormalisation of the operator $\vp^2_+$, as well as the phenomenon of operator mixing in interacting SdSET. 
We then use the operator-matching equation \eqref{eq::Phi2opmatch} already set up above to match the full-theory one-point function to a combination of EFT one-point functions, which includes $\cor{}{[\vp^2_+](t,\vec x)}$. The main result of this section is the determination of the matching coefficient $C_{20}$ and the anomalous dimension $\gamma^{\mu}_{20}$ at $\Lo(\kappa)$. The latter can be identified as an NNLO Kramers-Moyal coefficient according to \cite{Cohen:2021fzf} and is computed here for the first time.

\subsection{Matching in the free theory}

The matching of the free-theory operator $[\phi^2]$ onto SdSET is discussed in \appref{app:freematch}. The matching equation reads
\begin{equation}
[\phi^2](t,\vec x)=H^2\bigg[[\vp^2_+](t,\vec x)-\frac{1}{4\pi^2}\log\bigg(\frac{\mu}{\mu_f}\bigg)\un\bigg]\,.
\end{equation}
Taking the expectation value of this equation, and using 
\begin{equation}
\cor{}{[\vp_+^2](t,\vec x)}=-\frac{1}{4\pi^2}\log\bigg(\frac{\Lambda}{a(t)\mu}\bigg)\,,
\end{equation}
we reproduce the full-theory result given in \eqref{eq:treephisq},
\begin{equation}
\cor{}{[\phi^2](t,\vec x)}=-\frac{H^2}{4\pi^2}\log\bigg(\frac{\Lambda}{a(t)\mu_f}\bigg)\,.
\label{eq::Phi2free}
\end{equation}

\subsection{Computation of \texorpdfstring{$\cor{}{\vp^2_+(t,\vec x)}$ at $\Lo(\kappa)$, renormalisation of $\vp^2_+$}{{<φ₊²(η,x)> at O(ϰ), renormalisation of φ₊²}}}

The computation of the one-point function $\cor{}{\vp^2_+(t,\vec x)}$ parallels the computation of the one-loop SdSET power spectrum in \cite{Beneke:2026rtf}, since the one-point function considered here corresponds to the integration over the momentum $\vec{k}$ of the power spectrum. Therefore, all the effective couplings and initial conditions which contributed there appear again here as insertions in loop diagrams. 

Unlike in the previous cases, we refrain from distinguishing the couplings $c_{3,1}$ and $\kappa$ from each other, and directly set them equal for brevity. This choice simplifies the intermediate results of the present computation. 
Finally, we consider the combination $(a(t)\tmu)^{2\ve}\cor{}{\vp^2_+(t,\vec x)}$ to obtain a scaling- and mass-dimensionless result, which will prove convenient for the renormalisation of $\vp^2_+$.

\begin{figure}[t]
\centering
\begin{subfigure}{0.4\textwidth}
\centering
\includegraphics[width=0.65\textwidth]{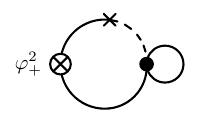}
\caption{}
\end{subfigure}%
\hspace{1cm}\begin{subfigure}{0.4\textwidth}
\centering
\includegraphics[width=0.65\textwidth]{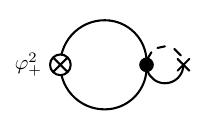}
\caption{}
\end{subfigure}
\caption{The non-vanishing (left panel) and vanishing (right panel) diagram topologies resulting from the insertion of the quartic SdSET vertex into the correlation function $\cor{}{\vp^2_+(t,\vec x)}$. The $\xi_{3,1}$ counterterm insertion leads to the same diagram topologies with the filled circle replaced by a small filled square.}
\label{fig:Phi20EFT1}
\end{figure}

\subsubsection{Quartic-vertex and \texorpdfstring{$\xi_{3,1}$}{ξ-3,1}-insertion}

We start by considering the contributions from the insertion of the quartic EFT Lagrangian interaction proportional to $c_{3,1}$ and initial-condition-counterterm $\xi_{3,1}$. They both lead to diagram topologies shown in the left panel of \figref{fig:Phi20EFT1}. We find
\begin{flalign}
&(a(t)\tmu)^{2\ve}\cor{}{\vp^2_+(t,\vec x)}_{|\,c_{3,1}+\xi_{3,1}}\nonumber\\
&=-\frac{\kappa}{12}(a(t)\tmu)^{2\ve}\frac{(a(t)\tmu)^{2\ve}-(a_*\tmu)^{2\ve}}{2\ve}\bigg[\int\frac{\der^{d-1}l}{(2\pi)^{d-1}}\frac{1}{l^3_{\Lambda}}\bigg]^2\nonumber\\
&=\frac{\kappa}{192\pi^4}\,\bigg(\frac{\Lambda}{a(t)\mu}\bigg)^{\!-4\ve}\log\bigg(\frac{a_*}{a(t)}\bigg)\,\Bigg\{\frac{1}{\ve^2}+\frac{1}{\ve}\log\bigg(\frac{a_*}{a(t)}\bigg)+\frac{2}{3}\log^2\bigg(\frac{a_*}{a(t)}\bigg)+\frac{\pi^2}{6}\,\Bigg\}\,,
\end{flalign}
where we dropped terms vanishing as $\Lambda\rightarrow0$, and kept the prefactor unexpanded. The diagrams with topology shown in the right panel vanish identically~\cite{Beneke:2026rtf}. 

\subsubsection{Mass-counterterm and \texorpdfstring{$\xi_{1,1}$}{ξ-1,1}-insertion}

\begin{figure}[t]
\centering
\includegraphics[width=0.22\textwidth]{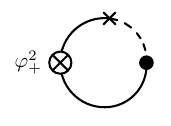}
\caption{Diagram topology resulting from the insertion of the SdSET mass counterterm into the correlation function $\cor{}{\vp^2_+(t,\vec x)}$.}
\label{fig:Phi20EFT2}
\end{figure}

The mass counterterm $c_{1,1}$ and the initial-condition counterterm $\xi_{1,1}$, were determined in \cite{Beneke:2026rtf} at $\Lo(\kappa)$. The result for the former can be found in \eqref{eq::c11}, and we restate the latter here, 
\begin{flalign}
\xi_{1,1}&=\underbrace{-\frac{\kappa}{16\pi^2}\bigg\{\frac{1}{\ve^2}-\frac{2}{\ve}\bigg[\frac{4}{3}-\log\bigg(\frac{e^{\gamma_E}\mu}{H}\bigg)\bigg]\bigg\}}_{\textrm{from }\Xi_{3,1}}-\underbrace{\frac{c_{3,1}}{8\pi^2\delta}\bigg[\frac{1}{2\ve}-\hat c_{1,1}\bigg]}_{\textrm{from }c_{3,1}+\xi_{3,1}}\,.
\label{eq::xi11}
\end{flalign}
Their insertions yield the diagram with topology shown in \figref{fig:Phi20EFT2}, which leads to the expression
\begin{flalign}
&(a(t)\tmu)^{2\ve}\cor{}{\vp^2_+(t,\vec x)}_{|\,c_{1,1}+\xi_{1,1}}\nonumber\\
&=-\frac{\kappa(a(t)\tmu)^{2\ve}(\nu a(t))^{2\delta}}{24\pi^2}\,\Bigg\{\frac{1}{2\delta}\bigg[-\frac{1}{2\ve}+\hat c_{1,1}\bigg]+\bigg(\frac{a_*}{a(t)}\bigg)^{\!2\delta}\,\bigg[\frac{1}{4\ve^2}-\frac{1}{2\ve}\bigg(\frac{4}{3}-\log\bigg(\frac{e^{\gamma_E}\mu}{H}\bigg)\bigg)\nonumber\\
&\quad+\frac{1}{2\delta}\bigg(\frac{1}{2\ve}-\hat c_{1,1}\bigg)\bigg]\Bigg\}\,\int\frac{\der^{d-1}l}{(2\pi)^{d-1}}\frac{1}{l^3_{\Lambda}}\nonumber\\
&=\frac{\kappa}{192\pi^4}\,\bigg(\frac{\Lambda}{a(t)\mu}\bigg)^{\!-2\ve}\,\Bigg\{-\frac{1}{2\ve^3}+\frac{1}{\ve^2}\bigg[\frac{4}{3}-\log\bigg(\frac{e^{\gamma_E}a_*\mu}{a(t)H}\bigg)\bigg]+\frac{1}{\ve}\bigg[2\hat c_{1,1}\log\bigg(\frac{a_*}{a(t)}\bigg)-\frac{\pi^2}{24}\bigg]\nonumber\\
&\quad-\frac{\pi^2}{12}\log\bigg(\frac{e^{\gamma_E}a_*\mu}{a(t)H}\bigg)+\frac{\pi^2}{9}+\frac{\zeta(3)}{6}\,\Bigg\}\,.
\end{flalign}
Note that while the intermediate result depends on the analytic regulator $\delta$, the final result does not. This happens by construction, since the explicit $\delta$-poles appearing in $\xi_{1,1}$ are chosen to cancel the ones generated by the time integral associated with the Lagrangian insertion of $c_{1,1}$. The momentum integral, which appears as a common multiplicative factor in both contributions, only generates poles in $\ve$, so the cancellation of the $\delta$-poles is exact in this case.

\subsubsection{\texorpdfstring{$\Xi_{3,1}$}{Ξ-3,1}-insertion}

\begin{figure}[t]
\centering
\begin{subfigure}{0.4\textwidth}
\centering
\includegraphics[width=0.8\textwidth]{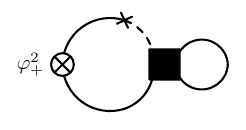}
\caption{}
\end{subfigure}%
\hspace{1cm}\begin{subfigure}{0.4\textwidth}
\centering
\includegraphics[width=0.8\textwidth]{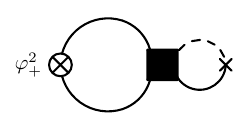}
\caption{}
\end{subfigure}
\caption{The non-vanishing (left panel) and vanishing (right panel) diagram topologies resulting from the insertion of the quartic initial condition $\Xi_{3,1}$ into the correlation function $\cor{}{\vp^2_+(t,\vec x)}$.}
\label{fig:Phi20EFT3}
\end{figure}

As was the case for the one-loop power spectrum in \cite{Beneke:2026rtf}, for the insertion of the renormalised initial condition $\Xi_{3,1}$ we only have to consider the diagram topology shown in the left panel of \figref{fig:Phi20EFT3}, since the topology shown on the right panel of this figure vanishes. We thus compute the expression
\begin{flalign}
&(a(t)\tmu)^{2\ve}\cor{}{\vp^2_+(t,\vec x)}_{|\,\Xi_{3,1}}\nonumber\\
&=\frac{\kappa(a(t)\tmu)^{2\ve}(a_*\tmu)^{2\ve}}{8}\int\frac{\der^{d-1}l_1\der^{d-1}l_2}{(2\pi)^{2d-2}}\frac{1}{l^3_{1\Lambda}l^3_{2\Lambda}}\,\Bigg\{\frac{2}{3}\log\bigg(\frac{2e^{\gamma_E}(l_{1\Lambda}+l_{2\Lambda})}{a_*H}\bigg)\nonumber\\
&\quad-\frac{1}{l^3_{12\Lambda}+l^3_{2\Lambda}}\bigg[\frac{l_{1\Lambda}^2l_{2\Lambda}^2}{2(l_{1\Lambda}+l_{2\Lambda})}+\frac{2}{9}\,\Big[7l^3_{1\Lambda}+3l^2_{1\Lambda}l_{2\Lambda}+3l_{1\Lambda}l^2_{2\Lambda}+7l_{2\Lambda}^3\Big]\bigg]\Bigg\}\,.
\end{flalign}
The details on the evaluation of these integrals, as well as the explicit results, can be found in \appref{app::Phi20loopsEFT}. Using these, we find
\begin{flalign}
&(a(t)\tmu)^{2\ve}\cor{}{\vp^2_+(t,\vec x)}_{|\,\Xi_{3,1}}\nonumber\\
&=\frac{\kappa}{192\pi^4}\,\bigg(\frac{\Lambda^2}{a(t)a_*\mu^2}\bigg)^{\!-2\ve}\Bigg\{\,\frac{3}{4\ve^3}+\frac{1}{\ve^2}\bigg[\log\!\bigg(\!\frac{e^{\gamma_E}\Lambda}{a_*H}\!\bigg)-\frac{4}{3}\bigg]+\frac{1}{\ve}\bigg[\frac{5}{4}-\frac{5\pi}{3\sqrt{3}}+\frac{\pi^2}{8}\bigg]+\frac{\pi^2}{6}\log\!\bigg(\!\frac{e^{\gamma_E}\Lambda}{a_*H}\!\bigg)
\nonumber\\
&\quad
+\frac{5}{2}+\frac{5\pi}{2\sqrt{3}}\bigg[\log(3)-\frac{4}{3}\bigg]-\frac{17\pi^2}{54}-\frac{3}{2}\zeta(3)+\frac{5}{36}\psi^{(1)}\bigg(\frac{1}{6}\bigg)-\frac{5}{18}\psi^{(1)}\bigg(\frac{1}{3}\bigg)\Bigg\}\,,
\end{flalign}
where $\psi^{(1)}(z)$ denotes the polygamma function of order one.

\subsubsection{\texorpdfstring{$\Xi_{1,1}$}{Ξ-1,1}-insertion}

The final contribution to the one-point function requires the initial condition  \cite{Beneke:2026rtf}
\begin{flalign}
\Xi_{1,1}(\vec k)&=-\frac{\kappa}{4\pi^2}\,\Bigg\{\bigg[\frac{4}{3}-\log\bigg(\frac{e^{\gamma_E}\mu}{H}\bigg)-\hat c_{1,1}\bigg]\log\bigg(\frac{2e^{\gamma_E}k}{a_*H}\bigg)\nonumber\\
&\quad+\bigg[1+\frac{1}{2}\log\bigg(\frac{e^{\gamma_E}\mu}{H}\bigg)\bigg]\log\bigg(\frac{e^{\gamma_E}\mu}{H}\bigg)+\frac{7\hat c_{1,1}}{3}-\frac{55}{36}-\frac{5\pi}{3\sqrt{3}}+\frac{7\pi^2}{48}\Bigg\}\,.
\label{eq::Xi11}
\end{flalign}
Its insertion into the one-point function leads to 
\begin{flalign}
&(a(t)\tmu)^{2\ve}\cor{}{\vp^2_+(t,\vec x)}_{|\,\Xi_{1,1}}=\frac{(a(t)\tmu)^{2\ve}}{6}\int\frac{\der^{d-1}l}{(2\pi)^{d-1}}\frac{\Xi_{1,1}(\vec l_{\Lambda})}{l^3_{\Lambda}}\\[-.1cm]
&=\frac{\kappa}{192\pi^4}\,\bigg(\frac{\Lambda}{a(t)\mu}\bigg)^{\!-2\ve}\,\bigg\{\frac{1}{\ve^2}\bigg[\log\bigg(\frac{e^{\gamma_E}\mu}{H}\bigg)+\hat c_{1,1}-\frac{4}{3}\bigg]
\nonumber\\
&\quad
+\frac{1}{\ve}\bigg[2\log\bigg(\frac{e^{\gamma_E}\Lambda}{a_*H}\bigg)\bigg(\log\bigg(\frac{e^{\gamma_E}\mu}{H}\bigg)+\hat c_{1,1}-\frac{4}{3}\bigg)
-\log^2\bigg(\frac{e^{\gamma_E}\mu}{H}\bigg)-\frac{8}{3}\hat c_{1,1}+\frac{7}{18}+\frac{10\pi}{3\sqrt{3}}-\frac{7\pi^2}{24}\bigg]
\nonumber\\
&\quad
+\frac{\pi^2}{12}\bigg[\frac{4}{3}-\log\bigg(\frac{e^{\gamma_E}\mu}{H}\bigg)-\hat c_{1,1}\bigg]\bigg\}\,.
\end{flalign}

\subsubsection{Renormalisation of \texorpdfstring{$\vp^2_+$}{φ₊²}}

We now have all the relevant contributions to $\cor{}{\vp^2_+(t,\vec x)}$. Summing them and expanding out the prefactors results in
\begin{flalign}
&(a(t)\tmu)^{2\ve}\cor{}{\vp^2_+(t,\vec x)}_{|\,c_{3,1}+\xi_{3,1}+\Xi_{3,1}+c_{1,1}+\xi_{1,1}+\Xi_{1,1}}\nonumber\\
&=\frac{\kappa}{192\pi^4}\,\Bigg\{\frac{1}{4\ve^3}+\frac{1}{\ve^2}\Bigg[\frac{1}{2}\log\bigg(\frac{e^{2\gamma_E}a_*\mu^2}{a(t)H^2}\bigg)+\hat c_{1,1}-\frac{4}{3}-\log\bigg(\frac{\Lambda}{a(t)\mu}\bigg)\Bigg]\nonumber\\
&\quad+\frac{1}{\ve}\Bigg[\log\bigg(\frac{\Lambda}{a(t)\mu}\bigg)\bigg[\log\bigg(\frac{\Lambda}{a(t)\mu}\bigg)-2\log\bigg(\frac{e^{\gamma_E}\mu a_*}{a(t)H}\bigg)+\frac{8}{3}\bigg]
\nonumber\\
&\qquad
+\log\bigg(\frac{e^{\gamma_E}\mu}{H}\bigg)\bigg[\log\bigg(\frac{e^{\gamma_E}\mu}{H}\bigg)+2\hat c_{1,1}-\frac{8}{3}\bigg]+\frac{1}{2}\log^2\bigg(\frac{a_*}{a(t)}\bigg)-\frac{8\hat c_{1,1}}{3}+\frac{59}{36}+\frac{5\pi}{3\sqrt{3}}-\frac{5\pi^2}{24}\Bigg]\nonumber\\
&\quad+\frac{2}{3}\log^3\bigg(\frac{e^{\gamma_E}\Lambda}{a(t)H}\bigg)+2\bigg[\log\bigg(\frac{e^{\gamma_E}a_*\mu}{a(t)H}\bigg)-\hat c_{1,1}-\frac{8}{3}\bigg]\log^2\bigg(\frac{e^{\gamma_E}\Lambda}{a(t)H}\bigg)\nonumber\\
&\quad+\Bigg[8\log\bigg(\frac{e^{\gamma_E}\mu}{H}\bigg)\bigg[2-\log\bigg(\frac{e^{\gamma_E}\mu a_*}{a(t)H}\bigg)\bigg]+\frac{16}{3}\log\bigg(\frac{a_*}{a(t)}\bigg)+\frac{16}{3}\hat c_{1,1}-\frac{52}{9}+\frac{\pi^2}{3}\Bigg]\log\bigg(\frac{e^{\gamma_E}\Lambda}{a(t)H}\bigg)\nonumber\\
&\quad-\frac{1}{3}\log^3\bigg(\frac{a_*}{a(t)}\bigg)+\bigg[2\log\bigg(\frac{e^{\gamma_E}\mu}{H}\bigg)-\frac{8}{3}\bigg]\log^2\bigg(\frac{a_*}{a(t)}\bigg)\nonumber\\
&\quad+\Bigg[\log\bigg(\frac{e^{\gamma_E}\mu}{H}\bigg)\bigg[6\log\bigg(\frac{e^{\gamma_E}\mu}{H}\bigg)-\frac{16}{3}\bigg]+\frac{5}{2}-\frac{10\pi}{3\sqrt{3}}+\frac{\pi^2}{6}\Bigg]\log\bigg(\frac{a_*}{a(t)}\bigg)\nonumber\\
&\quad+\frac{16}{3}\log^3\bigg(\frac{e^{\gamma_E}\mu}{H}\bigg)-2\bigg[\frac{16}{3}-\hat c_{1,1}\bigg]\log^2\bigg(\frac{e^{\gamma_E}\mu}{H}\bigg)+\bigg[\frac{52}{9}-\frac{\pi^2}{3}-\frac{16}{3}\hat c_{1,1}\bigg]\log\bigg(\frac{e^{\gamma_E}\mu}{H}\bigg)\nonumber\\
&\quad-\frac{\pi^2}{12}\hat c_{1,1}+\frac{5}{2}+\frac{5\pi}{2\sqrt{3}}\bigg[\log(3)-\frac{4}{3}\bigg]-\frac{5\pi^2}{54}-\frac{4}{3}\zeta(3)+\frac{5}{36}\psi^{(1)}\bigg(\frac{1}{6}\bigg)-\frac{5}{18}\psi^{(1)}\bigg(\frac{1}{3}\bigg)\Bigg\}\,,
\label{eq::20regresult}
\end{flalign}
where this result is organised as follows: always in descending order of powers, we first collect the pole terms in $\ve$, then the UV-finite but IR-divergent terms, which are all logarithmic, then the finite but time-dependent terms, which are also all logarithmic,  and  finally the constant terms, collecting powers of logarithms of $\mu/H$.
Note that the $\ve$-poles appearing above multiply IR-divergent terms. This is a consequence of operator mixing and is accounted for consistently in the operator renormalisation framework as will be seen below.

As was the case in \secref{sec::22q0EFT}, we can remove the UV-divergent terms by renormalising the operator $\vp^2_+$. We already set up the renormalisation equation for $\vp^2_+$ in \eqref{eq::vp2reneq} above. The relevant terms for the present discussion are
\begin{equation}
(a(t)\tmu)^{2\ve}\vp^2_+(t,\vec x)=(a(t)\tmu)^{2\ve}Z_{22}[\vp^2_+](t,\vec x)+Z_{20}\un+\Lo(\kappa^2)\,,
\end{equation}
and we take note of the fact that the unit operator must be kept here. 
The renormalisation equation for the one-point function can be obtained by taking the expectation value
\begin{equation}
(a(t)\tmu)^{2\ve}\cor{}{\vp^2_+(t,\vec x)}_{|\,\Lo(\kappa)}=\Big[(a(t)\tmu)^{2\ve}Z_{22}\cor{}{[\vp^2_+](t,\vec x)}+Z_{20}\Big]\Big|_{\Lo(\kappa)}\,.
\label{eq::20corren}
\end{equation}
The only unknown quantity in this equation is the $\Lo(\kappa)$-piece of $Z_{20}$, which must absorb the remaining UV divergences, and is determined by this requirement. After setting $c_{3,1}=\kappa$, $Z_{22}$ at $\Lo(\kappa)$, given in \eqref{eq::z22}, simplifies to 
\begin{equation}
Z_{22}=1+\frac{\kappa}{48\pi^2}\bigg\{\frac{1}{\ve^2}+\frac{2}{\ve}\bigg[\log\bigg(\frac{a_*}{a(t)}\bigg)+\log\bigg(\frac{e^{\gamma_E}\mu}{H}\bigg)-\frac{4}{3}\bigg]\bigg\}+\Lo(\kappa^2)\,.
\label{eq::Z22kappa}
\end{equation}
The $\Lo(\kappa)$-piece of $Z_{20}$ in the $\overline{\textrm{MS}}$ scheme is given by
\begin{equation}
Z^{(\kappa)}_{20}=\Big[(a(t)\tmu)^{2\ve}\cor{}{\vp^2_+(t,\vec x)}\Big|_{\Lo(\kappa)}-Z^{(\kappa)}_{22}(a(t)\tmu)^{2\ve}\cor{}{[\vp^2_+](t,\vec x)}\Big]\Big|_{\textrm{poles}}\,.
\label{eq::z20def}
\end{equation}
Since $Z_{22}^{(\kappa)}$ contains a double pole in $\ve$, one must use in \eqref{eq::z20def} the $d$-dimensional, renormalised one-point function $\cor{}{[\vp^2_+](t,\vec x)}$ at $\Lo(\kappa^0)$ 
expanded 
to $\mathcal{O}(\ve^2)$. It can be read off from first line of \eqref{eq:Z42phisq} as
\begin{flalign}
(a(t)\tmu)^{2\ve}\cor{}{[\vp^2_+](t,\vec x)}&=\frac{1}{8\pi^2}\bigg(\frac{\Lambda}{a(t)\mu}\bigg)^{-2\ve}e^{\ve\gamma_E}\Gamma(\ve)-\frac{1}{8\pi^2\ve}\nonumber\\
&=\frac{1}{4\pi^2}\,\Bigg\{-\log\bigg(\frac{\Lambda}{a(t)\mu}\bigg)+\ve\bigg[\frac{\pi^2}{24}+\log^2\bigg(\frac{\Lambda}{a(t)\mu}\bigg)\bigg]\nonumber\\
&\quad-\frac{\ve^2}{3}\bigg[2\log^3\bigg(\frac{\Lambda}{a(t)\mu}\bigg)+\frac{\pi^2}{4}\log\bigg(\frac{\Lambda}{a(t)\mu}\bigg)+\frac{\zeta(3)}{2}\bigg]\Bigg\}+\Lo(\ve^3)\,.
\label{eq::ddimrenvp2}
\end{flalign} 
We then obtain 
\begin{flalign}
Z^{(\kappa)}_{20}&=\frac{\kappa}{192\pi^4}\,\Bigg\{\frac{1}{4\ve^3}+\frac{1}{\ve^2}\bigg[\frac{1}{2}\log\bigg(\frac{a_*}{a(t)}\bigg)+\log\bigg(\frac{e^{\gamma_E}\mu}{H}\bigg)+\hat c_{1,1}-\frac{4}{3}\bigg]\nonumber\\
&\hspace{1.8cm}+\frac{1}{\ve}\bigg[\frac{1}{2}\log^2\bigg(\frac{a_*}{a(t)}\bigg)+\log^2\bigg(\frac{e^{\gamma_E}\mu}{H}\bigg)+\log\bigg(\frac{e^{\gamma_E}\mu}{H}\bigg)\bigg(2\hat c_{1,1}-\frac{8}{3}\bigg)\nonumber\\
&\hspace{2.8cm}-\frac{8\hat c_{1,1}}{3}+\frac{59}{36}+\frac{5\pi}{3\sqrt{3}}-\frac{\pi^2}{4}\bigg]\Bigg\}\,.
\end{flalign}
This leads to the final renormalised result
\begin{flalign}
&\cor{}{[\vp^2_+](t,\vec x)}_{|\,\Lo(\kappa)}
=(a(t)\tmu)^{2\ve}\cor{}{\vp^2_+(t,\vec x)}_{|\,\Lo(\kappa)}-Z^{(\kappa)}_{22}(a(t)\tmu)^{2\ve}\cor{}{[\vp^2_+](t,\vec x)}_{|\,\Lo(\kappa^0)}-Z^{(\kappa)}_{20}\nonumber\\
&=\frac{\kappa}{192\pi^4}\,\Bigg\{\frac{4}{3}\log^3\bigg(\frac{e^{\gamma_E}\Lambda}{a(t)H}\bigg)-\bigg[\frac{8}{3}+2\hat c_{1,1}+2\log\bigg(\frac{e^{\gamma_E}\mu}{H}\bigg)\bigg]\log^2\bigg(\frac{e^{\gamma_E}\Lambda}{a(t)H}\bigg)\nonumber\\
&\quad+\Bigg[\log\bigg(\frac{e^{\gamma_E}\mu}{H}\bigg)\bigg[\frac{32}{3}-2\log\bigg(\frac{e^{\gamma_E}\mu}{H}\bigg)-4\log\bigg(\frac{a_*}{a(t)}\bigg)\bigg]+\frac{16}{3}\log\bigg(\frac{a_*}{a(t)}\bigg)\nonumber\\
&\qquad+\frac{16}{3}\hat c_{1,1}-\frac{52}{9}+\frac{5\pi^2}{12}\Bigg]\log\bigg(\frac{e^{\gamma_E}\Lambda}{a(t)H}\bigg)-\frac{1}{3}\log^3\bigg(\frac{a_*}{a(t)}\bigg)+\bigg[2\log\bigg(\frac{e^{\gamma_E}\mu}{H}\bigg)-\frac{8}{3}\bigg]\log^2\bigg(\frac{a_*}{a(t)}\bigg)\nonumber\\
&\quad+\Bigg[\log\bigg(\frac{e^{\gamma_E}\mu}{H}\bigg)\bigg[4\log\bigg(\frac{e^{\gamma_E}\mu}{H}\bigg)-\frac{16}{3}\bigg]+\frac{5}{2}-\frac{10\pi}{3\sqrt{3}}+\frac{\pi^2}{12}\Bigg]\log\bigg(\frac{a_*}{a(t)}\bigg)\nonumber\\
&\quad+\frac{8}{3}\log^3\bigg(\frac{e^{\gamma_E}\mu}{H}\bigg)-2\Big[4-\hat c_{1,1}\Big]\log^2\bigg(\frac{e^{\gamma_E}\mu}{H}\bigg)+\bigg[\frac{52}{9}-\frac{\pi^2}{2}-\frac{16}{3}\hat c_{1,1}\bigg]\log\bigg(\frac{e^{\gamma_E}\mu}{H}\bigg)\nonumber\\
&\quad-\frac{\pi^2}{12}\hat c_{1,1}+\frac{5}{2}+\frac{5\pi}{2\sqrt{3}}\bigg[\log(3)-\frac{4}{3}\bigg]+\frac{\pi^2}{54}-\frac{7}{6}\zeta(3)+\frac{5}{36}\psi^{(1)}\bigg(\frac{1}{6}\bigg)-\frac{5}{18}\psi^{(1)}\bigg(\frac{1}{3}\bigg)\Bigg\}\,.
\label{eq::renvp2kappa}
\end{flalign}

\subsection{Matching}

This renormalised result can now be used to match the corresponding full-theory quantity. The latter is determined in \appref{app:Phi20}, and reads 
\begin{flalign}
\cor{}{[\phi^2](t,\vec x)}_{|\,\Lo(\kappa)}&=\frac{\kappa H^2}{192\pi^4}\,\Bigg\{\frac{4}{3}\log^3\bigg(\frac{e^{\gamma_E}\Lambda}{a(t)H}\bigg)-2\bigg[\log\bigg(\frac{e^{\gamma_E}\mu_f}{H}\bigg)+\frac{2}{3}+\delta\hat m^2_{\textrm{fin}}\bigg]\log^2\bigg(\frac{e^{\gamma_E}\Lambda}{a(t)H}\bigg)\nonumber\\
&\quad+\bigg[7\log\bigg(\frac{e^{\gamma_E}\mu_f}{H}\bigg)-\frac{16}{3}+\frac{2\pi^2}{3}+4\delta\hat m^2_{\textrm{fin}}\bigg]\log\bigg(\frac{e^{\gamma_E}\Lambda}{a(t)H}\bigg)\nonumber\\
&\quad-3\log^2\bigg(\frac{e^{\gamma_E}\mu_f}{H}\bigg)+\bigg[\frac{3}{2}-\frac{\pi^2}{3}-3\delta\hat m^2_{\textrm{fin}}\bigg]\log\bigg(\frac{e^{\gamma_E}\mu_f}{H}\bigg)\nonumber\\
&\quad-3+\frac{5\pi^2}{18}-\frac{\pi^2}{3}\delta\hat m^2_{\textrm{fin}}+\frac{11}{3}\zeta(3)\Bigg\}\,.
\label{eq::Phi2onept}
\end{flalign}
The matching equation is obtained by taking the expectation value of~\eqref{eq::Phi2opmatch}
\begin{equation}
\cor{}{[\phi^2](t,\vec x)}_{|\,\Lo(\kappa)}=H^2\Big[C_{20}+C_{22}\cor{}{[\vp^2_+](t,\vec x)}+C_{24}\cor{}{[\vp^4_+](t,\vec x)}\Big]\Big|_{\Lo(\kappa)}\,.
\label{eq::Phi20cormatch}
\end{equation}
From this equation, the operator matching coefficient $C_{20}$ is determined at $\Lo(\kappa)$ by
\begin{flalign}
C_{20}&=H^{-2}\cor{}{[\phi^2](t,\vec x)}_{|\,\Lo(\kappa)}-\cor{}{[\vp^2_+](t,\vec x)}_{|\,\Lo(\kappa)}-C^{(\kappa)}_{22}\cor{}{[\vp^2_+](t,\vec x)}_{|\,\Lo(\kappa^0)}\nonumber\\
&\quad-C^{(\kappa)}_{24}\cor{}{[\vp^4_+](t,\vec x)}_{|\,\Lo(\kappa^0)}+\Lo(\kappa^2)\,.
\label{eq::C20finalmatch}
\end{flalign}
All quantities appearing on the right-hand side are already known:  $C_{22}$ and $C_{24}$  have been determined to $\Lo(\kappa)$ in \secref{sec::22q0EFT}, and we restate them here for convenience,  
\begin{flalign}
C_{22}&=1+\frac{\kappa}{48\pi^2}\,\Bigg\{-4\log\bigg(\frac{a_*}{a(t)}\bigg)\bigg[\log\biggl(\frac{e^{\gamma_E}\mu}{H}\biggr)-\frac{4}{3}\bigg]-2\log^2\!\bigg(\frac{e^{\gamma_E}\mu}{H}\bigg)+\log\bigg(\frac{e^{\gamma_E}\mu}{H}\bigg)\nonumber\\
&\phantom{=}-\frac{5}{3}\log\bigg(\frac{\mu_f}{\mu}\bigg)+\frac{4}{3}\delta\hat m^2_{\textrm{fin}}-\frac{4}{9}-\frac{\pi^2}{4}\,\Bigg\}+\Lo(\kappa^2)\,,\\
C_{24}&=\frac{\kappa}{27}+\Lo(\kappa^2)\,.
\end{flalign}
In \secref{sec:freemix}, we determined the renormalised free-theory correlators appearing in \eqref{eq::C20finalmatch} 
\begin{flalign}
\cor{}{[\vp^2_+](t,\vec x)}_{|\,\Lo(\kappa^0)}&=-\frac{1}{4\pi^2}\log\bigg(\frac{\Lambda}{a(t)\mu}\bigg)\,,\\
\cor{}{[\vp^4_+](t,\vec x)}_{|\,\Lo(\kappa^0)}&=\frac{3}{16\pi^4}\log^2\bigg(\frac{\Lambda}{a(t)\mu}\bigg)\,.
\end{flalign}
Plugging all these expressions into \eqref{eq::C20finalmatch}, we find  
\begin{flalign}
C_{20}&=\frac{1}{4\pi^2}\log\bigg(\frac{\mu_f}{\mu}\bigg)+\frac{\kappa}{192\pi^4}\,\Bigg\{\frac{1}{3}\log^3\bigg(\frac{a_*}{a(t)}\bigg)-2\bigg[\log\bigg(\frac{e^{\gamma_E}\mu}{H}\bigg)-\frac{4}{3}\bigg]\log^2\bigg(\frac{a_*}{a(t)}\bigg)\nonumber\\
&\quad-\bigg[\frac{5}{2}-\frac{10\pi}{3\sqrt{3}}+\frac{\pi^2}{12}\bigg]\log\bigg(\frac{a_*}{a(t)}\bigg)-\frac{2}{3}\log^3\bigg(\frac{e^{\gamma_E}\mu}{H}\bigg)+2\bigg[\frac{4}{3}-\delta\hat m^2_{\textrm{fin}}\bigg]\log^2\bigg(\frac{e^{\gamma_E}\mu}{H}\bigg)\nonumber\\
&\quad-\bigg[\frac{23}{6}-\frac{5\pi^2}{12}-\delta\hat m^2_{\textrm{fin}}\bigg]\log\bigg(\frac{e^{\gamma_E}\mu}{H}\bigg)-3\log^2\bigg(\frac{\mu_f}{\mu}\bigg)\nonumber\\
&\quad+\log\bigg(\frac{\mu_f}{\mu}\bigg)\bigg[-2\log^2\bigg(\frac{e^{\gamma_E}\mu}{H}\bigg)+\log\bigg(\frac{e^{\gamma_E}\mu}{H}\bigg)-3\delta\hat m^2_{\textrm{fin}}+\frac{3}{2}-\frac{\pi^2}{4}\bigg]\nonumber\\
&\quad-\frac{\pi^2}{4}\delta\hat m^2_{\textrm{fin}}-\frac{11}{2}+\frac{5\pi}{2\sqrt{3}}\bigg[\frac{4}{3}-\log(3)\bigg]+\frac{7\pi^2}{27}+\frac{29}{6}\zeta(3)-\frac{5}{36}\psi^{(1)}\bigg(\frac{1}{6}\bigg)+\frac{5}{18}\psi^{(1)}\bigg(\frac{1}{3}\bigg)\Bigg\}\nonumber\\
&\quad+\Lo(\kappa^2)\,.
\label{eq::C20}
\end{flalign}
This expression is spatially local and independent of $\Lambda$, as it must be. This means that the complete $\Lambda$-dependence of the renormalised full-theory result is successfully reproduced by the SdSET correlators appearing in the matching equation.
The above represents the first two-loop matching computation of a composite-operator correlation function onto SdSET. The success of the matching procedure crucially relies on the self-consistency of the EFT framework, since several quantities that are determined from independent computations enter the matching equation \eqref{eq::Phi20cormatch}. 
As such, this particular computation may be the most stringent consistency check that we have been able to apply to SdSET so far. 
While local in space, $C_{20}$ contains explicit $a(t)$-dependence, which enters in the combination $a_*/a(t)$. The same feature was already observed for $C_{22}$, see the discussion at the end of \secref{sec::C22det}. Once again, there are no parametrically large logarithms when the matching is carried out at the common scale $\mu=\mu_f$, and setting $\mu=H e^{-\gamma_E}$ and $a_*=a(t)$.

\subsection{Anomalous dimensions}

From the above results for the SdSET operator-renormalisation factors $Z_{20}$ and $Z_{22}$ we can determine the anomalous dimensions $\gamma^{\mu}_{20}$ and $\gamma^{a_*}_{20}$ according to \eqref{eq::muADM} and \eqref{eq::asADM}. For the case of interest, 
\begin{flalign}
\gamma^{\mu}_{20}&=\sum_{l=0}^{\infty}Z^{-1}_{2l}\bigg[l\ve Z_{l0}-\frac{\der Z_{l0}}{\der\log(\mu)}\bigg]\,,\label{eq::gamma20Zmu}\\
\gamma^{a_*}_{20}&=-\sum_{l=0}^{\infty}Z^{-1}_{2l}\frac{\der Z_{l0}}{\der\log(a_*)}\label{eq::gamma20Zas}\,.
\end{flalign}

To compute $\gamma^{\mu}_{20}$ at $\Lo(\kappa)$ we truncate the sum in \eqref{eq::gamma20Zmu} after $l=2$, since $Z^{-1}_{2n}=\Lo(\kappa^2)$ for $n\geq 4$, and find
\begin{equation}
\gamma^{\mu}_{20}=2\ve Z^{(\kappa^0)}_{20}\Big[Z^{(\kappa^0)}_{22}-Z^{(\kappa)}_{22}\Big]+Z^{(\kappa^0)}_{22}\bigg[2\ve Z^{(\kappa)}_{20}-\frac{\der Z^{(\kappa)}_{20}}{\der\log(\mu)}\bigg]+\Lo(\kappa^2)\,.
\end{equation}
To determine $\der Z_{20}/\der\log(\mu)$ correctly, one must account for the running of the effective couplings~\eqref{eq::renCoup} and IC functions~\eqref{eq::renICs} entering $Z_{20}$ at the considered order in perturbation theory, 
\begin{flalign}
\frac{\der\hat c_{1,1}}{\der\log(\mu)}&=-1\,,\\
\frac{\der}{\der\log(\mu)}\,\Xi_{1,1}(\vec k)&=\frac{c_{3,1}}{4\pi^2}\bigg[\frac{4}{3}-\log\bigg(\frac{e^{\gamma_E}\mu}{H}\bigg)\bigg]\,.
\end{flalign}
The first equation follows from \eqref{eq::cRGE},  \eqref{eq::c11} noting that $\delta c_{1,1} = c^0_{1,1}|_{\rm pole\, part}$, and the tree-level running~\eqref{eq::c31Running} of $c_{3,1}$.  
The second one is obtained from~\eqref{eq::XimuRGE}, plugging in the expression for $\xi_{1,1}$~\eqref{eq::xi11} and using the tree-level running of $c_{3,1}$~\eqref{eq::c31Running} and of $\Xi_{3,1}$~\eqref{eq::Xi31Running}.
We then find the anomalous dimension
\begin{flalign}
\gamma^{\mu}_{20}&=\frac{1}{4\pi^2}+\frac{\kappa}{96\pi^4}\,\Bigg\{\log^2\bigg(\frac{a_*}{a(t)}\bigg)+2\log\bigg(\frac{a_*}{a(t)}\bigg)\bigg[\log\bigg(\frac{e^{\gamma_E}\mu}{H}\bigg)-\frac{4}{3}\bigg]+\log^2\bigg(\frac{e^{\gamma_E}\mu}{H}\bigg)\nonumber\\
&\quad+2\log\bigg(\frac{e^{\gamma_E}\mu}{H}\bigg)\bigg[-\frac{4}{3}+\delta\hat m^2_{\textrm{fin}}\bigg]+\frac{26}{9}-\frac{5\pi^2}{24}-\frac{8}{3}\delta\hat m^2_{\textrm{fin}}\Bigg\}+\Lo(\kappa^2)\,,
\label{eq::gamma20muZ}
\end{flalign}
where we incorporated the $\Lo(\kappa^0)$-piece determined from \eqref{eq::freegammamu}. Similarly, the formula for $\gamma^{a_*}_{20}$ at $\Lo(\kappa)$ reduces to
\begin{equation}
\gamma^{a_*}_{20}=-\frac{\der Z^{(\kappa)}_{20}}{\der\log(a_*)}+\Lo(\kappa^2)\,,
\end{equation}
and, also here, one must take into account the running of the effective couplings and IC-functions in $a_*$ to obtain the correct result. The renormalised couplings $c_{3,1}$, $c_{1,1}$ do not run with $a_*$, so neither does $\xi_{3,1}$ at tree level. The running of the remaining quantities contributing to $Z_{20}$ is given by
\begin{flalign}
\frac{\der}{\der\log(a_*)}\xi_{1,1}&=\frac{\kappa}{8\pi^2}\,\Bigg\{\frac{1}{\ve}-2\bigg[\frac{4}{3}-\log\bigg(\frac{e^{\gamma_E}\mu}{H}\bigg)\bigg]\Bigg\}+\Lo(\kappa^2)\,,\label{eq::xi11asRGE}\\
\frac{\der}{\der\log(a_*)}\Xi_{1,1}(\vec k)&=-\frac{\kappa}{4\pi^2}\bigg[\hat c_{1,1}+\log\bigg(\frac{e^{\gamma_E}\mu}{H}\bigg)-\frac{4}{3}\bigg]+\Lo(\kappa^2)\,,
\end{flalign}
and by \eqref{eq::asRunningXi31} for $\Xi_{3,1}$. The first equation is obtained from~\eqref{eq::xi11} and the $a_*$-evolution \eqref{eq::asRunningXi31} of $\Xi_{3,1}$, and the second one follows from~\eqref{eq::XiasRGE}, plugging in~\eqref{eq::xi11asRGE}.
With these results, we find
\begin{equation}
\gamma^{a_*}_{20}=\frac{\kappa}{192\pi^4}\,\Bigg\{\!-\log^2\bigg(\frac{a_*}{a(t)}\bigg)+4\log\bigg(\frac{a_*}{a(t)}\bigg)\bigg[\log\bigg(\frac{e^{\gamma_E}\mu}{H}\bigg)-\frac{4}{3}\bigg]+\frac{5}{2}-\frac{10\pi}{3\sqrt{3}}+\frac{\pi^2}{12}\,\Bigg\}+\Lo(\kappa^2)\,.
\label{eq::gamma20asZ}
\end{equation} 
The form of the anomalous dimension $\gamma_{20}^{\mu,a_*}$ is analogous to the previously determined $\gamma^{\mu,a_*}_{22}$, see~\eqref{eq::gamma22muZ} and~\eqref{eq::gamma22asZ}, with the same types of logarithms appearing. However, at the two-loop order, the anomalous dimensions feature logarithmic terms of the second power. The logarithms are not parametrically large if $\mu$ is chosen of order $H e^{-\gamma_E}$ and $a_*$ of order of the scale factor at correlation time $a(t)$.

The expressions obtained above can then be used to check the $\mu$- and $a_*$-dependence of the operator-matching coefficients as
\begin{flalign}
\gamma^{\mu}_{20}&=-\sum_{l=0}^{\infty}C^{-1}_{2l}\frac{\der C_{l0}}{\der\log(\mu)}\label{eq::gamma20Cmu}\,,\\
\gamma^{a_*}_{20}&=-\sum_{l=0}^{\infty}C^{-1}_{2l}\frac{\der C_{l0}}{\der\log(a_*)}\label{eq::gamma20Cas}\,.
\end{flalign}
At $\Lo(\kappa)$, the formula \eqref{eq::gamma20Cmu} reduces to
\begin{equation}
\gamma^{\mu}_{20}=-\bigg[\frac{\der C^{(\kappa^0)}_{20}}{\der\log(\mu)}+\frac{\der C^{(\kappa)}_{20}}{\der\log(\mu)}-C^{(\kappa)}_{22}\frac{\der C^{(\kappa^0)}_{20}}{\der\log(\mu)}-C^{(\kappa)}_{24}\frac{\der C^{(\kappa^0)}_{40}}{\der\log(\mu)}\bigg]+\Lo(\kappa^2)\,,
\label{eq::gamma20C}
\end{equation}
where we used that $C^{-1}_{22}=1+\Lo(\kappa)$, and the fact that $C_{n0}=\Lo(\kappa^2)$ for $n\geq 4$. Inserting \eqref{eq::C20} and the $\Lo(\kappa^0)$-results
\begin{equation}
\frac{\der C^{(\kappa^0)}_{20}}{\der\log(\mu)}=-\frac{1}{4\pi^2}\,,\qquad
\frac{\der C^{(\kappa^0)}_{40}}{\der\log(\mu)}=\frac{3}{8\pi^2}\log\bigg(\frac{\mu}{\mu_f}\bigg)\,,
\end{equation}
determined in \appref{app:freematch}, we find the same expression as in~\eqref{eq::gamma20muZ}, as it must be. Similarly, for $\gamma^{a_*}_{20}$ the formula \eqref{eq::gamma20Cas} reduces to
\begin{equation}
\gamma^{a_*}_{20}=-\frac{\der C_{20}}{\der\log(a_*)}+\Lo(\kappa^2)\,.
\end{equation}
In $d=4$, the $a_*$-running of all the finite parts of the initial conditions that enter $C_{20}$ simply reproduces the explicitly appearing logarithms of $a_*$. To compute $\gamma^{a_*}_{20}$, we only need to take the derivative of  \eqref{eq::C20} with respect to $\log(a_*)$, and find the same expression as \eqref{eq::gamma20asZ}. This verifies that the $\mu$- and $a_*$-dependence of the renormalised SdSET correlation functions entering the matching equation is correctly compensated by $C_{20}$ and $C_{22}$.

\section{Anomalous dimensions and Kramers-Moyal equation}
\label{sec:KM}

In \cite{Cohen:2021fzf}, it was proposed to extend the standard formalism of stochastic inflation \cite{Starobinsky:1994bd} by generalising the Fokker-Planck equation \eqref{eq::FP} for the SdSET field $\vp_+$ to a Kramers-Moyal equation of the form 
\begin{equation}
\frac{\p}{\p t}\rho_1(t,z)=\sum_{n=1}^{\infty}\frac{1}{n!}\frac{\p^n}{\p z^n}\bigg[\sum_{m=0}^{\infty}\frac{1}{m!}D_{nm}z^m\rho_1(t,z)\bigg]\,,
\label{eq::KM}
\end{equation}
where the $D_{nm}$ are local, $\kappa$-dependent coefficients. The one-point probability distribution function $\rho_1(t,z)$ is associated to the SdSET field $\vp_+$ via
\begin{equation}
\cor{}{[\vp^n_+](t,\vec x)}\equiv\int_{-\infty}^{\infty}\der z\;\rho_1(t,z)z^n\,.
\label{eq::rholink}
\end{equation}
Accounting for the different normalisation of the full and effective fields and time coordinates,\footnote{Recall that $t$ in SdSET conventionally refers to the dimensionless time variable $\hat{t}\equiv H t$.} the first two Kramers-Moyal coefficients are related to the (effective) potential and diffusion terms in  \eqref{eq::FP} by
\begin{align}
&V^\prime_{\rm eff}(\varphi) = \sum_{m=0}^\infty 
\frac{D_{1m}}{m!} \,3 H^{(3-m)}\varphi^m\,,
\\
&D=\frac{H^3}{2} D_{20}= \frac{H^3}{8\pi^2}\,\Big[1+
\mathcal{O}(\kappa)\ldots\Big]\,.
\end{align}

The Kramers-Moyal equation \eqref{eq::rholink} can be written as an equation of motion for the renormalised one-point functions $\cor{}{[\vp^n_+](t,\vec x)}$, reminiscent of a set of Schwinger-Dyson equations. Slightly adapting the notation of \cite{Cohen:2021fzf} to our setting and including $c_{1,1}$, one finds 
\begin{align}
\frac{\p}{\p t}\cor{}{[\vp^n_+](t,\vec x)}&=-\frac{n}{3}\sum_{m=0}^{\infty}\frac{c_{2m+1,1}}{(2m+1)!}\cor{}{[\vp^{n+2m}_+](t,\vec x)}+\binom{n}{2}\sum_{m=0}^{\infty}b_m\cor{}{[\vp^{n+2m-2}_+](t,\vec x)}
\nonumber\\
&\hspace*{-2cm}-\binom{n}{3}\sum_{m=0}^{\infty}d_m\cor{}{[\vp^{n+2m-2}_+](t,\vec x)}+\binom{n}{4}\sum_{m=0}^{\infty}e_m\cor{}{[\vp^{n+2m-4}_+](t,\vec x)} +\dots\,.
\label{eq:dtphin}
\end{align}
Here, the coefficients $c_{2m+1,1}$ are the renormalised couplings of the SdSET Lagrangian interactions, cf.\ \eqref{eq:SdSETLagrangian}, so that the drift term now reads 
\begin{equation}
V^\prime_{\rm eff}(z) = 
\sum_{n=0}^{\infty}\frac{c_{2n+1,1}}{(2n+1)!}\,z^{2n+1}\,,
\end{equation}
while the coefficients $b_m$, $d_m$, $e_m$ are related to the Kramers-Moyal coefficients $D_{2m}$, $D_{3m}$, $D_{4m}$, respectively. Specifically, 
\begin{equation}
  D_{20} = b_0\,, \qquad D_{22}=b_1\,. 
\end{equation}

It is worth noting that the Kramers-Moyal equation generates an expansion in $\sqrt{\kappa}$, since the effective field at late times counts as $\vp_+ \sim \kappa^{-1/4}$. This can be seen from the stationary late-time solution to the leading-order Fokker-Planck equation, which leads to one-point expectation values that scale as $\langle \phi^{2n}\rangle \sim 1/(\sqrt{\kappa})^n$.

In \cite{Cohen:2021fzf}, the Kramers-Moyal equation \eqref{eq:dtphin} appears as a consequence of the SdSET operator renormalisation-group equation, controlling the $t$-dependence of the operators $[\vp^n_+]$ via the factorisation scale $\mu$. As such, the coefficients $b_m$, $d_m$, $e_m$, ... should then be related to the $\mu$-anomalous dimensions $\gamma^{\mu}_{nm}$ of these operators. To compute these anomalous dimensions, the authors of \cite{Cohen:2021fzf} limited themselves to the contributions from the non-Gaussian initial conditions to the renormalisation of the $\vp^n_+$. They used the scaling dimension $\alpha$ 
of $\vp_+$ as UV regulator, which corresponds to the dimensional regulator $-\ve$ in our scheme. Their procedure amounts to setting $a_*=a(t)$ in our results. We therefore identify 
\begin{equation}
    b_1=\gamma^{\mu}_{22}\Big|_{\substack{a_*=a(t)}}\,,
    \qquad
    b_0=\gamma^{\mu}_{20}\Big|_{\substack{a_*=a(t)}}\,,
\end{equation}
and obtain from \eqref{eq::gamma22muZ}, \eqref{eq::gamma20muZ}
\begin{equation}
b_1=\frac{\kappa}{12\pi^2}\bigg[\log\bigg(\frac{e^{\gamma_E}\mu}{H}\bigg)-
\frac{4}{3}\bigg]\,,
\label{eq::gamma22muZa=a*}
\end{equation}
and
\begin{equation}
b_0=\frac{1}{4\pi^2}+\frac{\kappa}{96\pi^4}\,\Bigg\{\log^2\bigg(\frac{e^{\gamma_E}\mu}{H}\bigg)
+2\log\bigg(\frac{e^{\gamma_E}\mu}{H}\bigg) \,\bigg[
-\frac{4}{3}+\delta\hat m^2_{\textrm{fin}}\bigg]+\frac{26}{9}-\frac{5\pi^2}{24}-\frac{8}{3}\delta\hat m^2_{\textrm{fin}}\Bigg\}\,,\;
\label{eq::gamma20muZa=a*}
\end{equation}
respectively. Setting $\mu=H$, which is the implicit choice adopted in \cite{Cohen:2021fzf}, we reproduce the result for $b_1$ in that reference, while the one-loop correction to $b_0$ is new. In the context of the Kramers-Moyal expansion, $b_1$ is a next-to-leading order effect in the expansion in $\sqrt{\kappa}$, the first term $1/(4\pi^2)$ in $b_0$ produces the leading-order diffusion term. Since the Kramers-Moyal coefficients are related to UV quantities, the operator mixing anomalous dimensions, their expansion goes in powers of $\kappa$, and the loop correction to $b_0$ constitutes a NNLO effect, 
which appears in the Fokker-Planck equation \eqref{eq::FP} as a genuine quantum correction to the one-loop diffusion coefficient. 
Further NNLO effects have been given in \cite{Cohen:2021fzf}, among them a non-vanishing Kramers-Moyal coefficient $D_{31}$, which shows that at this order the Fokker-Planck ansatz is too restrictive. 

The results for $b_1$ and $b_0$ given above suggest that a convenient 
choice of the hard scale is $\mu=H e^{-\gamma_E}$ rather than 
$\mu=H$, which leads to a particularly simple expression. The 
dependence of the anomalous dimensions, resp.\ Kramers-Moyal coefficients on $\mu$ makes it evident that they do not have physical meaning by themselves, but depend on the choice of renormalisation scheme. This dependence must cancel when one assembles the full-theory correlation functions from the matching coefficients and SdSET correlation functions. 
This happens by construction in the perturbative evaluation of the (IR divergent) correlation functions, as checked in previous sections, but can also be verified on the non-perturbative solution of the Kramers-Moyal equation combined with matching coefficients. 

The situation simplifies at NLO in the $\sqrt{\kappa}$ expansion of the late-time asymptotic solution for the one-point expectation values $\langle \phi^{2n}\rangle$ obtained from the Kramer-Moyal equation, since the only relevant $\mu$-dependent matching coefficients are those of the free theory,
\begin{equation}
C_{2n,2n-2} = \binom{2n}{2}C_{20} = 
-\frac{1}{4\pi^2}\,\binom{2n}{2} \log\bigg(\frac{\mu}{\mu_f}\bigg)\,,
\label{eq:C2n2nm2tree}
\end{equation}
where the second equality follows from \eqref{eq:C22C20free} and the combinatorial factor is the same as in \eqref{eq::Znmfree}. The $\mu$-dependence must therefore cancel among the Kramers-Moyal and with the above $C_{2n, 2n-2}$ coefficients. 
To see how this cancellation occurs, we note that the Kramers-Moyal coefficient $b_1$ can be removed by a variable transformation and absorbed into the quadratic term of the effective potential \cite{Cohen:2021fzf} by the substitution
\begin{equation}
c_{1,1}\to c_{1,1}^\prime = c_{1,1}+\frac{3}{2}\, b_1\,.
\end{equation}
Combining $b_1$ from \eqref{eq::gamma22muZa=a*} with (Ref.~\cite{Beneke:2026rtf}, eqs.~(6.6) and (6.30))
\begin{equation}
c_{1,1} = \frac{\kappa}{8\pi^2}\left[\ln\bigg(\frac{\mu_f}{\mu}\bigg)+\delta\hat m^2_{\textrm{fin}}\right]\,,
\end{equation}
we find
\begin{equation}
c_{1,1}^\prime = \frac{\kappa}{8\pi^2}\left[\ln\bigg(\frac{e^{\gamma_E}\mu_f}{H}\bigg)-\frac{4}{3}+\delta\hat m^2_{\textrm{fin}}\right]\,, 
\end{equation}
which is indeed independent of the SdSET renormalisation scale $\mu$. 

While this renders the Kramers-Moyal coefficients, and therefore the solution for $\rho_1(z)$, scheme-independent at the next-to-leading order, the field redefinition amounts to a change of the integration variable $z$ that also affects the weight $z^n$ on the right-hand side of \eqref{eq::rholink} by a term that depends only on $b_1$ and not the scheme-invariant combination with $c_{1,1}$. One can show that the remaining scheme-dependence of the NLO expectation values can be expressed as  
\begin{equation} 
\langle[\vp_+^{2n}]\rangle|_{\rm NLO,\,\mu-dep.} = \frac{2}{\kappa}\cdot \frac{3}{2}\,b_1|_{\rm \mu-dep.}\, \binom{2n}{2}\langle[\vp_+^{2n-2}]\rangle|_{\rm LO}\,.
\label{eq::residualmu}
\end{equation}
The $\mu$-dependent terms of the $\mathcal{O}(\sqrt{\kappa})$ correction to the full-theory expectation value consists of the two terms\footnote{The following equation and cancellation of $\mu$-dependence within represents the analog of the cancellation of dependence on the parameter $\epsilon$, which separates the full-theory field $\phi$ into a long- and short-wavelength field, in Sec.~8.1.3 of \cite{Gorbenko:2019rza}, where the role of the operator-matching coefficient $C_{20}|_{\rm LO}$ is taken in their framework by the expectation value $\langle \phi_s^2\rangle$ of the short-distance part of the field.} 
\begin{equation}
H^{-2 n}\,\langle \phi^{2 n}\rangle|_{\rm NLO} = C_{2n,2n}|_{\rm LO}\,\times  \langle[\vp_+^{2n}]\rangle|_{\rm NLO} + C_{2n,2n-2}|_{\rm LO}\, \times\langle[\vp_+^{2n-2}]\rangle|_{\rm LO} +\ldots\,,
\end{equation}
where the dots refer to scheme-independent terms. Given $C_{2n,2n}|_{\rm LO}=1$ and the $\mu$-dependent logarithm in \eqref{eq::gamma22muZa=a*}, inserting \eqref{eq:C2n2nm2tree} and \eqref{eq::residualmu} into the previous equation, one finds that the $\mu$-dependence cancels. This provides a non-trivial consistency check of the construction of the Kramers-Moyal equation in SdSET. One is left with a logarithm of $e^{\gamma_E}\mu_f/H$ containing the full-theory UV renormalisation scale, which originates from operator mixing of the full-theory operator $\phi^2$.

 Starting from NNLO in the expansion in $\sqrt{\kappa}$, the matching coefficients computed in the present work contribute to the correlation functions and are required to restore scheme-independence. However, as not all NNLO terms are presently known with full $\mu$-dependence, this important consistency check cannot yet be carried out, and must be left for future work.


\section{Conclusion}

In this paper, building on the foundation laid in \cite{Beneke:2026rtf}, we constructed the formalism to renormalise composite SdSET operators of the form $\vp^n_+$, and demonstrated how to match their renormalised counterparts onto the corresponding renormalised full-theory operators $[\phi^n]$. The formalism was applied to several free-theory examples, and then to the problem of matching the one-loop bispectrum and the two-loop one-point function of $\phi^2$. In all of these cases, we verified that SdSET successfully reproduces all IR-sensitive terms of the full-theory, while the short-distance and early-time physics can be absorbed into IR-insensitive operator-matching coefficients, which can therefore be matched perturbatively. This work, together with \cite{Beneke:2026rtf}, places SdSET introduced in \cite{Cohen:2020php,Cohen:2021fzf} on a firm quantum-field-theoretic footing, and shows the self-consistency of this effective theory as a bona-fide QFT. It also allows us to conclude that, despite many unusual features, SdSET follows very similar rules to flat-space effective theories, in particular, non-relativistic effective field theory. Since it faithfully mirrors the region structure of the full theory, much of the familiar effective-field theory intuition carries over and can be applied. 

The pioneering works \cite{Cohen:2020php,Cohen:2021fzf} showed that the Fokker-Planck equation of the stochastic formalism must be generalised to the Kramers-Moyal equation starting from next-to-next-to-leading order (NNLO) in $\sqrt{\kappa}$ and computed NNLO effects within the SdSET that demonstrate the necessity of this generalisation. In the present work, we obtain two new NNLO corrections, which represent genuine quantum effects, in the sense that they constitute higher-order corrections to quantities that are already present in the LO Fokker-Planck equation:
\begin{itemize}
\item The classic diffusion term in the Fokker-Planck equation for stochastic inflation arises from the quantum fluctuations of the free short-wavelength scalar field modes. Drawing on the relation between SdSET operator anomalous dimensions and the Kramers-Moyal (KM) coefficients \cite{Cohen:2021fzf}, we computed the first correction from the scalar field self-interaction to the diffusion coefficient, which now reads 
\begin{equation}
D=\frac{H^3}{8\pi^2}\Bigg[1+\frac{\kappa}{24\pi^2}\,\bigg\{L_\mu^2
+L_\mu\,\bigg[
-\frac{8}{3}+2\delta\hat m^2_{\textrm{fin}}\bigg]+\frac{26}{9}-\frac{5\pi^2}{24}-\frac{8}{3}\delta\hat m^2_{\textrm{fin}}\bigg\}+\mathcal{O}(\kappa^2)\Bigg]\,,\;
\label{eq:1loopdiffusion}
\end{equation}
where $L_\mu = \ln(e^{\gamma_E}\mu/H)$. The diffusion coefficient is a short-distance quantity and depends on the renormalisation scale $\mu$ that separates subhorizon and superhorizon modes. The $\mu$-dependence is cancelled in observables by other NNLO effects that are not yet completely known.
\item The Fokker-Planck equation emerges as the effective equation of motion for the superhorizon modes, which allows one to compute the late-time limit of one-point functions of the SdSET superhorizon field operator $[\vp_+^n]$. The quantities of interest are, however, the expectation values of the full-theory operator $[\phi^n]$. The relation is provided by short-distance matching coefficients that encode subhorizon physics not captured by the SdSET field. In this work, we obtain the $\mathcal{O}(\kappa)$ loop correction to the coefficients relevant to the coincident two-point function $\langle \phi^2\rangle$, see \eqref{eq::C22final} and \eqref{eq::C20}.
\end{itemize}
While \cite{Cohen:2020php,Cohen:2021fzf} provided convincing arguments for the Kramer-Moyal extension, it would still be interesting to use the framework set up in \cite{Beneke:2026rtf} and the present work to provide a rigorous derivation of the Kramers-Moyal equation
from operator renormalisation in the dimensional regularisation scheme. Likewise, it would be useful to clarify the connection of the two anomalous-dimension matrices $\gamma^{\mu}$ and $\gamma^{a_*}$ introduced in \secref{sec:EFTopmatch} to the KM coefficients and the role of the choice $a_*=a(t)$ that was required to obtain the latter.   
The authors of \cite{Cohen:2020php,Cohen:2021fzf} used the intuition that the flow of time and the RG-flow in SdSET are intimately related to each other to calculate KM coefficients 
by extracting UV poles from composite-operator correlation functions. However, a general proof of this statement is still missing and does not emerge immediately in the present regularisation. 
We aim to fill these gaps in a future publication.

\subsubsection*{Acknowledgement}

We thank Tim Cohen for many useful discussions and continued support. 
This work has been supported in part by the Excellence Cluster ORIGINS funded by the Deutsche Forschungsgemeinschaft under Grant No.~EXC - 2094 - 390783311 and by the Cluster of Excellence Precision Physics, Fundamental Interactions, and Structure of Matter (PRISMA$^+$ EXC 2118/1) funded by the German Research Foundation (DFG) within the German Excellence Strategy (Project ID 390831469). The work of AFS is supported by the grants
EUR2024.153549, CNS2024-154834 and PID2022-139466NB-C21 (FEDER/UE) funded by
the Spanish Research Agency (MICIU/AEI/10.13039/501100011033). 


\appendix
\addtocontents{toc}{\protect\setcounter{tocdepth}{1}}

\section{Derivation of the formula for \texorpdfstring{$Z^{-1}_{nm}$}{Z⁻¹ₙₘ} in free SdSET}
\label{app::Zinv}

In this Appendix, we derive \eqref{eq::Z0inv} for the inverse operator-renormalisation matrix $Z$. The elements of the matrix $N^l$ with $N$ defined in \eqref{eq::Zfreedecomp} are given by 
\begin{equation}
N^l_{nm}=\sum_{n>k_1>k_2>...>k_{l-1}>m}Z_{nk_1}Z_{k_1k_2}...Z_{k_{l-1}m}\,.
\label{eq::Nlsum1}
\end{equation}
If the constraint on the summation indices cannot be satisfied, $N^l_{nm}$ vanishes, since $N$ is lower triangular with no diagonal entries. 
Eq.~\eqref{eq::Z0} for the $Z_{nm}$ implies the simplification 
\begin{equation}
Z_{nk_1}Z_{k_1k_2}...Z_{k_{l-1}m}=\frac{n!}{2^{\frac{n-m}{2}}m!}\bigg(\frac{1}{8\pi^2\ve}\bigg)^{\frac{n-m}{2}}\frac{1}{(\frac{n-k_1}{2})!(\frac{k_1-k_2}{2})!...(\frac{k_{l-1}-m}{2})!}
\label{eq::weightfactor}
\end{equation}
for the summands. The sum in \eqref{eq::Nlsum1} is cumbersome to work with, since the $l-1$ summation indices are interlinked. To write it more transparently, it is useful to think of each $Z_{ij}$ as a bucket containing $(i-j)/2$ balls, representing the difference between the indices, which is always an integer. Using this picture, the sum  \eqref{eq::Nlsum1} runs over all possibilities of distributing $(n-m)/2$ balls into $l-1$ buckets, such that each bucket contains at least one ball, and each possible configuration is then weighted by the factor \eqref{eq::weightfactor}. We can use this picture and \eqref{eq::weightfactor} to rewrite \eqref{eq::Nlsum1} as
\begin{equation}
N^l_{nm}=\frac{n!}{2^{\frac{n-m}{2}}m!}\bigg(\frac{1}{8\pi^2\ve}\bigg)^{\frac{n-m}{2}}\hspace*{-0.5cm}\sum_{\substack{k_1>0\,,...\,,k_{l-1}>0\\ k_1+k_2+...+k_{l-1}=\frac{n-m}{2}}}\prod_{j=1}^{l-1}\frac{1}{k_j!}\,.
\label{eq::Nlsum2}
\end{equation}
This reduces the problem of computing $N^l_{nm}$ to evaluating the  combinatorial sum, which can be achieved by the observation 
\begin{equation}
\sum_{\substack{k_1>0\,,...\,,k_{l-1}>0\\ k_1+...+k_{l-1}=\frac{n-m}{2}}}\prod_{j=1}^{l-1}\frac{1}{k_j!}=\frac{1}{(\frac{n-m}{2})!}\frac{\der^{\frac{n-m}{2}}}{\der x^{\frac{n-m}{2}}}\bigg|_{x=0}(e^x-1)^{l-1}\,.
\end{equation}
The factor $(e^x-1)^r$ is the generating function of the Stirling numbers of the second kind $S(n,k)$ via
\begin{equation}
(e^x-1)^r=r!\sum_{n=r}^{\infty}S(n,r)\frac{x^n}{n!}\,.
\end{equation}
Combining these two expressions, we find
\begin{equation}
\sum_{\substack{k_1>0\,,...\,,k_{l-1}>0\\ k_1+k_2+...+k_{l-1}=\frac{n-m}{2}}}\prod_{j=1}^{l-1}\frac{1}{k_j!}=\frac{(l-1)!}{(\frac{n-m}{2})!}S\bigg(\frac{n-m}{2},l-1\bigg)\,.
\end{equation}
Inserting this formula into \eqref{eq::Nlsum2} gives 
\begin{flalign}
N^l_{nm}&=\frac{n!}{2^{\frac{n-m}{2}}m!}\bigg(\frac{1}{8\pi^2\ve}\bigg)^{\frac{n-m}{2}}\,\frac{(l-1)!}{(\frac{n-m}{2})!}\,S\bigg(\frac{n-m}{2},l-1\bigg)
\nn\\
&=Z_{nm}\,(l-1)!\,S\bigg(\frac{n-m}{2},l-1\bigg)\,.
\end{flalign}
Finally, armed with this result, we can use \eqref{eq::Z0inv}, letting $m\rightarrow n-m$, to compute
\begin{flalign}
Z^{-1}_{nm}&=\delta_{nm}+\sum_{l=1}^{\frac{n-m}{2}}(-1)^lN^l_{nm}
=Z_{nm}\sum_{l=0}^{\frac{n-m}{2}}(-1)^ll!\,S\bigg(\frac{n-m}{2},l\bigg)
\nonumber\\
&=(-1)^{\frac{n-m}{2}}Z_{nm}\,,
\end{flalign}
where the last step uses a standard identity for the Stirling numbers of the second kind.

\section{Inversion of the operator renormalisation matrix in the interacting theory}
\label{app:intZinv}

In this Appendix, we show how to invert the operator-renormalisation matrix $Z$ in the interacting theory. 
We build on the considerations in \secref{sec:freemix} and write
\begin{equation}
Z=Z_{\textrm{free}}+\sum_{n=1}^{\infty}\kappa^n Z^{(n)}_{\textrm{int}}\,,
\end{equation}
where $Z_{\textrm{free}}$ is the known counterterm matrix of the free theory and the sum collects the correction from interactions expanded in $\kappa$. 
To obtain the inverse of $Z$, we factor out $Z_{\textrm{free}}$ from the remainder, and write
\begin{equation}
Z=Z_{\textrm{free}}\bigg[\un+Z_{\textrm{free}}^{-1}\sum_{n=1}^{\infty}\kappa^n Z^{(n)}_{\textrm{int}}\bigg]\,.
\end{equation}
Its inverse 
\begin{equation}
Z^{-1}=\bigg[\un+Z_{\textrm{free}}^{-1}\sum_{n=1}^{\infty}\kappa^n Z^{(n)}_{\textrm{int}}\bigg]^{-1}Z^{-1}_{\textrm{free}}\,,
\end{equation}
can then be computed perturbatively in $\kappa$. 

To determine $Z^{-1}$ at $\Lo(\kappa)$, one can truncate the resulting series after the second summand,
\begin{equation}
Z^{-1}=Z_{\textrm{free}}^{-1}-\kappa Z^{-1}_{\textrm{free}}Z^{(1)}_{\textrm{int}}Z^{-1}_{\textrm{free}}+\Lo(\kappa^2)\,.
\end{equation}
We can now use the decomposition
\begin{equation}
Z_{\textrm{free}}=\un+N\,,\qquad Z^{-1}_{\textrm{free}}=\un+\sum_{l=1}^{\infty}(-1)^lN^l\,,
\end{equation}
 of $Z_{\textrm{free}}$ and its inverse introduced in \eqref{eq::Zfreedecomp}, \eqref{eq::Zfreeinv}
 to write
\begin{flalign}
Z^{-1}&=Z_{\textrm{free}}^{-1}-\kappa \bigg[Z^{(1)}_{\textrm{int}}+\sum_{l=1}^{\infty}(-1)^lN^lZ^{(1)}_{\textrm{int}}+Z^{(1)}_{\textrm{int}}\sum_{l=1}^{\infty}(-1)^lN^l\nonumber\\
&\quad+\sum_{l=1}^{\infty}(-1)^lN^lZ^{(1)}_{\textrm{int}}\sum_{m=1}^{\infty}(-1)^mN^m\bigg]+\Lo(\kappa^2)\,.
\end{flalign}
Since the matrix $N^m$ is lower triangular, and its main diagonal and first $m-1$ lower diagonals are vanishing, the calculation of any particular entry $(n,n-m)$ of $Z^{-1}$ requires only finitely many summands
\begin{flalign}
Z^{-1}_{n,n-m}&=(Z_{\textrm{free}}^{-1})_{n,n-m}-\kappa \bigg[Z^{(1)}_{\textrm{int}}+\sum_{l=1}^{m}(-1)^lN^lZ^{(1)}_{\textrm{int}}+Z^{(1)}_{\textrm{int}}\sum_{l=1}^{m}(-1)^lN^l\nonumber\\
&\quad+\sum_{l=1}^{m}(-1)^lN^lZ^{(1)}_{\textrm{int}}\sum_{k=1}^{m}(-1)^kN^k\bigg]_{n,n-m}\,.
\end{flalign}
This procedure can be carried out systematically in the same fashion to any order in $\kappa$.

\section{Operator-matching coefficients in the free theory}
\label{app:freematch}

In this Appendix, we illustrate the structure of the matching coefficients in the free theory, and how they can be used to determine $\gamma^{\mu}_{nm}$.

In the free theory, the correlation functions of the full theory and the EFT differ only by powers of $H$ and the replacement $\mu_f\leftrightarrow\mu$, so we can recycle the examples from \secref{sec:freemix}. 
According to the matching equation \eqref{matchdiffmu}, we have
\begin{equation}
\cor{}{[\phi^2](t,\vec x)}=H^2\Big[C_{22}\cor{}{[\vp^2_+](t,\vec x)}+C_{20}\Big]\,.
\label{eq:appmatch1}
\end{equation}
Using the results of \secref{sec:freemix}, the two relevant one-point functions are given by
\begin{equation}
\cor{}{[\phi^2](t,\vec x)}=-\frac{H^2}{4\pi^2}\log\bigg(\frac{\Lambda}{a(t)\mu_f}\bigg)\,,\qquad \cor{}{[\vp_+^2](t,\vec x)}=-\frac{1}{4\pi^2}\log\bigg(\frac{\Lambda}{a(t)\mu}\bigg)\,.
\label{eq:treephisq}
\end{equation}
Inserting them into \eqref{eq:appmatch1} yields
\begin{equation}
-\frac{H^2}{4\pi^2}\log\bigg(\frac{\Lambda}{a(t)\mu_f}\bigg)=-\,C_{22}\frac{H^2}{4\pi^2}\log\bigg(\frac{\Lambda}{a(t)\mu}\bigg)+H^2\,C_{20}\,,
\end{equation}
which fixes
\begin{equation}
C_{22}=1\,,\qquad C_{20}=-\frac{1}{4\pi^2}\log\bigg(\frac{\mu}{\mu_f}\bigg)\,.
\label{eq:C22C20free}
\end{equation}

Next, we consider the matching equation
\begin{equation}
[\phi^3](t,\vec x)=H^3\Big[C_{33}[\vp^3_+](t,\vec x)+C_{31}\vp_+(t,\vec x)\Big]\,,
\label{eq:appmatch2}
\end{equation}
which, at the level of correlation functions, implies
\begin{equation}
\cor{}{[\phi^3](t,\vec x)\phi(t,\vec x_1)}=H^4\Big[C_{33}\cor{}{[\vp^3_+](t,\vec x)\vp_+(t,\vec x_1)}+C_{31}\cor{}{\vp_+(t,\vec x)\vp_+(t,\vec x_1)}\Big]\,.
\end{equation}
Evaluating the correlation functions in the free theory with the help of \eqref{eq:phi3tophi1} results in  
\begin{flalign}
\cor{}{[\phi^3](t,\vec x)\phi(t,\vec x_1)}&=-\frac{3H^2}{4\pi^2}\log\bigg(\frac{\Lambda}{a(t)\mu_f}\bigg)\cor{}{\phi(t,\vec x)\phi(t,\vec x_1)}\,,\\
\cor{}{[\vp_+^3](t,\vec x)\vp_+(t,\vec x_1)}&=-\frac{3}{4\pi^2}\log\bigg(\frac{\Lambda}{a(t)\mu}\bigg)\cor{}{\vp_+(t,\vec x)\vp_+(t,\vec x_1)}\,.
\end{flalign}
The matching equation then reads
\begin{flalign}
&-\frac{3H^2}{4\pi^2}\log\bigg(\frac{\Lambda}{a(t)\mu_f}\bigg)\cor{}{\phi(t,\vec x)\phi(t,\vec x_1)}\nonumber\\
&=H^4\bigg[-C_{33}\frac{3}{4\pi^2}\log\bigg(\frac{\Lambda}{a(t)\mu}\bigg)\cor{}{\vp_+(t,\vec x)\vp_+(t,\vec x_1)}+C_{31}\cor{}{\vp_+(t,\vec x)\vp_+(t,\vec x_1)}\bigg]\,.
\end{flalign}
Accounting for the factor $H$ relating the full and effective fields,
we get
\begin{equation}
C_{33}=1\,,\qquad C_{31}=-\frac{3}{4\pi^2}\log\bigg(\frac{\mu}{\mu_f}\bigg)\,.
\end{equation}

Similarly, from the matching equation 
\begin{equation}
[\phi^4](t,\vec x)=H^4\Big[C_{44}[\vp^4_+](t,\vec x)+C_{42}[\vp^2_+](t,\vec x)+C_{40}\un\Big]
\label{eq:match4}
\end{equation}
we obtain
\begin{align}
\cor{}{[\phi^4](t,\vec x)\phi(t,\vec x_1)\phi(t,\vec x_2)}&=H^6\Big[C_{44}\cor{}{[\vp_+^4](t,\vec x)\vp_+(t,\vec x_1)\vp_+(t,\vec x_2)}\nonumber\\
&\hspace*{-3cm}+\,C_{42}\cor{}{[\vp^2_+](t,\vec x)\vp_+(t,\vec x_1)\vp_+(t,\vec x_2)}+C_{40}\cor{}{\vp_+(t,\vec x_1)\vp_+(t,\vec x_2)}\Big]\,.\qquad
\end{align}
Matching the connected part of the left- and right-hand side of this equation results in
\begin{flalign}
&-\frac{3H^2}{2\pi^2}\log\bigg(\frac{\Lambda}{a(t)\mu_f}\bigg)
\cor{}{\phi(t,\vec x)\phi(t,\vec x_1)}\cor{}{\phi(t,\vec x)\phi(t,\vec x_2)}\nonumber\\
&=H^6\bigg[-\frac{3C_{44}}{2\pi^2}\log\bigg(\frac{\Lambda}{a(t)\mu}\bigg)+C_{42}\bigg]\cor{}{\vp_+(t,\vec x)\vp_+(t,\vec x_1)}\cor{}{\vp_+(t,\vec x)\vp_+(t,\vec x_2)}\,.
\end{flalign}
Accounting for the different normalisation of the full and effective fields, we obtain
\begin{equation}
C_{44}=1\,,\qquad C_{42}=-\frac{3}{2\pi^2}\log\bigg(\frac{\mu}{\mu_f}\bigg)\,.\label{eq::C42free}
\end{equation}
Taking the expectation value of \eqref{eq:match4} and inserting the already determined values of $C_{44}$, $C_{42}$ as well as  
the expressions
\begin{equation}
\cor{}{[\phi^4](t,\vec x)}=\frac{3H^4}{16\pi^4}\log^2\bigg(\frac{\Lambda}{a(t)\mu_f}\bigg)\,,\qquad
\cor{}{[\vp_+^4](t,\vec x)}=\frac{3}{16\pi^4}\log^2\bigg(\frac{\Lambda}{a(t)\mu}\bigg)\,,
\end{equation} 
we get
\begin{flalign}
C_{40}&=H^{-4}\cor{}{[\phi^4](t,\vec x)}-\cor{}{[\vp_+^4](t,\vec x)}+\frac{3}{2\pi^2}\log\bigg(\frac{\mu}{\mu_f}\bigg)\cor{}{[\vp^2_+](t,\vec x)}\nonumber\\
&=\frac{3}{16\pi^4}\log^2\bigg(\frac{\mu}{\mu_f}\bigg)\,.
\end{flalign}

To summarise, we determined the upper $5\times5$-block of the infinite-dimensional matrix $C$, which reads
\begin{equation}
C|_{\,\textrm{upper }5\times5}=\left(\begin{array}{ccccc}
1 & 0 & 0 & 0 & 0\\
0 & 1 & 0 & 0 & 0\\
-\frac{L}{4\pi^2} & 0 & 1 & 0 & 0\\
0 & -\frac{3L}{4\pi^2} & 0 & 1 & 0\\
\frac{3L^2}{16\pi^4} & 0 & -\frac{3L}{2\pi^2} & 0 & 1\\
\end{array}\right)\,,
\end{equation}
where we abbreviated 
\begin{equation}
L\equiv\log\bigg(\frac{\mu}{\mu_f}\bigg)\,.
\end{equation}
Since this is a lower-triangular matrix, it can be inverted independently of the unwritten part of the matrix, and we find
\begin{equation}
C^{-1}|_{\,\textrm{upper }5\times5}=\left(\begin{array}{ccccc}
1 & 0 & 0 & 0 & 0\\
0 & 1 & 0 & 0 & 0\\
\frac{L}{4\pi^2} & 0 & 1 & 0 & 0\\
0 & \frac{3L}{4\pi^2} & 0 & 1 & 0\\
\frac{3L^2}{16\pi^4} & 0 & \frac{3L}{2\pi^2} & 0 & 1\\
\end{array}\right)\,.
\end{equation}
Using these results in \eqref{eq::CADM} one finds 
\begin{align}
\gamma^{\mu}_{20}&=-\sum_{l=0}^{\infty}C^{-1}_{2l}\frac{\der C_{l0}}{\der\log(\mu)}=\frac{1}{4\pi^2}\,, &
\gamma^{\mu}_{31}&=-\sum_{l=0}^{\infty}C^{-1}_{3l}\frac{\der C_{l1}}{\der\log(\mu)}=\frac{3}{4\pi^2}\,,\nonumber\\
\gamma^{\mu}_{42}&=-\sum_{l=0}^{\infty}C^{-1}_{4l}\frac{\der C_{l2}}{\der\log(\mu)}=\frac{3}{2\pi^2}\,, &
\gamma^{\mu}_{40}&=-\sum_{l=0}^{\infty}C^{-1}_{4l}\frac{\der C_{l0}}{\der\log(\mu)}=0\,,
\end{align}
in agreement with the general formula \eqref{eq::freegammamu} obtained from the $Z$-factor matrix. 

\section{Late-time limit of the one-loop bispectrum of the operator \texorpdfstring{$\phi^2$}{φ²}}
\label{app:22q0}

\begin{figure}[t]
    \centering
    \includegraphics[width=0.45\textwidth]{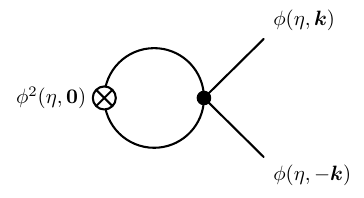}
    \caption{The 1PI diagram topology for the bispectrum of the operator $\phi^2$ at $\Lo(\kappa)$.}
    \label{fig:Phi2PhiPhi}
\end{figure}

In this Appendix, we compute the 1PI component of the correlation function
\begin{equation}
\cor{}{\phi^2(\eta,\vec 0)\phi(\eta,\vec k)\phi(\eta,-\vec k)}'
\end{equation}
at the one-loop order, see diagram in Fig.~\ref{fig:Phi2PhiPhi}. The same correlation function with general external momenta ($\vec{q}\not= 0$)  was computed in \cite{Beneke:2023wmt} employing dimensional regularisation for both the UV- and IR-divergences. Here, we repeat the computation using the comoving $\Lambda$-regulator for the IR, which simplifies the disentanglement of the two types of divergences. The starting expression reads
\begin{flalign}
&\cor{}{\phi^2(\eta,\vec 0)\phi(\eta,\vec k)\phi(\eta,-\vec k)}'_{|\,\Lo(\kappa)\textrm{, 1PI}}\nonumber\\
&=\frac{\tmu_f^{4-d}\kappa H^d(-H\eta)^{2d-8}}{8k^6_{\Lambda}}\,\Im\,\Bigg[\int_{-\infty}^{\eta}\frac{\der\eta'}{(-\eta')^{8-d}}\int\frac{\der^{d-1}l}{(2\pi)^{d-1}}\frac{e^{2i(k_{\Lambda}+l_{\Lambda})(\eta'-\eta)}}{l_{\Lambda}^6}(-i+l_{\Lambda}\eta)^2(i+l_{\Lambda}\eta')^2\nonumber\\
&\quad\times(-i+k_{\Lambda}\eta)^2(i+k_{\Lambda}\eta')^2\Bigg]\,.
\label{eq::22start}
\end{flalign}
As in \cite{Beneke:2023wmt}, we use the method of regions to compute the late-time asymptotics of this integral directly. The structure of the computation is analogous to the one presented there, and some of the results obtained can be recycled. It is necessary to introduce an auxiliary analytic regulator
\begin{equation}
(-\nu\eta')^{-2\delta}
\end{equation}
in \eqref{eq::22start} to render all region-integrals  well-defined.

\subsection{Late-time regions}

\subsubsection{Soft momentum}

In this region, we expand the integrand of \eqref{eq::22start} assuming both $-l_{\Lambda}\eta^{(\prime)}\ll1$ and $-k_{\Lambda}\eta^{(\prime)}\ll1$, and keep only the terms which do not vanish in the late-time limit after integration. The analytic regulator is not needed here, so we may set $\delta$ to zero from the start. Doing so, we find
\begin{flalign}
&\cor{}{\phi^2(\eta,\vec 0)\phi(\eta,\vec k)\phi(\eta,-\vec k)}'_{|\,\textrm{late }\eta,\,\textrm{soft }l}\nonumber\\
&=-\frac{\tmu_f^{4-d}\kappa H^d(-H\eta)^{2d-8}}{8k_{\Lambda}^6}\frac{(-\eta)^{-2\ve}}{2\ve(3+2\ve)}\bigg[2\int\frac{\der^{d-1}l}{(2\pi)^{d-1}}\frac{1}{l^3_{\Lambda}}+2k_{\Lambda}^3\int\frac{\der^{d-1}l}{(2\pi)^{d-1}}\frac{1}{l^6_{\Lambda}}\bigg]\nonumber\\
&=-\frac{\kappa H^4(-H\eta)^{-4\ve}}{96\pi^2k_{\Lambda}^6}\,\bigg(\frac{e^{\gamma_E}\mu_f}{H}\bigg)^{\!2\ve} (-e^{\gamma_E}\Lambda\eta)^{-2\ve}\nonumber\\
&\phantom{=}\times\,\bigg[\,\frac{1}{\ve^2}-\frac{2}{3\ve}+\frac{4}{9}+\frac{\pi^2}{12}
+\frac{\pi k_{\Lambda}^3}{\Lambda^3}\,\bigg(\frac{1}{8\ve}+\frac{1}{6}-\frac{1}{4}\log 2 \bigg)\bigg]\,.
\end{flalign}
To obtain the last equality, we expanded the integrals in $\ve$, and dropped the terms that vanish as $\Lambda\rightarrow0$. The UV divergences appear as poles in $\ve$, while the IR divergences are both power-like and logarithmic in $\Lambda$. Since inverse powers of $\Lambda$ appear, the expansion of $k_{\Lambda}$ in $\Lambda$ would generate additional, non-vanishing terms as $\Lambda\rightarrow0$. 
However, since the full $k$- and $\Lambda$-dependence must be reproduced by the EFT, these additional terms drop out in the matching.
Therefore, we replace $k_{\Lambda}$ by $k$ at the end of the computations in the following equations, keeping in mind that these terms are present.

\subsubsection{Hard momentum}

In this region we still assume $-k_{\Lambda}\eta^{(\prime)}\ll1$, but now $l\gg k$, which implies $-l\eta^{(\prime)}\sim 1$, so we cannot expand the integrand of \eqref{eq::22start} in $-l\eta^{(\prime)}$. The assumption that $l\gg k$ also implies $l\gg\Lambda$, and we therefore let $\Lambda\rightarrow0$ in the integrand. The only leftover $\Lambda$-dependence thus arises from the external momenta, but, as discussed above, we do not write this out explicitly in the final result. 
The late-time, hard-momentum region is therefore unchanged with respect to the result obtained in \cite{Beneke:2023wmt}, and reads 
\begin{flalign}
&\cor{}{\phi^2(t,\vec 0)\phi(t,\vec k)\phi(t,-\vec k)}'_{|\,\textrm{late $\eta$, hard $l$}}\nonumber\\
&=\frac{\tmu_f^{4-d}\kappa H^d(-H\eta)^{2d-8}}{8k_{\Lambda}^6}\,\nu^{-2\delta}\,\Im\,\Bigg[\int_{-\infty}^{\eta}\frac{\der \eta'}{(-\eta')^{8-d+2\delta}}\int\frac{\der^{d-1}l}{(2\pi)^{d-1}}\frac{e^{2il(\eta'-\eta)}(-i+l\eta)^2(i+l\eta')^2}{l^6}\Bigg]\nonumber\\
&=\frac{\kappa H^4(-H\eta)^{-4\ve}}{96\pi^2k^6}\,\bigg(\frac{e^{\gamma_E}\mu_f}{H}\bigg)^{\!2\ve}(-\nu\eta)^{-2\delta}\,\bigg[\frac{1}{\delta}\bigg(\frac{1}{\ve}-\frac{8}{3}\bigg)-\frac{13}{6\ve}+\frac{56}{9}-\frac{2\pi^2}{3}\bigg]\,.
\end{flalign}
Since $\Lambda$ was set to zero before the integration, the IR is now regulated by the presence of $\delta$ and $\ve$, as in \cite{Beneke:2023wmt}.

\subsection{Early-time regions}

\subsubsection{Soft-momentum}

Here we assume that $-k\eta\sim-l\eta\ll1$, and expand the integrand of \eqref{eq::22start} in these quantities, but $-k\eta'\sim-l\eta'\sim 1$, 
so the full dependence of the integrand on these combinations must be kept. Under these assumptions the upper boundary of the time integral can be set to zero, which gives the expression
\begin{flalign}
&\cor{}{\phi^2(\eta,\vec 0)\phi(\eta,\vec k)\phi(\eta,-\vec k)}'|_{\,\textrm{early $\eta$, soft $l$}}\nonumber\\[0.2cm]
&=\frac{\kappa \tmu_f^{4-d} H^d(-H\eta)^{2d-8}}{8k_{\Lambda}^6}\,\nu^{-2\delta}\,\Im\,\Bigg[\int\frac{\der^{d-1}l}{(2\pi)^{d-1}}\frac{1}{l^6_{\Lambda}}\int_{-\infty}^0\frac{\der \eta'}{(-\eta')^{8-d+2\delta}}\;e^{2i(k_{\Lambda}+l_{\Lambda})\eta'}(i+l_{\Lambda}\eta')^2\nonumber\\
&\quad\times(i+k_{\Lambda}\eta')^2\Bigg]\nonumber\\
&=\frac{\kappa H^4(-H\eta)^{-4\ve}}{96\pi^2k^6}\bigg(\frac{e^{\gamma_E}\mu_f}{H}\bigg)^{\!2\ve}\bigg(\frac{e^{\gamma_E}\Lambda}{\nu}\bigg)^{\!2\delta}\,\Bigg\{\frac{1}{\ve^2}-\frac{1}{\delta}\bigg[\frac{1}{\ve}-\frac{8}{3}\bigg]+\frac{1}{\ve}\frac{\pi k^3}{8\Lambda^3}-\frac{\pi k^3}{4\Lambda^3}\bigg[\frac{4}{3}-\log\bigg(\frac{k}{\Lambda}\bigg)\bigg]\nonumber\\
&\quad-\frac{9\pi k}{4\Lambda}+\log\bigg(\frac{2k}{\Lambda}\bigg)\bigg[2\log\bigg(\frac{2k}{\Lambda}\bigg)+\frac{1}{3}\bigg]-\frac{17}{3}+\frac{11\pi^2}{12}\Bigg\}\,.
\end{flalign}

\subsubsection{Hard momentum}

As in the late-time, hard-momentum region considered above, in this region we let $\Lambda\rightarrow0$ in the integrand. This renders all integrals appearing in this region scaleless and vanishing, as was found for the analogous region in \cite{Beneke:2023wmt}. 

\subsection{Sum of the regions, renormalisation of the operator \texorpdfstring{$\phi^2$}{φ²}}

Summing the regions determined above, we find the regularised correlator
\begin{flalign}
&\cor{}{\phi^2(\eta,\vec 0)\phi(\eta,\vec k)\phi(\eta,-\vec k)}'_{|\,\Lo(\kappa)\textrm{, 1PI}}\nonumber\\
&=\frac{\kappa H^4(-H\eta)^{-4\ve}}{96\pi^2k^6}\,\bigg(\frac{e^{\gamma_E}\mu_f}{H}\bigg)^{\!2\ve}\,\Bigg\{-\frac{3}{2\ve}+2\log(-2e^{\gamma_E}k\eta)\bigg[-\log(-2e^{\gamma_E}k\eta)+2\log\bigg(\frac{2k}{\Lambda}\bigg)+2\bigg]\nonumber\\
&\quad-\frac{11}{3}\log\bigg(\frac{2k}{\Lambda}\bigg)-\frac{\pi k^3}{4\Lambda^3}\bigg[2-\log(-2e^{\gamma_E}k\eta)\bigg]-\frac{9\pi k}{4\Lambda}+\frac{1}{9}+\frac{\pi^2}{6}\Bigg\}\,.
\label{eq::Phi2reg}
\end{flalign}
The only leftover pole in $\ve$, which originates from the UV divergence of this correlation function, matches the UV pole determined in \cite{Beneke:2023wmt}, while the IR divergences are now captured by the $\Lambda$-dependence of the result. 

To subtract the UV divergence
we define the renormalised operator $[\phi^2](\eta,\vec x)$ via
\begin{equation}
\tmu_f^{2\ve}\phi^2(\eta,\vec x)=Z^{\phi}_{22}\,\tmu_f^{2\ve}[\phi^2](\eta,\vec x)+H^2Z^{\phi}_{20 }\,\un+\textrm{ higher powers of }\phi\,.
\label{eq::Phi2reneq}
\end{equation}
The pole in \eqref{eq::Phi2reg} is removed by 
\begin{equation}
Z^{\phi}_{22}=1-\frac{\kappa}{48\pi^2}\frac{3}{2\ve}+\Lo(\kappa^2)\,,
\end{equation}
which then defines the four-dimensional renormalised correlation function
\begin{flalign}
&\cor{}{[\phi^2](\eta,\vec 0)\phi(\eta,\vec k)\phi(\eta,-\vec k)}'_{|\,\Lo(\kappa)\textrm{, 1PI}}\nonumber\\
&=\frac{\kappa H^4}{96\pi^2k^6}\,\Bigg\{-3\log\bigg(\frac{e^{\gamma_E}\mu_f}{H}\bigg)+2\log(-2e^{\gamma_E}k\eta)\bigg[-\log(-2e^{\gamma_E}k\eta)+2\log\bigg(\frac{2k}{\Lambda}\bigg)+2\bigg]\nonumber\\
&\quad-\frac{11}{3}\log\bigg(\frac{2k}{\Lambda}\bigg)-\frac{\pi k^3}{4\Lambda^3}\bigg[2-\log(-2e^{\gamma_E}k\eta)\bigg]-\frac{9\pi k}{4\Lambda}+\frac{1}{9}+\frac{\pi^2}{6}\Bigg\}\,.
\end{flalign}
This is the result that is used in the main text for the matching computation.


\section{Loop integrals for the one-loop bispectrum of the composite operator \texorpdfstring{$\vp^2_+$}{φ₊²}}
\label{app::22q0loops}

In this Appendix, we collect the explicit expressions for the evaluation of the non-trivial loop integrals resulting from the insertion of the non-Gaussian initial condition $\Xi_{3,1}$ into the SdSET one-loop bispectrum of the composite operator $\vp_+^2$ discussed in \secref{sec::22q0EFT}. 

To compute the integrals involving the logarithm of $l_{\Lambda}+k_{\Lambda}$ we first write
\begin{equation}
\log\bigg(\frac{2e^{\gamma_E}(l_{\Lambda}+k_{\Lambda})}{a_*H}\bigg)=\frac{\der}{\der u}\bigg|_{u=0}\bigg(\frac{2e^{\gamma_E}(l_{\Lambda}+k_{\Lambda})}{a_*H}\bigg)^{\!u}\,,
\end{equation}
and pull the $u$-derivative out of the integrals. We use the Mellin-Barnes representation \cite{Smirnov:2012gma}
\begin{equation}
\frac{1}{(l_{\Lambda}+k_{\Lambda})^{-u}}=\frac{1}{\Gamma(-u)}\frac{1}{2\pi i}\int_{-i\infty}^{i\infty}\der z\;\Gamma(-u+z)\Gamma(-z)\frac{l_{\Lambda}^z}{k_{\Lambda}^{-u+z}}
\end{equation}
to simplify the $\vec l$-integration. After performing this integration, the $z$-integration can be carried out using the residue theorem. We only need to keep the residues that give non-vanishing terms in the limit $\Lambda\rightarrow0$. Finally, we take the $u$-derivative of the resulting expression, and then set $u=0$, obtaining the three results
\begin{flalign}
&\bigg(\frac{e^{\gamma_E}\Lambda^2}{4\pi}\bigg)^{\!\ve}\int\frac{\der^{d-1}l}{(2\pi)^{d-1}}\frac{1}{l^6_{\Lambda}}\log\bigg(\frac{2e^{\gamma_E}(l_{\Lambda}+k_{\Lambda})}{a_*H}\bigg)\nonumber\\
&\quad=\frac{1}{6\pi^2}\,\Bigg\{\frac{1}{k^3}\bigg[\log\bigg(\frac{2k}{\Lambda}\bigg)-\frac{2}{3}\bigg]-\frac{3\pi}{8k^2\Lambda}+\frac{1}{k\Lambda^2}+\frac{3\pi}{16\Lambda^3}\log\bigg(\frac{e^{\gamma_E}\Lambda}{a_*H}\bigg)\Bigg\}\,,\\[.3cm]
&\bigg(\frac{e^{\gamma_E}\Lambda^2}{4\pi}\bigg)^{\!\ve}\int\frac{\der^{d-1}l}{(2\pi)^{d-1}}\frac{1}{l^3_{\Lambda}}\log\bigg(\frac{2e^{\gamma_E}(l_{\Lambda}+k_{\Lambda})}{a_*H}\bigg)\nonumber\\
&\quad=\frac{1}{8\pi^2}\,\Bigg\{\frac{1}{\ve^2}+\frac{2}{\ve}\bigg[1+\log\bigg(\frac{e^{\gamma_E}\Lambda}{a_*H}\bigg)\bigg]+2\log^2\bigg(\frac{2k}{\Lambda}\bigg)-4\log\bigg(\frac{2k}{\Lambda}\bigg)+4+\frac{5\pi^2}{12}\Bigg\}\,,\\[.3cm]
&\bigg(\frac{e^{\gamma_E}\Lambda^2}{4\pi}\bigg)^{\!\ve}\int\frac{\der^{d-1}l}{(2\pi)^{d-1}}\frac{1}{l^4_{\Lambda}(l_{\Lambda}+k_{\Lambda})}=\frac{1}{2\pi^2}\,\Bigg\{\frac{1}{k^2}\bigg[1-\log\bigg(\frac{2k}{\Lambda}\bigg)\bigg]+\frac{\pi}{4k\Lambda}\Bigg\}\,.
\end{flalign}
To carry out the analytic continuation of the Mellin-Barnes integrands that allows one to compute the expansion in $\ve$ in terms of residues, we used the \texttt{Mathematica} package \texttt{MB} \cite{Czakon:2005rk}. The residues which do not vanish as $\Lambda\rightarrow0$ were identified by hand and computed using the built-in \texttt{Mathematica} function.


\section{Late-time limit of the two-loop one-point function of the operator \texorpdfstring{$\phi^2$}{φ²}}
\label{app:Phi20}

In this Appendix, we compute  in the full, minimally coupled massless $\phi^4$ theory the one-point function of the operator $\phi^2(\eta,\vec x)$ at $\Lo(\kappa)$, defined as the momentum integral over the power spectrum of $\phi$,
\begin{equation}
\cor{}{\phi^2(\eta,\vec x)}=\int\frac{\der^{d-1}l}{(2\pi)^{d-1}}\cor{}{\phi(\eta,\vec l)\phi(\eta,-\vec l)}'\,.
\label{eq::20def}
\end{equation}
In the absence of the IR regulator $\Lambda$, and if UV and IR singularities were both regularised dimensionally or by another analytic regulator, the regularised one-point function $\cor{}{\phi^2(\eta,\vec x)}$ would vanish to all orders in $\kappa$, since the integral appearing in \eqref{eq::20def} would then be scaleless. However, the one-point function of the renormalised operator $[\phi^2]$ would be given by the IR poles and all terms that are generated by them, and would thus be non-vanishing. In the scheme we are using, $\Lambda$ provides a scale for the integral  \eqref{eq::20def}, and therefore $\cor{}{\phi^2(\eta,\vec x)}$ does not vanish. This allows us to more easily extract its UV poles, to renormalise the composite operator $\phi^2$ according to \eqref{eq::Phi2reneq}, and to define the renormalised one-point function $\cor{}{[\phi^2](t,\vec x)}$. This computation also provides an example of operator mixing in the full theory, since we will see that the appropriate counterterm to subtract the UV poles is $Z^{\phi}_{20}$.

\subsection{\texorpdfstring{$\cor{}{\phi^2(\eta,\vec x)}$}{<φ²(η,x)>} in the free theory}

In the free theory, \eqref{eq::20def} corresponds to the one-loop diagram shown in the left panel of \figref{fig:Phi20diags} below. Evaluating it, we obtain\footnote{In the following we always multiply the bare, $d$-dimensional one-point function by a factor $\tmu^{4-d}_f$ to obtain a quantity of mass dimension two for all $d$.}
\begin{flalign}
\tmu_f^{2\ve}\cor{}{\phi^2(\eta,\vec x)}_{|\,\Lo(\kappa^0)}&=\frac{H^2(-H\eta)^{-2\ve}\tmu_f^{2\ve}}{2}\int\frac{\der^{d-1}l}{(2\pi)^{d-1}}\frac{1+l^2_{\Lambda}{\eta}^2}{l^3_{\Lambda}}\nonumber\\
&=\frac{H^2}{8\pi^2}\,\bigg(\frac{-\Lambda H\eta}{\mu_f}\bigg)^{\!-2\ve}e^{\ve\gamma_E}\Gamma(\ve)\bigg[1-\frac{(-\Lambda\eta)^2}{2(1-\ve)}\bigg]\,.
\label{eq:freephi2vev}
\end{flalign}
There is a logarithmic UV and IR divergence, and a quadratic UV divergence, which is proportional to $(-\Lambda\eta)^2$. The latter can be ignored, since we take $\eta\rightarrow0$, and we are left with the pole term stemming from the logarithmic divergence. Taking the expectation value of the operator renormalisation equation \eqref{eq::Phi2reneq} and neglecting the terms involving higher powers of $\phi$, we find
\begin{equation}
\tmu_f^{2\ve}\cor{}{\phi^2(\eta,\vec x)}=Z^{\phi}_{22}\,\tmu_f^{2\ve}\cor{}{[\phi^2](\eta,\vec x)}+H^2Z^{\phi}_{20}\,,
\end{equation}
which needs to be evaluated at $\Lo(\kappa^0)$. In the free theory, we have $Z^{\phi,(\kappa^0)}_{22}=1$, which allows us to define the renormalised one-point function as
\begin{equation}
\cor{}{[\phi^2](\eta,\vec x)}_{|\,\Lo(\kappa^0)}=\tmu_f^{2\ve}\cor{}{\phi^2(\eta,\vec x)}_{|\,\Lo(\kappa^0)}-H^2Z^{\phi,(\kappa^0)}_{20}\,,
\end{equation}
where we dropped the factor $\tmu_f^{2\ve}$ on the left-hand side, since, by assumption, the correlation function is UV finite and the limit $\ve\rightarrow0$ can be taken. 
From \eqref{eq:freephi2vev} the counterterm that removes the UV divergence of $\cor{}{\phi^2(\eta,\vec x)}$ is
\begin{equation}
Z^{\phi}_{20}=\frac{1}{8\pi^2\ve}+\Lo(\kappa)\,,
\end{equation}
which leads to the renormalised one-point function
\begin{flalign}
\cor{}{[\phi^2](\eta,\vec x)}_{|\,\Lo(\kappa^0)}&=\frac{H^2}{8\pi^2}\bigg[\bigg(\frac{-\Lambda H\eta}{\mu_f}\bigg)^{\!-2\ve}e^{\ve\gamma_E}\Gamma(\ve)-\frac{1}{\ve}\bigg]\label{eq::renPhi20free}\\
&=-\frac{H^2}{4\pi^2}\log\bigg(\frac{-\Lambda H\eta}{\mu_f}\bigg)+\Lo(\ve)\,.
\label{eq::Phi20free}
\end{flalign}
For the present computation, we can drop the higher-order terms in $\ve$, but they will be needed below, when we turn to $\Lo(\kappa)$.

\subsection{One-loop renormalisation of \texorpdfstring{$\Lambda$}{Λ}}

\begin{figure}[t]
\centering
\begin{subfigure}{0.33\textwidth}
\centering
\includegraphics[width=0.72\textwidth]{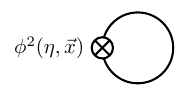}
\caption{}
\end{subfigure}%
\begin{subfigure}{0.33\textwidth}
\centering
\includegraphics[width=0.88\textwidth]{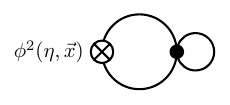}
\caption{}
\end{subfigure}%
\begin{subfigure}{0.33\textwidth}
\centering
\includegraphics[width=0.76\textwidth]{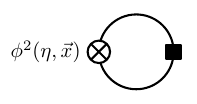}
\caption{}
\end{subfigure}
\caption{The diagrams contributing to $\cor{}{\phi^2(\eta,\vec x)}$ in the free theory (a), and at $\Lo(\kappa)$, (b) and (c).}
\label{fig:Phi20diags}
\end{figure}

Before proceeding with the computation of $\cor{}{\phi^2(\eta,\vec x)}$ at $\Lo(\kappa)$, we must address the renormalisation of the IR regulator $\Lambda$, which becomes necessary for the first time when computing the power spectrum of $\phi$ at $\Lo(\kappa)$. This quantity was already considered in \cite{Beneke:2023wmt}, but the renormalisation of $\Lambda$ was not discussed in this reference, since the final result for the power spectrum at $\Lo(\kappa)$ does not depend on it. However, the renormalisation of $\Lambda$ must be taken into account when computing UV-divergent correlation functions which contain the one-loop power spectrum as a subdiagram, as happens at the two-loop order (see Fig.~\ref{fig:Phi20diags}, diagram (b)).

The insertion of the quartic vertex into the power spectrum leads to the expression
\begin{flalign}
&\cor{}{\phi(\eta,\vec k)\phi(\eta,-\vec k)}'_{|\,\kappa}\nonumber\\
&=\frac{\kappa H^{d-2}}{8}\,(-H\eta)^{d-4}\,\Im\,\Bigg[\int_{-\infty}^{\eta}\frac{\der\eta'}{(-\eta')^{8-d}}\frac{e^{2i k_{\Lambda}(\eta'-\eta)}(i-k_{\Lambda}\eta)^2(i+k_{\Lambda}\eta')^2}{k_{\Lambda}^6}\Bigg]\nonumber\\
&\quad\times\tmu_f^{4-d}\int\frac{\der^{d-1}l}{(2\pi)^{d-1}}\frac{1+l^2_{\Lambda}{\eta'}^2}{l_{\Lambda}^3}\,.
\label{eq::onelooppwrvert}
\end{flalign}
The momentum integral is
\begin{equation}
I(\eta')\equiv\tmu_f^{2\ve}\int\frac{\der^{d-1}l}{(2\pi)^{d-1}}\frac{1+(l^2+\Lambda^2){\eta'}^2}{[l^2+\Lambda^2]^{\frac{3}{2}}}=\frac{1}{4\pi^2}\bigg(\frac{\Lambda}{\mu_f}\bigg)^{\!-2\ve}e^{\ve\gamma_E}\Gamma(\ve)\bigg[1-\frac{(\Lambda\eta')^2}{2(1-\ve)}\bigg]\,.
\label{eq::tadint}
\end{equation}
There are two types of UV divergent terms, originating from the logarithmic and quadratic divergences of the integral, respectively. Since $\Lambda$ is the only scale appearing in the integral in question, the quadratic divergence is necessarily proportional to $\Lambda^2$, which then combines with ${\eta'}^2$ to give a dimensionless quantity. The explicit time-dependence of this divergence signals the breaking of the de Sitter-invariance of the theory by the presence of the comoving regulator $\Lambda$. 
Even though for the quantity of interest the quadratic divergence vanishes in the late-time limit after the time integration has been carried out, as well as in the limit $\Lambda\rightarrow0$, it is still a UV divergence that must be subtracted by an appropriate counterterm. The action of the $\phi^4$ theory, expressed in terms of renormalised quantities, contains the following counterterms quadratic in $\phi$ \cite{Beneke:2026rtf},
\begin{equation}
S\supset-\frac{1}{2}\int\frac{\der^dx}{(-H\eta)^d}\bigg[\delta m^2\phi(\eta,\vec x)^2+(-H\eta)^2\Lambda^2\delta\Lambda^2\phi(\eta,\vec x)^2\bigg]\,.
\end{equation}
The mass counterterm $\delta m^2$ serves to subtract the logarithmic divergence. It was already determined in \cite{Beneke:2023wmt}, and reads 
\begin{equation}
\delta m^2=\frac{\kappa H^2}{8\pi^2}\bigg[-\frac{1}{2\ve}+\delta\hat m^2_{\textrm{fin}}\bigg]+\Lo(\kappa^2)\,.
\label{eq::dm}
\end{equation}
The insertion of $\delta \Lambda^2$ into the power spectrum leads to the expression
\begin{flalign}
&\cor{}{\phi(\eta,\vec k)\phi(\eta,-\vec k)}'_{|\,\delta\Lambda^2}\nonumber\\
&=\frac{H^2\delta\Lambda^2}{2}\,(-H\eta)^{d-4}\,\Im\,\Bigg[\int_{-\infty}^{\eta}\frac{\der \eta'}{(-\eta')^4}\frac{e^{2ik_{\Lambda}(\eta'-\eta)}(i-k_{\Lambda}\eta)^2(i+k_{\Lambda}\eta')^2}{k_{\Lambda}^6}(-\Lambda\eta')^2\Bigg]\,.
\end{flalign}
Comparing this expression to \eqref{eq::onelooppwrvert}, we notice that it has the right structure to subtract the UV divergence proportional to $(-\Lambda\eta')^2$ generated by the momentum integral \eqref{eq::tadint}. The divergent part of the counterterm is unambiguous, and we further choose its finite part such that it subtracts the finite $\Lo(\Lambda^2)$ terms which stem from the $\ve$-expansion of the integral \eqref{eq::tadint} as well. This corresponds to the scheme defined in \cite{Beneke:2026rtf}, and leads to
\begin{equation}
\delta\Lambda^2=\frac{\kappa}{32\pi^2}\bigg[\frac{1}{\ve}+1\bigg]+\Lo(\kappa^2)\,.
\label{eq::dL}
\end{equation}

\subsection{Relevant diagrams and regions at \texorpdfstring{$\Lo(\kappa)$}{O(ϰ)}}

We now turn on the interactions and compute $\cor{}{\phi^2(\eta,\vec x)}$ at $\Lo(\kappa)$. The correlation function receives contributions from the insertion of the $\phi^4$ vertex, which corresponds to the diagram (b) shown in \figref{fig:Phi20diags}, as well as from the two counterterms $\delta m^2$ and $\delta\Lambda^2$, depicted in diagram (c).
Both together give 
\begin{flalign}
&\tmu^{4-d}_f\cor{}{\phi^2(\eta,\vec x)}_{|\,\Lo(\kappa)}\nonumber\\
&=(-H\eta)^{d-4}\tmu^{4-d}_f\int\frac{\der^{d-1}k}{(2\pi)^{d-1}}\,\Bigg\{\frac{\kappa H^{d-2}}{8}\,\Im\,\Bigg[\int_{-\infty}^{\eta}\frac{\der\eta'}{(-\eta')^{8-d}}\frac{e^{2ik_{\Lambda}(\eta'-\eta)}(i-k_{\Lambda}\eta)^2(i+k_{\Lambda}\eta')^2}{k_{\Lambda}^6}\Bigg]\nonumber\\
&\quad\times\tmu^{4-d}\int\frac{\der^{d-1}l}{(2\pi)^{d-1}}\frac{1+l_{\Lambda}^2{\eta'}^2}{l_{\Lambda}^3}\nonumber\\
&\quad+\frac{1}{2}\,\Im\,\Bigg[\int_{-\infty}^{\eta}\frac{\der\eta'}{(-\eta')^4}\frac{e^{2ik_{\Lambda}(\eta'-\eta)}(i-k_{\Lambda}\eta)^2(i+k_{\Lambda}\eta')^2}{k_{\Lambda}^6}\Big[\delta m^2+(-H\eta')^2\Lambda^2\delta\Lambda^2\Big]\Bigg]\Bigg\}\,.
\end{flalign}
We have already determined the relevant counterterms at $\Lo(\kappa)$, see \eqref{eq::dm} and \eqref{eq::dL}. 

We apply the method of regions to simplify the calculation. 
The momentum integral over $\vec{l}$ has only one region, namely 
$\vec{l}\sim \Lambda$, and there is no further simplification. It is given directly by \eqref{eq::tadint}. However, unlike in the case of the power spectrum, the $\Lo(\Lambda^2)$-piece of $I(\eta')$ cannot be dropped at this point in the computation, since it can combine with an $\Lo(\Lambda^{-2})$ term from the remaining integrations to yield a finite remainder when $\Lambda\rightarrow0$. 
Indeed, we demonstrate that this happens in one of the regions below. 
For the same reason, the counterterm $\delta\Lambda^2$ is relevant, since it is required to consistently subtract the $\Lo(\Lambda^2/\ve)$ subdivergence of this integral.

Thus, we need to determine the regions of the integrals 
\begin{flalign}
\tmu^{4-d}_f\cor{}{\phi^2(\eta,\vec x)}_{|\,\kappa}&=\frac{\tmu_f^{4-d}\kappa H^{d-2}}{8}\,(-H\eta)^{d-4}\,\Im\,\Bigg[\int\frac{\der^{d-1}k}{(2\pi)^{d-1}}\int_{-\infty}^{\eta}\frac{\der\eta'}{(-\eta')^{8-d}}\frac{e^{2ik_{\Lambda}(\eta'-\eta)}}{k_{\Lambda}^6}(i-k_{\Lambda}\eta)^2\nonumber\\[-0.2cm]
&\quad\times(i+k_{\Lambda}\eta')^2\,I(\eta')\Bigg]\,,
\label{eq:2loopI}
\end{flalign}
and 
\begin{flalign}
\tmu^{4-d}_f\cor{}{\phi^2(\eta,\vec x)}_{|\,\textrm{c.t.}}&=\frac{(-H\eta)^{d-4}}{2}\tmu^{4-d}_f\,\Im\,\Bigg[\int\frac{\der^{d-1}k}{(2\pi)^{d-1}}\int_{-\infty}^{\eta}\frac{\der\eta'}{(-\eta')^4}\Big[\delta m^2+(-H\eta')^2\Lambda^2\delta\Lambda^2\Big]\nonumber\\
&\quad\times\frac{e^{2ik_{\Lambda}(\eta'-\eta)}(i-k_{\Lambda}\eta)^2(i+k_{\Lambda}\eta')^2}{k_{\Lambda}^6}\Bigg]\,.
\label{eq:2loopCT}
\end{flalign}
In the region decomposition of the $\vec k$- and $\eta'$-integrals, there is only one external scale, the correlation time $\eta$. We have the two regions
\begin{equation}
\eta'\ll \eta\quad\textrm{(early)}\,,\qquad \eta'\sim \eta\quad\textrm{(late)}
\end{equation}
for $\eta'$, and
\begin{equation}
k\sim -\frac{1}{\eta}\quad\textrm{(hard)}\,,\qquad k\ll-\frac{1}{\eta}\quad\textrm{(soft)}
\end{equation}
for $\vec k$. The regulator $\Lambda$ is treated as a quantity that scales like a soft comoving momentum. As in the calculation of the bispectrum of $\phi^2$ in \appref{app:22q0}, we find integrals that are not fully regularised by the regulators already in place. 
The origin of these unregulated divergences is twofold. 
First, the time integral for the counterterm-insertion diagrams is accidentally $d$-independent in the adopted regularisation scheme  \cite{Beneke:2023wmt}. 
Second, both in the late-time, hard-momentum region and the early-time, soft-momentum region, the combination $-k\eta'$ scales like $-k\eta'\sim1$ for different reasons, which, as was also discussed in the aforementioned reference, leads to unregulated divergences~\cite{Beneke:2023wmt}.
Therefore, as in \appref{app:22q0}, we introduce the analytic regulator
\begin{equation}
(-\nu\eta')^{-2\delta}
\end{equation}
in both integrands above, and after integration, we take the limits $\delta\rightarrow0$, $\ve\rightarrow0$, in this order, as well as the limit $\Lambda\rightarrow0$. 
We expect the appearance of poles in $\delta$ in the respective regions, which then cancel in their sum.

\subsection{Late-time regions}
\subsubsection{Soft momentum}

In this region, we have
\begin{equation}
-k\eta\sim-k\eta'\ll1
\end{equation}
and we can therefore expand the numerator of the integrals 
\eqref{eq:2loopI}, \eqref{eq:2loopCT} into the series 
\begin{equation}
e^{2ik_{\Lambda}(\eta'-\eta)}(i-k_{\Lambda}\eta)^2(i+k_{\Lambda}\eta')^2=1+k_{\Lambda}^2(\eta^2+{\eta'}^2)+\frac{2ik_{\Lambda}^3}{3}({\eta'}^3-\eta^3)+\Lo(k_{\Lambda}^4)\,.
\end{equation}
We truncate the expansion $\Lo(k_{\Lambda}^3)$ since the neglected terms result in positive powers of $\Lambda$ after integration, which can be dropped as $\Lambda\rightarrow0$. We then find
\begin{flalign}
&\tmu_f^{2\ve}\cor{}{\phi^2(\eta,\vec x)}_{|\,\kappa,\textrm{ late-$\eta$, soft-$k$}}\nonumber\\[0.2cm]
&=\frac{\tmu_f^{2\ve}\kappa H^{2-2\ve}}{8}\,(-H\eta)^{-2\ve}\,\Im\,\Bigg[\int\frac{\der^{d-1}k}{(2\pi)^{d-1}}\int_{-\infty}^{\eta}\frac{\der\eta'}{(-\eta')^{8-d}}\frac{(-\nu\eta')^{-2\delta}}{k_{\Lambda}^6}\bigg(1+k_{\Lambda}^2(\eta^2+{\eta'}^2)\nonumber\\
&\quad+\frac{2ik_{\Lambda}^3}{3}({\eta'}^3-\eta^3)\bigg)I(\eta')\Bigg]\nonumber\\
&=-\frac{\kappa H^2}{384\pi^4}\bigg(\frac{\mu_f}{H}\bigg)^{\!4\ve}e^{2\ve\gamma_E}(-\Lambda\eta)^{-2\ve}(-e^{\gamma_E}\Lambda\eta)^{-2\ve}\,\bigg[\,\frac{1}{\ve^3}-\frac{2}{3\ve^2}+\frac{1}{\ve}\bigg(\frac{4}{9}+\frac{\pi^2}{6}\bigg)\nonumber\\
&\quad-\frac{8}{27}-\frac{\pi^2}{9}-\frac{2\zeta(3)}{3}\,\bigg]
\end{flalign}
for the vertex-insertion diagram, where $\zeta(z)$ denotes the Riemann zeta function, and
\begin{flalign}
&\tmu_f^{2\ve}\cor{}{\phi^2(\eta,\vec x)}_{|\,\textrm{c.t.},\textrm{ late-$\eta$, soft-$k$}}\nonumber\\[0.2cm]
&=\frac{\tmu_f^{2\ve}}{2}\,(-H\eta)^{-2\ve}\,\Im\,\Bigg[\int\frac{\der^{d-1}k}{(2\pi)^{d-1}}\int_{-\infty}^{\eta}\frac{\der\eta'}{(-\eta')^4}\frac{(-\nu\eta')^{-2\delta}}{k_{\Lambda}^6}\Big[\delta m^2+(-H\eta')^2\Lambda^2\delta\Lambda^2\Big]\nonumber\\
&\quad\times\bigg(1+k_{\Lambda}^2(\eta^2+{\eta'}^2)+\frac{2ik_{\Lambda}^3}{3}({\eta'}^3-\eta^3)\bigg)\Bigg]\nonumber\\
&=-\frac{\kappa H^2}{192\pi^4}\,\bigg(\frac{\mu_f}{H}\bigg)^{2\ve}(-\Lambda\eta)^{-2\ve}(-\nu\eta)^{-2\delta}\,\Bigg\{-\frac{1}{\ve^2}\bigg[\frac{1}{2\delta}-\frac{1}{3}\bigg]+\frac{\delta\hat m^2_{\textrm{fin}}}{\ve}\bigg[\frac{1}{\delta}-\frac{2}{3}\bigg]-\frac{\pi^2}{24\delta}+\frac{\pi^2}{36}\,\Bigg\}
\end{flalign}
for the counterterm diagrams. The presence of poles in $\delta$ in the latter result signals the necessity of the analytic regulator $(-\nu\eta')^{-2\delta}$, due to the $d$-independence of the time integral. For the former result it is not needed, and one can set $\delta\rightarrow0$ already before evaluating the integrals. For both of these terms, the $\Lo(\Lambda^2)$-pieces of the tadpole integral and counterterms, respectively, are irrelevant, since they remain $\Lo(\Lambda^2)$ after the integration.

\subsubsection{Hard momentum}
In this region, we have
\begin{equation}
-k\eta\sim-k\eta'\sim 1
\end{equation}
so the numerator of the integrand stays unchanged. However, since $k\gg\Lambda$, we must set $\Lambda\rightarrow0$ in the integrand of the $\vec k$-integral, and every divergence is regularised dimensionally or by the analytic regulator. We find
\begin{flalign}
&\tmu_f^{2\ve}\cor{}{\phi^2(\eta,\vec x)}_{|\,\kappa,\textrm{ late-$\eta$, hard-$k$}}\nonumber\\[0.2cm]
&=\frac{\tmu_f^{2\ve}\kappa H^{2-2\ve}}{8}\,(-H\eta)^{-2\ve}\,\Im\,\Bigg[\int\frac{\der^{d-1}k}{(2\pi)^{d-1}}\int_{-\infty}^{\eta}\frac{\der\eta'}{(-\eta')^{8-d}}\,(-\nu\eta')^{-2\delta}\,\frac{(i-k\eta)^2(i+k\eta')^2}{k^6}\nonumber\\
&\quad\times e^{2ik(\eta'-\eta)}I(\eta')\Bigg]\nonumber\\
&=\frac{\kappa H^2}{384\pi^4}\,\bigg(\frac{\mu_f}{H}\bigg)^{\!4\ve}e^{2\ve\gamma_E}(-\Lambda\eta)^{-2\ve}(-\nu\eta)^{-2\delta}\,\Bigg\{\,\frac{1}{\ve^2}\bigg[\frac{1}{\delta}-\frac{13}{6}\bigg]+\frac{1}{\ve}\bigg[-\frac{8}{3\delta}+\frac{56}{9}-\frac{2\pi^2}{3}\bigg]\nonumber\\
&\quad+\frac{1}{\delta}\bigg[\frac{52}{9}-\frac{\pi^2}{3}\bigg]-30+\frac{5\pi^2}{2}+8\zeta(3)\Bigg\}
\end{flalign}
for the vertex-insertion diagram, and for the counterterm diagrams
\begin{flalign}
&\tmu_f^{2\ve}\cor{}{\phi^2(\eta,\vec x)}_{|\,\textrm{c.t.},\textrm{ late-$\eta$, hard-$k$}}\nonumber\\[0.2cm]
&=\frac{\tmu_f^{2\ve}}{2}\,(-H\eta)^{-2\ve}\,\Im\,\Bigg[\int\frac{\der^{d-1}k}{(2\pi)^{d-1}}\int_{-\infty}^{\eta}\frac{\der\eta'}{(-\eta')^4}\,(-\nu\eta')^{-2\delta}\,\frac{e^{2ik(\eta'-\eta)}(i-k\eta)^2(i+k\eta')^2}{k^6}\nonumber\\
&\phantom{=}\times \Big[\delta m^2+(-H\eta')^2\Lambda^2\delta\Lambda^2\Big]\bigg]\nonumber\\
&=\frac{\kappa H^2}{192\pi^4}\,\bigg(\frac{\mu_f}{H}\bigg)^{\!2\ve}(-\Lambda\eta)^{-2\ve}(-e^{\gamma_E}\Lambda\eta)^{2\ve}\,\Bigg\{\,\frac{1}{2\ve^3}-\frac{1}{\ve^2}\bigg[\frac{1}{4}+\delta\hat m^2_{\textrm{fin}}\bigg]+\frac{1}{2\ve}\bigg[\frac{\pi^2}{4}+\delta\hat m^2_{\textrm{fin}}\bigg]\nonumber\\
&\quad-\frac{\pi^2}{4}\bigg[\frac{1}{4}+\delta\hat m^2_{\textrm{fin}}\bigg]+\frac{\zeta(3)}{6}\,\Bigg\}\,.
\end{flalign}
Notice the absence of the subscript $\Lambda$ in both integrands. Also for these terms, the $\Lo(\Lambda^2)$-pieces of the tadpole and counterterm, respectively, remain $\Lo(\Lambda^2)$ after the integration and thus their contribution vanishes. The appearance of the $\delta$-poles in the vertex-insertion contribution is expected for the reasons discussed above. However, curiously, we find no $\delta$-poles in the counterterm-insertion contribution, even though, naively, one would have expected them.

\subsection{Early-time regions}

\subsubsection{Soft momentum}
In these regions, we have
\begin{equation}
-k\eta\ll1\,,\quad -k\eta'\sim1\,,
\end{equation}
so we can approximate
\begin{equation}
e^{2ik_{\Lambda}(\eta'-\eta)}(i-k_{\Lambda}\eta)^2(i+k_{\Lambda}\eta')^2=-e^{2ik_{\Lambda}\eta'}(i+k_{\Lambda}\eta')^2+\Lo\bigg(\frac{\eta}{\eta'}\bigg)
\end{equation}
and integrate over $\eta'\in(-\infty,0)$. We find
\begin{flalign}
&\tmu_f^{2\ve}\cor{}{\phi^2(\eta,\vec x)}_{|\,\kappa,\textrm{ early-$\eta$, soft-$k$}}\nonumber\\[0.2cm]
&=\frac{\tmu_f^{2\ve}\kappa H^{2-2\ve}}{8}\,(-H\eta)^{-2\ve}\,\Im\,\Bigg[\int\frac{\der^{d-1}k}{(2\pi)^{d-1}}\int_{-\infty}^0\frac{\der\eta'}{(-\eta')^{8-d}}\,(-\nu\eta')^{-2\delta}\,\frac{e^{2ik_{\Lambda}\eta'}(i+k_{\Lambda}\eta')^2}{k_{\Lambda}^6}\,I(\eta')\Bigg]\nonumber\\
&=\frac{\kappa H^2}{384\pi^4}\,\bigg(\frac{\mu_f}{H}\bigg)^{\!4\ve}\!e^{2\ve\gamma_E}(-\Lambda\eta)^{-2\ve}\bigg(\frac{\nu}{e^{\gamma_E}\Lambda}\bigg)^{\!-2\delta}\,\Bigg\{\frac{1}{\ve^3}-\frac{1}{\delta\ve^2}+\frac{1}{\ve}\bigg[\frac{8}{3\delta}-\frac{72}{18}+\frac{\pi^2}{2}\bigg]-\frac{1}{\delta}\bigg[\frac{52}{9}-\frac{\pi^2}{3}\bigg]\nonumber\\
&\quad+\frac{1361}{54}-\frac{20\pi^2}{9}-\zeta(3)\Bigg\}
\end{flalign}
for the vertex-insertion diagram, and for the counterterm diagrams
\begin{flalign}
&\tmu_f^{2\ve}\cor{}{\phi^2(\eta,\vec x)}|_{\,\textrm{c.t.},\textrm{ early-$\eta$, soft-$k$}}\nonumber\\[0.2cm]
&=\frac{\tmu_f^{2\ve}}{2}\,(-H\eta)^{-2\ve}\,\Im\,\Bigg[\int\frac{\der^{d-1}k}{(2\pi)^{d-1}}\int_{-\infty}^0\frac{\der\eta'}{(-\eta')^4}\,(-\nu\eta')^{-2\delta}\,\frac{e^{2ik_{\Lambda}\eta'}(i+k_{\Lambda}\eta')^2}{k^6_{\Lambda}}\nonumber\\
&\quad\times\Big[\delta m^2+(-H\eta')^2\Lambda^2\delta\Lambda^2\Big]\Bigg]\nonumber\\
&=\frac{\kappa H^2}{192\pi^4}\bigg(\frac{\mu_f}{H}\bigg)^{\!2\ve}(-\Lambda\eta)^{-2\ve}\bigg(\frac{\nu}{e^{\gamma_E}\Lambda}\bigg)^{\!-2\delta}\,\Bigg\{-\frac{1}{2\ve^3}+\frac{1}{\ve^2}\bigg[-\frac{1}{2\delta}+\frac{4}{3}+\delta\hat m^2_{\textrm{fin}}\bigg]\nonumber\\
&\quad-\frac{1}{\ve}\bigg[\frac{\delta\hat m^2_{\textrm{fin}}}{\delta}-\frac{3}{4}+\frac{\pi^2}{24}-\frac{8}{3}\delta\hat m^2_{\textrm{fin}}\bigg]-\frac{\pi^2}{24\pi^2\delta}-\frac{3}{4}+\frac{\pi^2}{9}-\frac{\zeta(3)}{3}-\frac{\pi^2}{12}\delta\hat m^2_{\textrm{fin}}\Bigg\}\,.
\end{flalign}
For both of these expressions, the $\Lo(\Lambda^2)$-pieces of $I(\eta)$ and $\delta\Lambda^2$ matter, since the time integral generates IR-divergent $\Lo(\Lambda^{-2})$ terms, which promote them to $\Lo(\Lambda^0)$, and hence they survive in the limit $\Lambda\rightarrow0$ taken at the end of the computation. The presence of the $\Lo(\Lambda^2)$-term in the mass counterterm ensures that these terms are subtracted consistently and do not get erroneously identified as pure UV-poles associated with the renormalisation of the composite operator, since they stem from two overlapping UV and IR divergences. In practice, the presence or absence of these terms impacts the coefficient of the single-logarithmic term in $\Lambda$ in the final, renormalised result, so to ensure that we find the correct IR-divergent structure of the renormalised correlator, it is crucial to take this point into account. 

\subsubsection{Hard momentum}

In these regions, we have
\begin{equation} 
-k\eta\sim1\,,\quad -k\eta'\gg 1
\end{equation}
and we can replace the integrand with
\begin{equation}
\frac{e^{2ik_{\Lambda}(\eta'-\eta)}(i-k_{\Lambda}\eta)^2(i+k_{\Lambda}\eta')^2}{k_{\Lambda}^6}\rightarrow \frac{e^{2ik\eta'}(k\eta')^2}{k^6}\,.
\end{equation}
Therefore, after momentum (or time, whichever is taken first) integration, the remaining time (momentum) integral is scaleless and vanishes. 

\subsection{Sum of the regions, renormalisation of the operator \texorpdfstring{$\phi^2$}{φ²}}
Summing up the three non-vanishing regions for each contribution, we find
\begin{flalign}
&\tmu_f^{2\ve}\cor{}{\phi^2(\eta,\vec x)}_{|\,\kappa}\nonumber\\
&=\frac{\kappa H^2}{384\pi^4}\,\bigg(\frac{e^{\gamma_E}\mu_f}{H}\bigg)^{\!4\ve}(-e^{\gamma_E}\Lambda\eta)^{-2\ve}\Bigg\{-\frac{3}{2\ve^2}+\frac{1}{\ve}\bigg[-2\log^2(-e^{\gamma_E}\Lambda\eta)+4\log(-e^{\gamma_E}\Lambda\eta)\nonumber\\
&\quad+\frac{3}{2}-\frac{\pi^2}{3}\bigg]+\frac{4}{3}\log^3(-e^{\gamma_E}\Lambda\eta)+\frac{4}{3}\log^2(-e^{\gamma_E}\Lambda\eta)-\bigg(\frac{32}{3}+\pi^2\bigg)\log(-e^{\gamma_E}\Lambda\eta)\nonumber\\
&\quad-\frac{9}{2}+\frac{7\pi^2}{18}+\frac{23\zeta(3)}{3}\,\Bigg\}
\end{flalign}
for the vertex-insertion contribution, and from the counterterms
\begin{flalign}
&\tmu_f^{2\ve}\cor{}{\phi^2(\eta,\vec x)}_{|\,\textrm{c.t.}}\nonumber\\
&=\frac{\kappa H^2}{384\pi^4}\,\bigg(\frac{e^{\gamma_E}\mu_f}{H}\bigg)^{\!2\ve}(-e^{\gamma_E}\Lambda\eta)^{-2\ve}\,\Bigg\{\,\frac{3}{2\ve^2}+\frac{1}{\ve}\bigg[2\log^2(-e^{\gamma_E}\Lambda\eta)-\log(-e^{\gamma_E}\Lambda\eta)-\frac{3}{2}+\frac{\pi^2}{3}\nonumber\\
&\quad-3\delta\hat m^2_{\textrm{fin}}\bigg]+\frac{4}{3}\log^3(-e^{\gamma_E}\Lambda\eta)-(1+4\delta\hat m^2_{\textrm{fin}})\log^2(-e^{\gamma_E}\Lambda\eta)+\bigg(\frac{\pi^2}{3}+2\delta\hat m^2_{\textrm{fin}}\bigg)\log(-e^{\gamma_E}\Lambda\eta)\nonumber\\
&\quad-\frac{3}{2}+\frac{\pi^2}{24}-\frac{2\pi^2}{3}\delta\hat m^2_{\textrm{fin}}-\frac{\zeta(3)}{3}\,\Bigg\}\,.
\end{flalign}
Notice that all $\delta$-poles cancel in both results. The double and single poles in $\ve$ are of UV origin and stem from the subdivergences of the tadpole integral $I(\eta')$ and from the counterterms, respectively, as well as from the additional UV divergence generated by the momentum integral in \eqref{eq::20def}. Summing the two contributions and collecting logarithms of $\Lambda$, we find
\begin{flalign}
&\tmu_f^{2\ve}\Big[\cor{}{\phi^2(\eta,\vec x)}_{|\,\kappa}+\cor{}{\phi^2(\eta,\vec x)}_{|\,\textrm{c.t.}}\Big]\nonumber\\
&=\frac{\kappa H^2}{384\pi^4}\,\bigg(\frac{e^{\gamma_E}\mu_f}{H}\bigg)^{\!2\ve}(-e^{\gamma_E}\Lambda\eta)^{-2\ve}\,\Bigg\{\,\frac{3}{\ve}\,\bigg[-\delta\hat m^2_{\textrm{fin}}+\log\bigg(-\frac{\Lambda H\eta}{\mu_f}\bigg)\bigg]+\frac{8}{3}\log^3(-e^{\gamma_E}\Lambda\eta)\nonumber\\
&\quad+\bigg[\frac{1}{3}-4\delta\hat m^2_{\textrm{fin}}-4\log\bigg(\frac{e^{\gamma_E}\mu_f}{H}\bigg)\bigg]\log^2(-e^{\gamma_E}\Lambda\eta)+2\,\bigg[-\frac{16}{3}+\delta\hat m^2_{\textrm{fin}}+\frac{2\pi^2}{3}\nonumber\\
&\quad+4\log\bigg(\frac{e^{\gamma_E}\mu_f}{H}\bigg)\bigg]\log(-e^{\gamma_E}\Lambda\eta)-3\log^2\bigg(\frac{e^{\gamma_E}\mu_f}{H}\bigg)+\bigg[3-\frac{2\pi^2}{3}\bigg]\log\bigg(\frac{e^{\gamma_E}\mu_f}{H}\bigg)\nonumber\\
&\quad-6+\frac{31\pi^2}{72}-\frac{2\pi^2}{3}\delta\hat m^2_{\textrm{fin}}+\frac{22}{3}\zeta(3)\Bigg\}\,.
\label{eq:phi2bare}
\end{flalign}
Only a single pole is left in the sum due to the consistent cancellation of the subdivergences. The remaining overall UV divergence is generated by the operator $\phi^2$. The pole contains a term proportional to $\log(-\Lambda\eta H/\mu)$, which is expected due to operator mixing, and can be understood from the renormalisation equation
\begin{equation}
\tmu_f^{2\ve}\cor{}{\phi^2(\eta,\vec x)}=\tmu_f^{2\ve}Z^{\phi}_{22}\,\cor{}{[\phi^2](\eta,\vec x)}+H^2Z^{\phi}_{20}
\label{eq::Phi20kappa}
\end{equation}
for this one-point function. The $\Lo(\kappa)$ terms are
\begin{equation}
\tmu_f^{2\ve}\cor{}{\phi^2(\eta,\vec x)}_{|\,\Lo(\kappa)}=\tmu_f^{2\ve}\Big[Z^{\phi,(\kappa^0)}_{22}\cor{}{[\phi^2](\eta,\vec x)}_{|\,\Lo(\kappa)}+Z^{\phi,(\kappa)}_{22}\cor{}{[\phi^2](\eta,\vec x)}_{|\,\Lo(\kappa^0)}\Big]+H^2Z^{\phi,(\kappa)}_{20}.
\end{equation}
Terms involving renormalised operators containing higher powers of $\phi$ can be dropped, since they would all be at least $\Lo(\kappa^2)$. In \appref{app:22q0} we already determined the renormalisation factor
\begin{equation}
Z^{\phi}_{22}=1-\frac{\kappa}{32\pi^2\ve}+\Lo(\kappa^2)\,,
\end{equation}
and we can therefore define the renormalised one-point function at $\Lo(\kappa)$ by
\begin{equation}
\cor{}{[\phi^2](\eta,\vec x)}|_{\,\Lo(\kappa)}=\tmu_f^{2\ve}\Big[\cor{}{\phi^2(\eta,\vec x)}|_{\,\Lo(\kappa)}-Z^{\phi,(\kappa)}_{22}\cor{}{[\phi^2](\eta,\vec x)}|_{\,\Lo(\kappa^0)}\Big]-H^2Z^{\phi,(\kappa)}_{20}\,,
\label{eq::Phi20kappacor}
\end{equation}
where we took the limit $\ve\rightarrow0$ on the left-hand side of the equation. We already determined the free-theory result for $\cor{}{[\phi^2](\eta,\vec x)}$ above in \eqref{eq::renPhi20free}, 
and expanding it up to $\Lo(\ve)$ gives
\begin{flalign}
\cor{}{[\phi^2](\eta,\vec x)}_{|\,\Lo(\kappa^0)}&=\frac{H^2}{4\pi^2}\,\Bigg\{-\log\bigg(-\frac{\Lambda H\eta}{\mu_f}\bigg)+\ve\bigg[\log^2\bigg(-\frac{\Lambda H\eta}{\mu_f}\bigg)+\frac{\pi^2}{24}\bigg]\Bigg\}\,.
\end{flalign}
With the above results, we see that the product
\begin{equation}
Z^{\phi,(\kappa)}_{22}\cor{}{[\phi^2](\eta,\vec x)}_{|\,\Lo(\kappa^0)}=\frac{\kappa H^2}{128\pi^4}\,\Bigg\{\frac{1}{\ve}\log\bigg(-\frac{\Lambda\eta H}{\mu_f}\bigg)-\bigg[\log^2\bigg(-\frac{\Lambda\eta H}{\mu_f}\bigg)+\frac{\pi^2}{24}\bigg]\Bigg\}\,,
\end{equation}
 when plugged into \eqref{eq::Phi20kappacor},
correctly subtracts the pole term proportional to the logarithm of $(-\Lambda\eta)$ in \eqref{eq:phi2bare}. The remaining pole term is local and can therefore be subtracted by choosing
\begin{equation}
Z^{\phi,(\kappa)}_{20}=\frac{\kappa}{128\pi^4\ve}\,\Big[-\delta\hat m^2_{\textrm{fin}}\Big]\,.
\end{equation}
This results in the renormalised $\phi^2$ expectation value
\begin{flalign}
\cor{}{[\phi^2](\eta,\vec x)}_{|\,\Lo(\kappa)}&=\frac{\kappa H^2}{384\pi^4}\,\Bigg\{\,\frac{8}{3}\log^3(-e^{\gamma_E}\Lambda\eta)-4\,\bigg[\log\bigg(\frac{e^{\gamma_E}\mu_f}{H}\bigg)+\frac{2}{3}+\delta\hat m^2_{\textrm{fin}}\bigg]\log^2(-e^{\gamma_E}\Lambda\eta)\nonumber\\
&\quad+2\,\bigg[7\log\bigg(\frac{e^{\gamma_E}\mu_f}{H}\bigg)-\frac{16}{3}+\frac{2\pi^2}{3}+4\delta\hat m^2_{\textrm{fin}}\bigg]\log(-e^{\gamma_E}\Lambda\eta)\nonumber\\
&\quad-6\log^2\bigg(\frac{e^{\gamma_E}\mu_f}{H}\bigg)+\bigg[3-\frac{2\pi^2}{3}-6\delta\hat m^2_{\textrm{fin}}\bigg]\log\bigg(\frac{e^{\gamma_E}\mu_f}{H}\bigg)\nonumber\\
&\quad-6+\frac{5\pi^2}{9}-\frac{2\pi^2}{3}\delta\hat m^2_{\textrm{fin}}+\frac{22}{3}\zeta(3)\,\Bigg\}\,,
\label{eq::phi2renrfull}
\end{flalign}
which is matched to SdSET in \secref{sec::vp20} of the main text.

We can use this result to make contact with the formalism of stochastic inflation~\cite{Starobinsky:1994bd}, as well as with its extension \cite{Cohen:2021fzf}. Like all correlation functions computed in this paper, the expression for the one-point function $\cor{}{[\phi^2](t,\vec x)}$ used in the matching computation is UV-finite, but IR-divergent and secularly growing. It is thus unsuitable to be interpreted as a physically meaningful quantity. The appropriate way to compute such one-point functions is by solving the Kramers-Moyal equation of the stochastic formalism non-perturbatively. Nevertheless, the perturbative result can be linked to the stochastic formalism by collecting the leading-logarithmic (LL) terms in $\Lambda/(a(t)\mu_f)$ of the renormalised result at $\Lo(\kappa^0)$ and $\Lo(\kappa)$ from \eqref{eq::Phi2free} and \eqref{eq::Phi2onept}, respectively, into 
\begin{equation}
\cor{}{[\phi^2](t,\vec x)}_{|\,\textrm{LL}}=-\frac{H^2}{4\pi^2}\log\bigg(\frac{\Lambda}{a(t)\mu_f}\bigg)+\frac{\kappa H^2}{144\pi^4}\log^3\bigg(\frac{\Lambda}{a(t)\mu_f}\bigg)+\Lo(\kappa^2)\,.
\label{eq::Phi20LL}
\end{equation}
This expression agrees with the one obtained by  solving the Fokker-Planck equation of stochastic inflation perturbatively~\cite{Kamenshchik:2020yyn,Cespedes:2023aal}. The exact equilibrium solution of said equation generates a non-perturbative result for the late-time limit of $\cor{}{[\phi^2](t,\vec x)}$ which is IR-finite and time-independent. In the literature, this is interpreted as a procedure to ``resum" the IR-divergent logarithms, see e.g. \cite{Prokopec:2007ak}. 


\section{Loop integrals for the one-point function of the operator \texorpdfstring{$\vp^2_+$}{φ₊²}}
\label{app::Phi20loopsEFT}

In this Appendix, we collect the explicit expressions for the evaluation of the loop integrals resulting from the insertion of the non-Gaussian initial conditions $\Xi_{3,1}$ into the SdSET one-point function $\cor{}{\vp^2_+(t,\vec x)}$ discussed in \secref{sec::vp20}. 

To evaluate the integrals containing the logs, we write
\begin{equation}
\log\bigg(\frac{2e^{\gamma_E}(l_{1\Lambda}+l_{2\Lambda})}{a_*H}\bigg)=\frac{\der}{\der u}\bigg|_{u=0}\bigg(\frac{2e^{\gamma_E}(l_{1\Lambda}+l_{2\Lambda})}{a_*H}\bigg)^{\!u}
\end{equation}
and then we use the Mellin-Barnes representation \cite{Smirnov:2012gma}
\begin{equation}
\frac{1}{(l_{1\Lambda}+l_{2\Lambda})^{-u}}=\frac{1}{\Gamma(-u)}\frac{1}{2\pi i}\int_{-i\infty}^{i\infty}\der z\;\Gamma(-u+z)\Gamma(-z)\frac{l^z_{2\Lambda}}{l^{-u+z}_{1\Lambda}}
\label{eq::MBid}
\end{equation}
to simplify the $\vec l_{1,2}$-integrations.

To compute the integrals containing the rational functions of $l_{1,2}$, we simplify the denominator structure algebraically by 
\begin{flalign}
&-\frac{1}{l^3_{1\Lambda}+l^3_{2\Lambda}}\bigg[\frac{l_{1\Lambda}^2l_{2\Lambda}^2}{2(l_{1\Lambda}+l_{2\Lambda})}+\frac{2}{9}\Big[7l^3_{1\Lambda}+3l^2_{1\Lambda}l_{2\Lambda}+3l_{1\Lambda}l^2_{2\Lambda}+7l_{2\Lambda}^3\Big]\bigg]\nonumber\\
&=-\frac{14}{9}-\frac{l^2_{1\Lambda}}{6(l_{1\Lambda}+l_{2\Lambda})^2}+\frac{l_{1\Lambda}(l^2_{1\Lambda}-6l_{1\Lambda}l_{2\Lambda}-4l^2_{2\Lambda})}{6(l^3_{1\Lambda}+l^3_{2\Lambda})}\,,
\end{flalign}
and then employ the Mellin-Barnes identity \eqref{eq::MBid}. 
We find the four non-trivial integrals
\begin{flalign}
&\bigg(\frac{e^{\gamma_E}\Lambda^2}{4\pi}\bigg)^{2\ve}\int\frac{\der^{d-1}l_1\der^{d-1}l_2}{(2\pi)^{2d-2}}\frac{1}{l^3_{1\Lambda}l^3_{2\Lambda}}\log\bigg(\frac{2e^{\gamma_E}(l_{1\Lambda}+l_{2\Lambda})}{a_*H}\bigg)\nonumber\\
&=\frac{1}{192\pi^4}\,\bigg\{\frac{9}{\ve^3}+\frac{12}{\ve^2}\bigg[1+\log\bigg(\frac{e^{\gamma_E}\Lambda}{a_*H}\bigg)\bigg]+\frac{3}{2\ve}\Big[8+\pi^2\Big]\nonumber\\
&\quad+2\pi^2\log\bigg(\frac{e^{\gamma_E}\Lambda}{a_*H}\bigg)+24-4\pi^2-18\zeta(3)\bigg\}\,,\\[.3cm]
&\bigg(\frac{e^{\gamma_E}\Lambda^2}{4\pi}\bigg)^{2\ve}\int\frac{\der^{d-1}l_1\der^{d-1}l_2}{(2\pi)^{2d-2}}\frac{1}{l^3_{1\Lambda}l_{2\Lambda}(l_{1\Lambda}+l_{2\Lambda})^2}=\frac{1}{32\pi^4}\bigg[\frac{1}{\ve^2}-\frac{2}{\ve}-4+\frac{7\pi^2}{6}\bigg]\,,\\[.3cm]
&\bigg(\frac{e^{\gamma_E}\Lambda^2}{4\pi}\bigg)^{2\ve}\int\frac{\der^{d-1}l_1\der^{d-1}l_2}{(2\pi)^{2d-2}}\frac{1}{l^3_{2\Lambda}(l^3_{1\Lambda}+l^3_{2\Lambda})}=\frac{1}{32\pi^4}\bigg[\frac{1}{\ve^2}+\frac{\pi^2}{6}\bigg]\,,\\[.3cm]
&\bigg(\frac{e^{\gamma_E}\Lambda^2}{4\pi}\bigg)^{2\ve}\int\frac{\der^{d-1}l_1\der^{d-1}l_2}{(2\pi)^{2d-2}}\frac{1}{l_{1\Lambda}l^2_{2\Lambda}(l^3_{1\Lambda}+l^3_{2\Lambda})}=\frac{1}{192\pi^4}\bigg[\frac{8\pi}{\sqrt{3}\ve}+\frac{16\pi}{\sqrt{3}}-\frac{23\pi^2}{9}-4\sqrt{3}\pi\log(3)\nonumber\\
&\quad-\frac{2}{3}\psi^{(1)}\bigg(\frac{1}{6}\bigg)+\frac{4}{3}\psi^{(1)}\bigg(\frac{1}{3}\bigg)\bigg]\,,
\end{flalign}
Here $\psi^{(1)}(z)$ denotes the polygamma function of order one. 
The Mellin-Barnes integrals are evaluated as described at the end of App.~\ref{app::22q0loops}. 

\bibliography{Bibliography}{}
\end{document}